\documentclass[twocolumn,floatfix]{revtex4-2}
\usepackage{amsfonts}
\usepackage{hyperref}
\usepackage{mathtools}
\usepackage{multirow}
\usepackage{graphicx}
\usepackage[format=hang,justification=centerlast]{caption}
\usepackage{subcaption}
\usepackage{dcolumn}
\usepackage{amsmath} 
\usepackage{mathrsfs}
 
\begin{document}
\preprint{APS/123-QED}
\title{Constructing maximal extensions of the Vaidya metric in Israel coordinates:\texorpdfstring{\\}{\textbackslash}II. The completeness of  Israel coordinates}
\author{Sheref Nasereldin}
 \email{16sna1@queensu.ca}
\author{Kayll Lake}
 \email{lakek@queensu.ca}
\affiliation{
 Department of Physics, Queen's University, Kingston, Ontario, Canada, K7L3N6
}
\date{\today}
\begin{abstract}
We present the results of an analysis of three maximal extensions of the Vaidya metric in  Israel coordinates, a spherically symmetric solution to the Einstein field equations for the energy momentum tensor of pure radiation in the high-frequency approximation. This metric is necessary for various applications, such as describing the exterior geometry of a radiating star in astrophysics and studying possible formation of naked singularities in the geometry of spacetime. Contrary to the common Eddington-Finkelstein-Like (EFL) coordinates, these maximal extensions, in Israel coordinates, are complete and cover the entirety of the Vaidya manifold. We develop three mass functions, one for each extension, and consider the qualitative characteristics of the three mass models and the surfaces of constant (dynamical) radius. We demonstrate that each maximal extension is null geodesically complete, which we assess by solving the radial null geodesics equation and forming the Penrose conformal diagram for each extension.  
\end{abstract}
\maketitle
\section{Introduction}\label{Sec:Introduction}
The Vaidya metric \cite{Vaidya_I,Vaidya_II,Vaidya_III} was initially introduced as a generalization of the Schwarzschild vacuum metric \cite{Schwarzschild1916} which accommodates a varying mass function. This metric has been widely utilized for classically studying the geometry around spherically symmetric stellar objects when radiation is effective, e.g. \cite{PhysRevD.31.233,Castagnino_Umerez,PhysRevD.25.2527,PhysRevD.22.2305,PhysRevD.19.2838, Hamity_Gleiser,PhysRevD.24.3019,adams1994analytic,1988MNRAS.231.1011D,PhysRevD.70.084004,PhysRevD.74.044001}. In a semi-classical context, the metric is indispensable for research into black holes evaporation through Hawking's radiation \cite{Hawking_BH_evaporating}, as evidenced by \cite{1982NCimB..70..201B,Kuroda,zhen_Zhao,beciu1984evaporating,PhysRevD.23.2823,Sung_Won,PhysRevD.23.2813,PhysRevD.41.1356,PhysRevD.63.041503,2006GReGr..38..425F,2013IJMPA..2850050K,PhysRevD.91.044020}. The EFL coordinates, frequently used to express the Vaidya metric, were criticized by Lindquist and Israel \cite{Lindquist1965,Israel1967} for having only partial coverage of the Vaidya manifold. However, they did not provide a proof of such incompleteness problem. The proof is introduced in \cite{paper1}.

The most successful attempt to find maximal analytical coverings of the Vaidya manifold, apart from Israel coordinates \cite{Israel1967}, was demonstrated in \cite{Vaidya_DoubleNull}. Waugh and Lake were able to construct double null covering $(u,v,\theta,\phi)$, where both  $u$ and $v$ are null coordinates, of the Vaidya manifold. However, the spacetime line element is only available when the mass function is linear. 
The Vaidya metric in double-null coordinates \cite{Vaidya_DoubleNull} has been used to study Quasi-Normal Modes (QNMs) in order to gain deeper insights into the gravitational excitations of black holes (see \cite{abdalla2006quasinormal} for an example). That said, the mass models used in this study were semi-analytical. It will be shown shortly that the mass function appearing in the Vaidya metric in Israel coordinates, as reexamined in \cite{paper1}, does not require a linear form, thus rendering the metric explicit for general cases. This will be useful in revisiting the applications of the Vaidya metric, as it removes the need to restrict ourselves to any particular ``linear" mass model, or to rely on not explicit mass functions. The Vaidya line element in Israel coordinates is given by
\begin{equation}
\begin{split}\label{Israel_ext_general}
     ds^{2} =& \left(\frac{w^2}{2m(u)r(u,w)}+\frac{4hm^{'}(u)}{U(u)}\right) du^2+2hdudw\\&+r(u,w)^{2}\left(d\theta^2+\sin^{2}\theta d\phi^{2}\right),
\end{split}
\end{equation}
where $U(u) = \int_{0}^{u} \frac{hdx}{4m(x)} du$, $ r(u,w) = U(u)w+2m(u)$, and $m(u)$ is always positive. This is a spherically symmetric solution to the Einstein field equations for the energy momentum tensor of pure radiation given in the eikonal form. This models a unidirectional radial flow of unpolarized radiation, 
\begin{equation}\label{EMT}
\begin{split}
T^{\alpha\beta} &= \Phi k^{\alpha}k^{\beta}\\
    &= \frac{1}{8\pi}\left(\frac{2hm^{'}(u)}{U(u)r(u,w)^2}\right)\delta^{\alpha}_{w} \delta^{\beta}_{w},
\end{split}
\end{equation}
with the radiation directed along $u=\text{const}$, and tangents to radial null geodesics are affinely parametrized by $w$. In \cite{paper1},  Israel coordinates were algorithmically constructed by directly integrating the field equations and introducing appropriate choices for the characterizing functions that arise as a consequence of integrating the field equations. Additionally, we have resolved the ambiguity in the definition of the $U(u)$ function and have demonstrated that the sign of $h$ does not solely specify the direction of radiation as in the EFL coordinates. In this paper, we investigate three possible maximal extensions of the Israel metric given an explicit form of the mass function $m(u)$ that is well-defined for $u\in(-\infty,\infty)$. To achieve this, we specify the energy-momentum tensor in the region $u\geq0$, the unextended Vaidya metric, and the region $u<0$. The energy-momentum tensor depends on the definition of $\Phi$ (\ref{EMT}), which includes the first derivative of the mass function $m(u)$ and the function $U(u)$. Thus, the mass function $m(u)$ plays the most important role in determining the nature of any maximal extension to the Vaidya metric. For instance, if the mass function is chosen to be constant over some parts of its domain and then monotonically increasing for the rest, the resulting maximal extension has a particular interpretation. Conversely, if the mass function is monotonically decreasing for some parts and then becomes constant on the remaining parts, the resulting maximal extension will have a different interpretation from the former. Finally, if the mass function is not monotonic, the resulting maximal extension will have a different nature than the previous two. Thus, it  should be clear now that due to the freedom in choosing the mass model, the maximal extension of the Vaidya manifold, given in terms of  Israel coordinates, is destined to be not unique.
This paper is organised as follows. In the next section, we discuss the criteria that the mass function must fulfil in order for it to be considered as a valid candidate for a maximal extension of the Vaidya metric in Israel coordinates. In Section \ref{Sec: First Extension}, we undertake a comprehensive analysis of the first maximal extension; beginning with introducing a mass function and deriving the relevant $U(u)$ function, and studying both the causal nature of the surfaces of constant radius and the surfaces of dynamical radius. The same procedure is repeated for the second and third extensions in Sections \ref{Sec: Second Extension} and \ref{Sec: Third Extension} respectively. In Section \ref{Sec: Completeness}, we use specific instances of the mass functions and $U(u)$ to solve the radial null geodesic equations numerically within the three extensions. Through studying the behaviour of the radial null geodesics and constructing the Penrose diagrams, we demonstrate the completeness of Israel coordinates. 
The final section summarises the main results of the paper and considers potential applications of this work.
\section{The Mass Function}\label{Sec: Mass}
By recognizing that the mass function is the foundation on which maximal extensions are built, we will establish a set of requirements for the mass function that ensure that both the metric and the energy-momentum tensor are well-behaved at all points within the domain of the mass function. These requirements are:
\begin{enumerate} \label{mass_requirements}
\item $m(u)>0$
\item $m^{'}(0)=0$,
\item $m^{''}(0)=\text{const}$, where the constant does not have to be zero,
\item $m(u)$ of class $C^{2}$.
\end{enumerate}
The first requirement is reasonable since matter fields with negative mass functions violate one or more of the energy conditions. This can lead to inconsistencies in the solutions, as the Weak Energy Condition (WEC) is assumed to be satisfied. As for the second requirement, it is clear from the definition of $\Phi$, i.e., $8\pi \Phi = \frac{2hm^{'}(u)}{U(u)r(u,w)^2}>0$ that this quantity must always remain positive and finite. However, it should be noted that $\Phi|_{u=0}$ is not finite because $U(0) = 0$. To avoid any pathological behavior of this quantity, it is necessary to stipulate that $m^{'}(0) = 0$ \footnote{$\underset{u\rightarrow 0}{\lim} \frac{m^{'}(u)}{U(u)}$ can be proven to exist by using L'Hospitals rule.}. 
 There are three types of Vaidya models. In the first, we start with a Schwarzschild vacuum solution $\big( m^{'}(u)=0, m(u)=\text{const}=M_{0}>0~\text{for}~u<u_{1}<0\big)$ and add an outflux of radiation $\big(m^{'}(u)<0~\text{for}~u_{1}\leq u <0\big)$; this outflux is then terminated, resulting in a distinct Schwarzschild vacuum solution $\big(m^{'}(u)=0, m(u)=\text{const}=M_{1}>0~\text{for}~ u\geq0\big)$ whose mass is less than that of the initial Schwarzschild vacuum solution, $M_{0}$. In the second model, we start with a Schwarzschild vacuum solution $\big(m^{'}(u)=0, m(u)=\text{const}=M_{0}>0$ for $u\leq0 \big)$ and add an influx of radiation $\big(m^{'}(u)>0 ~\text{for}~ 0<u\leq u_{1}\big)$; after the influx is terminated, we end up with a different Schwarzschild vacuum solution $\big(m^{'}(u)=0, m(u)=\text{const}=M_{1}>M_{0} ~\text{for}~u>u_{1}\big)$. In the third model, we start with a Schwarzschild vacuum solution $\big( m^{'}(u)=0, m(u)=\text{const}=M_{0}>0~\text{for}~u<u_{1}<0\big)$ and add an outflux of radiation $\big(m^{'}(u)<0~\text{for}~u_{1}\leq u <0\big)$, and then we add an influx or radiation $\big(m^{'}(u)>0 ~\text{for}~ 0<u\leq u_{2}\big)$; after the influx is terminated, we end up with a different Schwarzschild vacuum solution $\big(m^{'}(u)=0, m(u)=\text{const}=M_{1} ~\text{for}~u>u_{2}\big)$.
\section{First Maximal Extension} \label{Sec: First Extension}
 In this section, we explore a mass function that leads to a maximal extension of the Israel metric. The mass function is monotonically decreasing for $u<0$ and constant for $u\geq0$, thereby defining an Israel metric that is irradiated with streams of outgoing radiation. Ultimately, this solution evolves to the Schwarzschild solution once the radiation is no longer present. We can conclude that outgoing radiation Israel solution represents a white hole, while the ingoing radiation Schwarzschild solution represents a black hole. This is shown below.
\subsection{ The Mass Function and the \texorpdfstring{$U(u)$}{LG} Function }
\begin{figure}[htb]
\includegraphics[width=\linewidth]{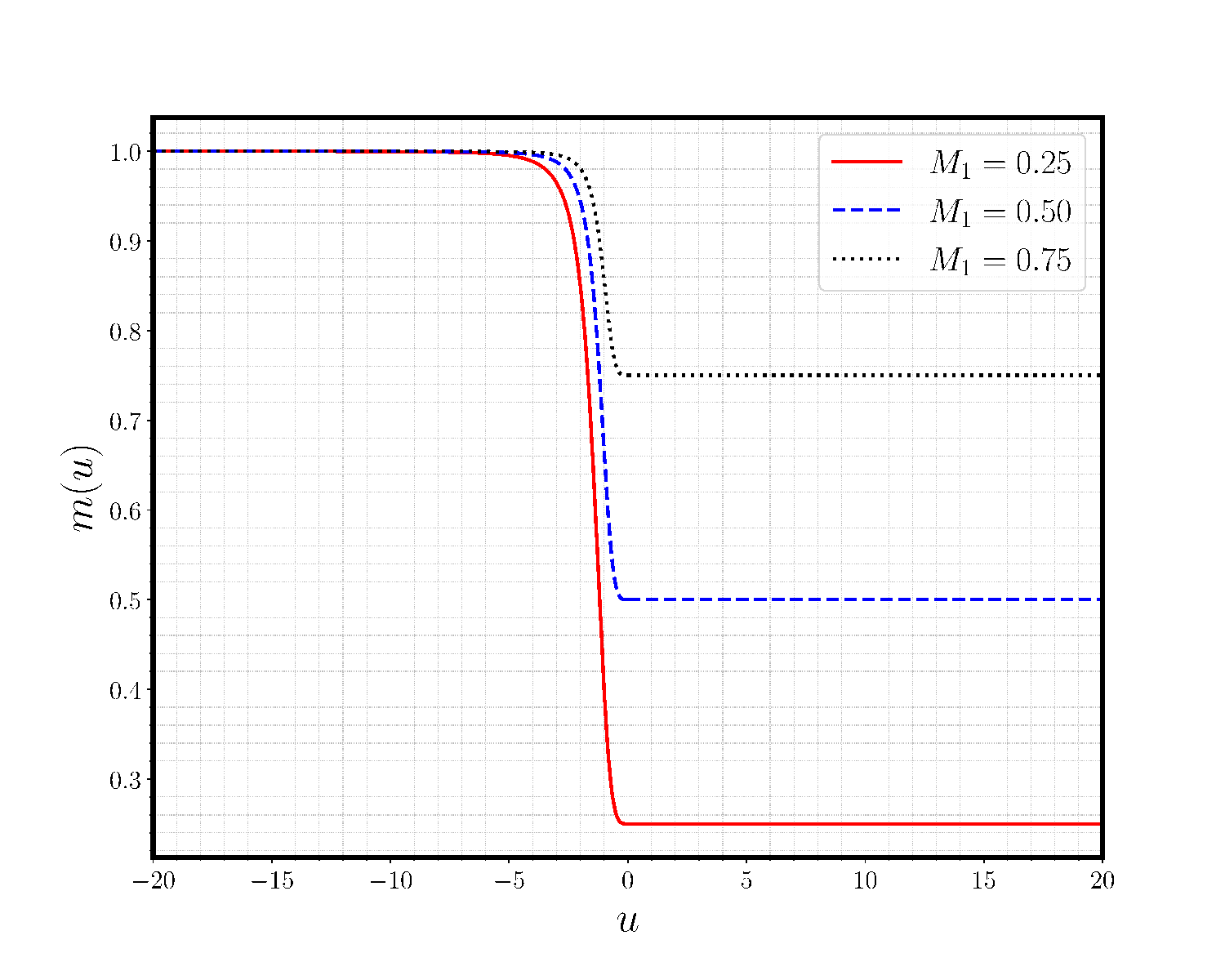}
\caption{ A plot of the mass function (\ref{mass_outgoing_F1}) that gives rise to the first maximal extension to the Vaidya metric. There are three values used for the parameter $M_{1}<M_{0}$, where $M_{0}$ is taken to equal unity.}
\label{Ext1_m1}
\end{figure}

\begin{figure}[htb]
  \begin{subfigure}{0.85\columnwidth}
    \includegraphics[width=\linewidth]{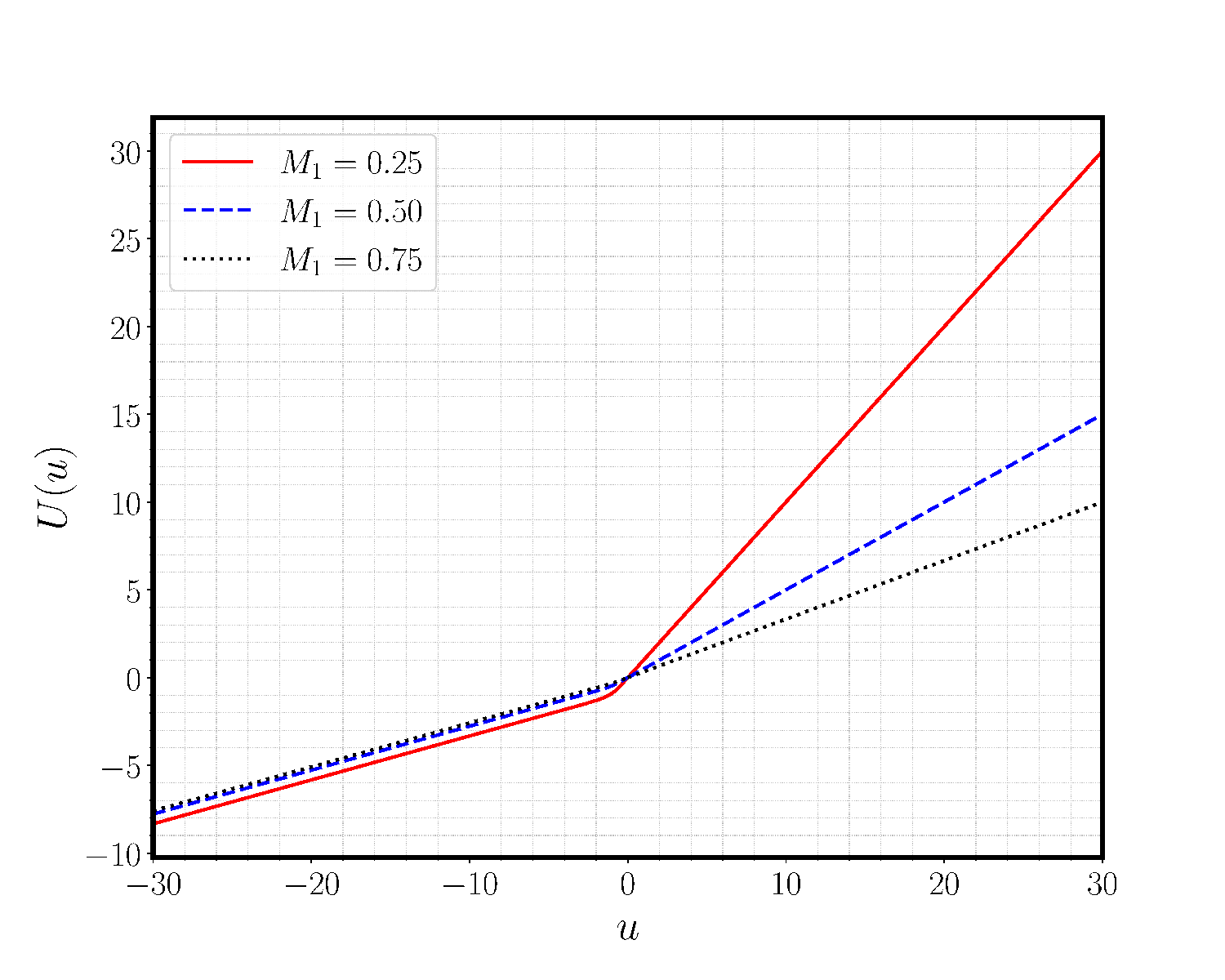}
    \caption{$h = +1$}
    \label{h_p}
  \end{subfigure}
  \begin{subfigure}{0.85\columnwidth}
    \includegraphics[width=\linewidth]{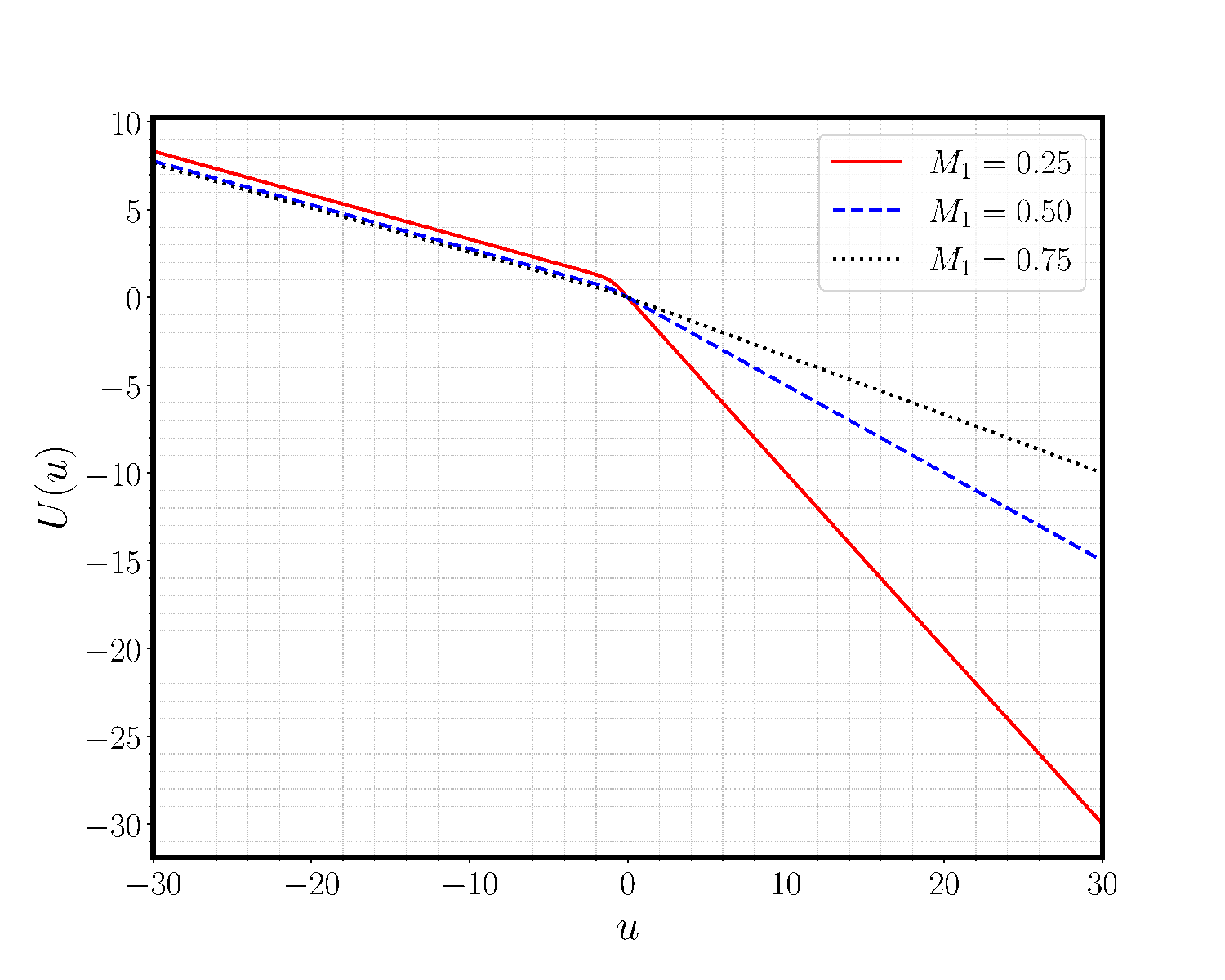}
    \caption{$h = -1$}
    \label{h_n}
  \end{subfigure}
  \caption{Two illustrations of the function $U(u)$ (\ref{Ex1_U1_eq}) for two different values of $h$ are presented.}
\label{Ext1_U_func_figs}
\end{figure}

One particular example of a mass function $m(u)$ that necessarily satisfies the previously discussed requirements may explicitly be written as
\begin{equation}
  \label{mass_outgoing_full}
   m(u)= \begin{dcases}
          M_{0}, &  u \leq u_{1} \\
         \frac{a_{n}u^{2n}+a_{0}}{b_{n}u^{2n}+b_{0}},& u_{1}\leq u < 0 \\
         M_{1}, & u\geq 0 \\
      \end{dcases}
\end{equation}
where $n\geq 2$ and $a_n, a_0, b_n$, and $b_0$ are positive real numbers. The mass function $m(u)$ is decreasing for all values $(u_1 \leq u \leq 0)$ and is constant ($M_0$ or $M_1$) otherwise. We can demand that the mass function $m(u)$ is only constant, $m(u) = M_0$, asymptotically \footnote{The primary motivation for this assumption is to simplify the analysis of the maximal extension that results. Consequently, if we adhere to (\ref{mass_outgoing_full}), we will need to analyze three solutions that comprise the maximal extension rather than two.}, we thus write the mass function in the following form:
\begin{equation}
  \label{mass_outgoing_reduced}
   m(u)= \begin{dcases}
         \frac{a_{n}u^{2n}+a_{0}}{b_{n}u^{2n}+b_{0}}, &  u < 0 \\
         M_{1}, & u\geq 0 \\
      \end{dcases}
\end{equation}
 where it should be noted that the previous form implies the following constraint on the choice of the parameters 
\begin{equation}\label{param_constr_I}
\lim_{u\rightarrow-\infty} m(u)=\frac{a_{n}}{b_{n}}=M_{0}.
\end{equation}
We can further benefit from the fourth requirement to impose additional restrictions on the selection of the parameters of the mass function
\begin{equation}
\label{param_constr_II}
\lim_{u\rightarrow 0^{-}}m(u)=\frac{a_0}{b_0} = m(0)=M_{1}<M_{0}. 
\end{equation}
Now that we have introduced an explicit form of the mass function (\ref{mass_outgoing_reduced}), we need to write the function $U(u)$ so that it becomes defined at every point of the manifold. An explicit expression that gives $U(u)$ for $u\in(-\infty,\infty)$ can now be provided,
 \begin{equation} \label{U_ext1_expression}
     U(u)=\begin{dcases}
      h\int_{0}^{u<0}\frac{dx}{4\left(\frac{a_{n}x^{2n}+a_{0}}{b_{n}x^{2n}+b_{0}}\right)}, &  u < 0 \\
      h\int_{0}^{u>0} \frac{dx}{4 M_{1}}. & u\geq 0 \\
     \end{dcases}
 \end{equation}
 The constraints (\ref{param_constr_I}) and (\ref{param_constr_II}) can be utilized to introduce a more specific form of the mass function for later integration of the radial null geodesics equation, as will be seen in Section \ref{Sec: Completeness}. 
 In accordance with the previously established requirements on the mass function, we consider the following values for the parameters: $n=2$, $a_{2} = a_{0} = b_{2} = M_{1}$, and $b_{0} = 1$. The mass function,(\ref{mass_outgoing_reduced}), then becomes
\begin{equation}\label{mass_outgoing_F1}
m(u)= \begin{dcases}
         \frac{M_{1}\left(u^{4}+1\right)}{M_{1}u^{4}+1}, &  u < 0 \\
         M_{1}. & u\geq 0 \\
\end{dcases}
\end{equation}
It can be seen that, in accordance with the requirement that the mass function be asymptotic to Schwarzschild's mass in the past, the mass function (\ref{mass_outgoing_F1}) has a horizontal asymptote of $m(u)=1$ as $u\rightarrow -\infty$. Over a certain section of its domain, the function slope turns negative (as seen in Fig. \ref{Ext1_m1}), thereby giving rise to the outgoing Israel metric. Then, the function remains constant for the remainder of its domain (resulting in another Schwarzschild vacuum solution with a distinct mass, i.e., $M_1$).
The function $U(u)$, with the aid Table. II in \cite{paper1}, can be written as 
\begin{equation}
\label{Ex1_U1_eq}
\scriptstyle U(u) = \begin{dcases}
\scriptstyle -h \left(\frac{u}{4}+\frac{\sqrt{2}\left(M_1-1\right)}{16 M_1} \left(-\tanh ^{-1} \Gamma -\tan^{-1}\Delta+\tan ^{-1} \Psi\right)\right),  & \scriptstyle u < 0 \\
 \scriptstyle h \left(\frac{u}{4M_{1}}\right),& \scriptstyle u\geq 0 \\
\end{dcases}
\end{equation}
where $\Gamma = \frac{\sqrt{2} u}{u^2+1}$, $\Delta = \sqrt{2} u+1$, and $\Psi = 1-\sqrt{2} u$.
In order to obtain the previous equation, we have substituted (\ref{mass_outgoing_F1}) in (\ref{U_ext1_expression}), and $h$ can either be $1$ or $-1$. The graphs of the function $U(u)$, see Figs. \ref{Ext1_U_func_figs}, show that for the choice $h = +1$, $U(u<0)<0$ and $U(u>0)>0$; this is reversed if $h = -1$ is chosen. Moreover, a reflection on (\ref{Ex1_U1_eq}) and inspection of Figs. \ref{h_p} and \ref{h_n} shows that as the parameter $M_{1}$ approaches unity or the coordinate $u$ decreases significantly, the linear term, $\frac{u}{4}$, dominates over the terms $\text{arctan}\left(1 \pm \sqrt{2} u\right)$ and $\text{arctanh} \left( \frac{\sqrt{2} u}{u^2+1}\right)$, making $U(u<<0)$ linear. 
It is noteworthy that, due to the assumption that the mass function $m(u)$ is asymptotically constant in the past, see (\ref{param_constr_I}), the total flux 
\begin{equation}
    \int_{-\infty}^{0} \Phi du = \frac{1}{4\pi}\int_{-\infty}^{0}\frac{hm^{'}(u)}{U(u)\big(U(u)w+2m(u)\big)^2}du,
\end{equation}
is not finite. This will also be the case in the subsequent two extensions. The only way to save the flux from diverging is by relaxing the condition that the mass function is only constant asymptotically and thus the previous integration can be rewritten as 
\begin{align}
    \int_{-\infty}^{0} \Phi du &=\int_{-\infty}^{u^{*}} (0)du+\frac{1}{4\pi}\int_{u^{*}}^{0}\frac{hm^{'}(u)du}{U(u)\big(U(u)w+2m(u)\big)^2},\nonumber\\ &= \frac{1}{4\pi}\int_{u^{*}}^{0}\frac{hm^{'}(u)}{U(u)\big(U(u)w+2m(u)\big)^2}du,
\end{align}
where $-\infty<u^{*}<0$.
\subsection{Surfaces of Constant Radius}\label{Surfaces_const_radii}
Now that we have constructed explicit expressions for both the mass function $m(u)$ and the function $U(u)$, we can analyze various features related to the surfaces $r(u,w) = \text{const} = \alpha M_{1}$ with $\alpha \in \mathbb{N}$, illustrated in Fig. \ref{Ext1_1_ConstR} below. Notably, these surfaces can be mathematically expressed as
\begin{equation}
\label{Ext1_1_ConR_eq}
    w= \begin{dcases}
         \frac{\alpha M_{1}-2m(u)}{U(u)}, &  u < 0 \\
         \frac{4(\alpha-2)M_1^2}{hu}. & u\geq 0 \\
      \end{dcases}
\end{equation}
The exploration of the causal nature of (\ref{Ext1_1_ConR_eq}) may be achieved by studying the associated Lagrangian, confining attention only to the submanifold $\theta=\phi=\text{const}$, of (\ref{Israel_ext_general})
\begin{equation}
\label{Ext1_Lagrangian}
    2\mathscr{L}= \begin{dcases}
         \frac{-w}{\alpha M_{1} U(u)} \dot{u}^2, &  u < 0 \\
         \frac{16(2-\alpha)M_{1}^2}{\alpha h^2 u^2}\dot{u}^2. & u\geq 0 \\
      \end{dcases}
\end{equation}
Thus, when $u\geq 0$, for $\alpha<2$ the surfaces of constant radius are spacelike; for $\alpha=2$ they are null; and for $\alpha>2$ they are timelike. For $u<0$, depending on the sign of $w$ and the sign of $U(u)$, the surfaces of constant radius in quadrant II \footnote{In this paper, the $u-w$ plane is divided into four quadrants: quadrant I ($u>0$ and $w>0$), quadrant II ($u<0$ and $w>0$), quadrant III ($u<0$ and $w<0$), and quadrant IV ($u>0$ and $w<0$).}  ($u<0$ and $w>0$) are spacelike, and those in the quadrant III ($u<0$ and $w<0$) are timelike, with $h = +1$.  However, if  $h=-1$, the second quadrant surfaces are timelike and the third quadrant ones are spacelike. As the Lagrangian (\ref{Ext1_Lagrangian}) is negative when $w>0$ and positive for $w<0$, these surfaces can be identified as null when they cross the $u$-axis ($w=0$), although this is only true when $m(u)=\text{const.}$
\begin{figure*}[htb]
  \centering
  \subcaptionbox{$M_1 = 0.25$}[.32\linewidth][c]{%
    \includegraphics[width=1\linewidth]{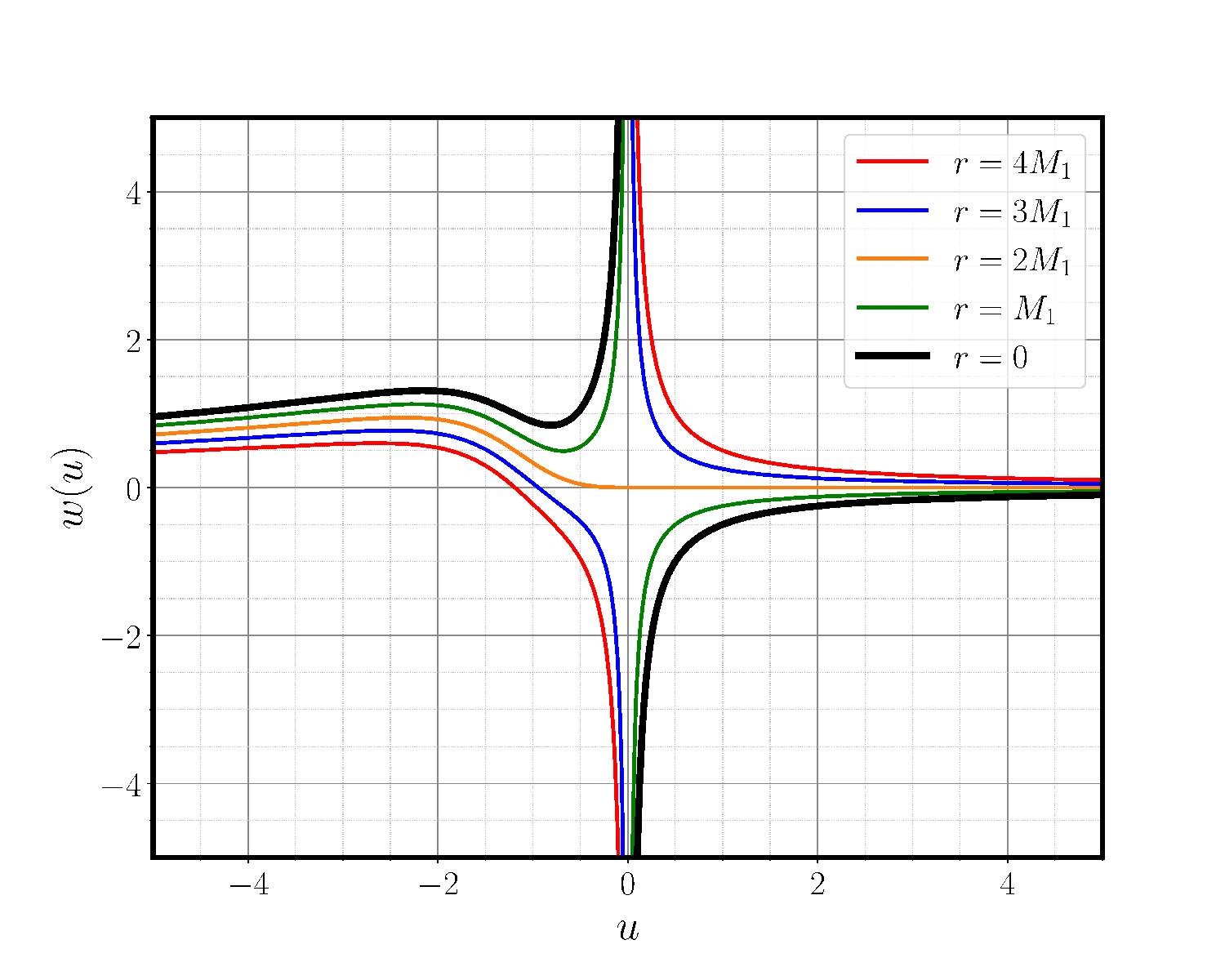}}\quad
  \subcaptionbox{$M_1 = 0.5$}[.32\linewidth][c]{%
    \includegraphics[width=1\linewidth]{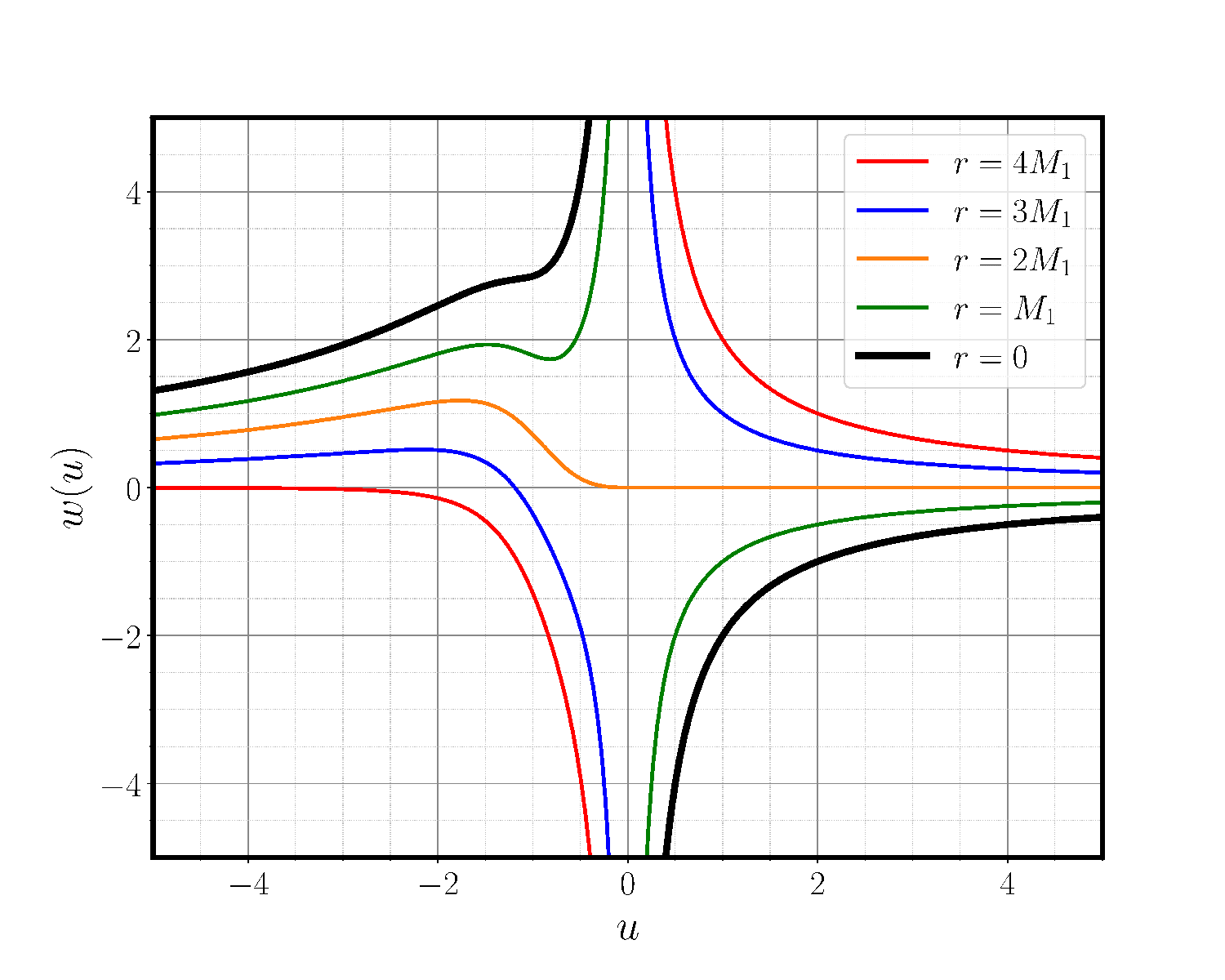}}\quad
  \subcaptionbox{$M_1 = 0.75$}[.32\linewidth][c]{%
    \includegraphics[width=1\linewidth]{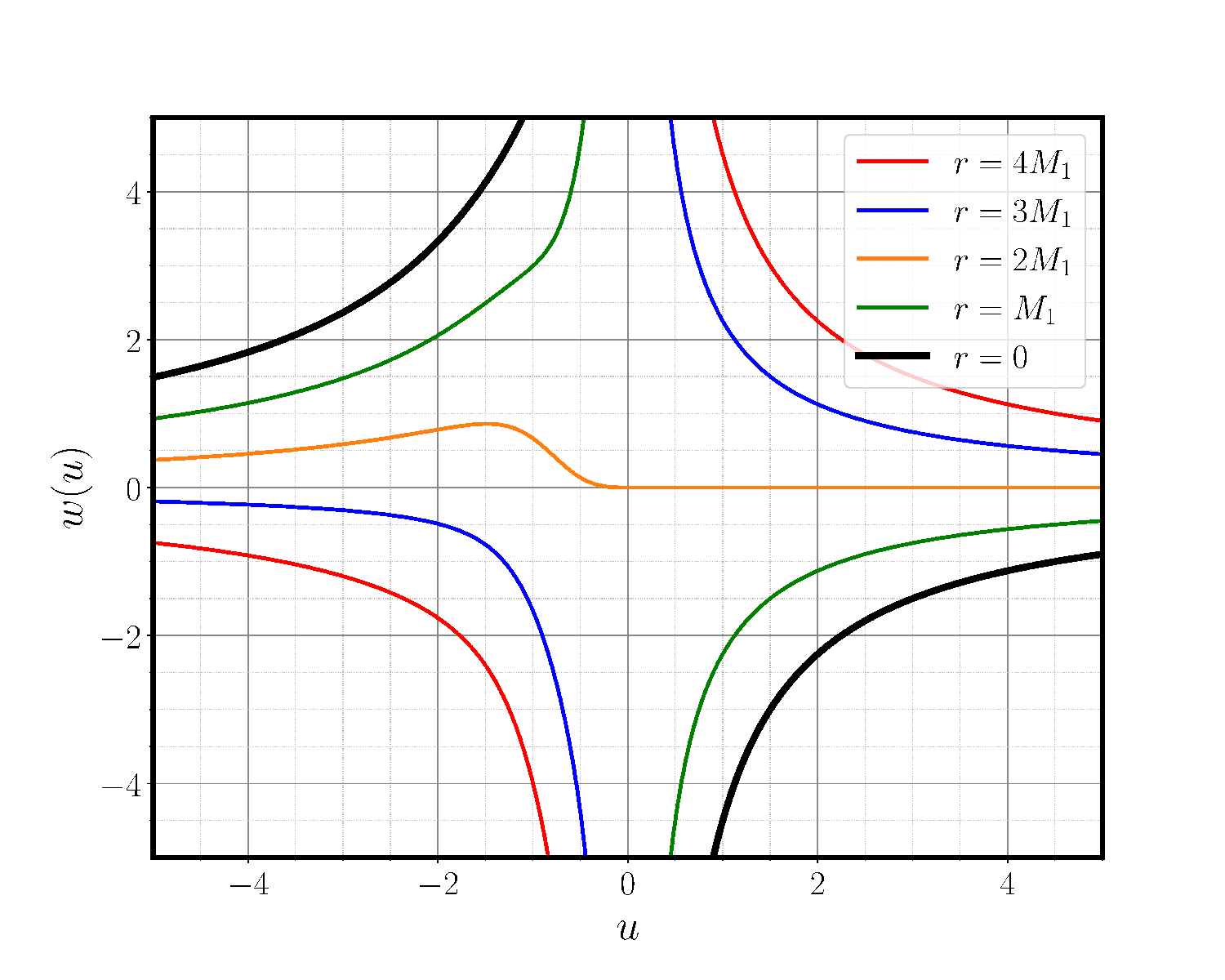}}
  \subcaptionbox{$M_1 = 0.25$}[.32\linewidth][c]{%
    \includegraphics[width=1\linewidth]{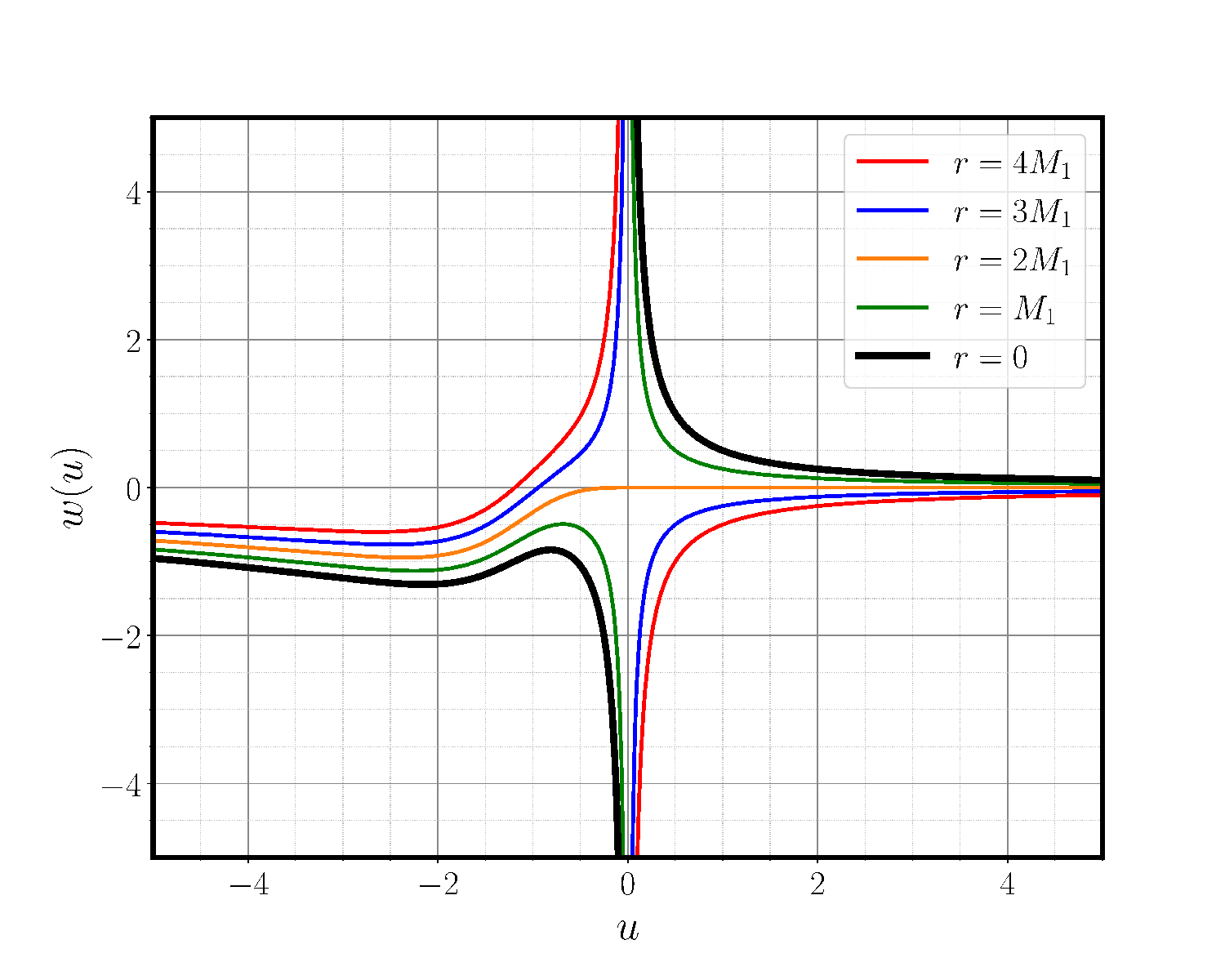}}\quad
  \subcaptionbox{$M_1 = 0.5$}[.32\linewidth][c]{%
    \includegraphics[width=1\linewidth]{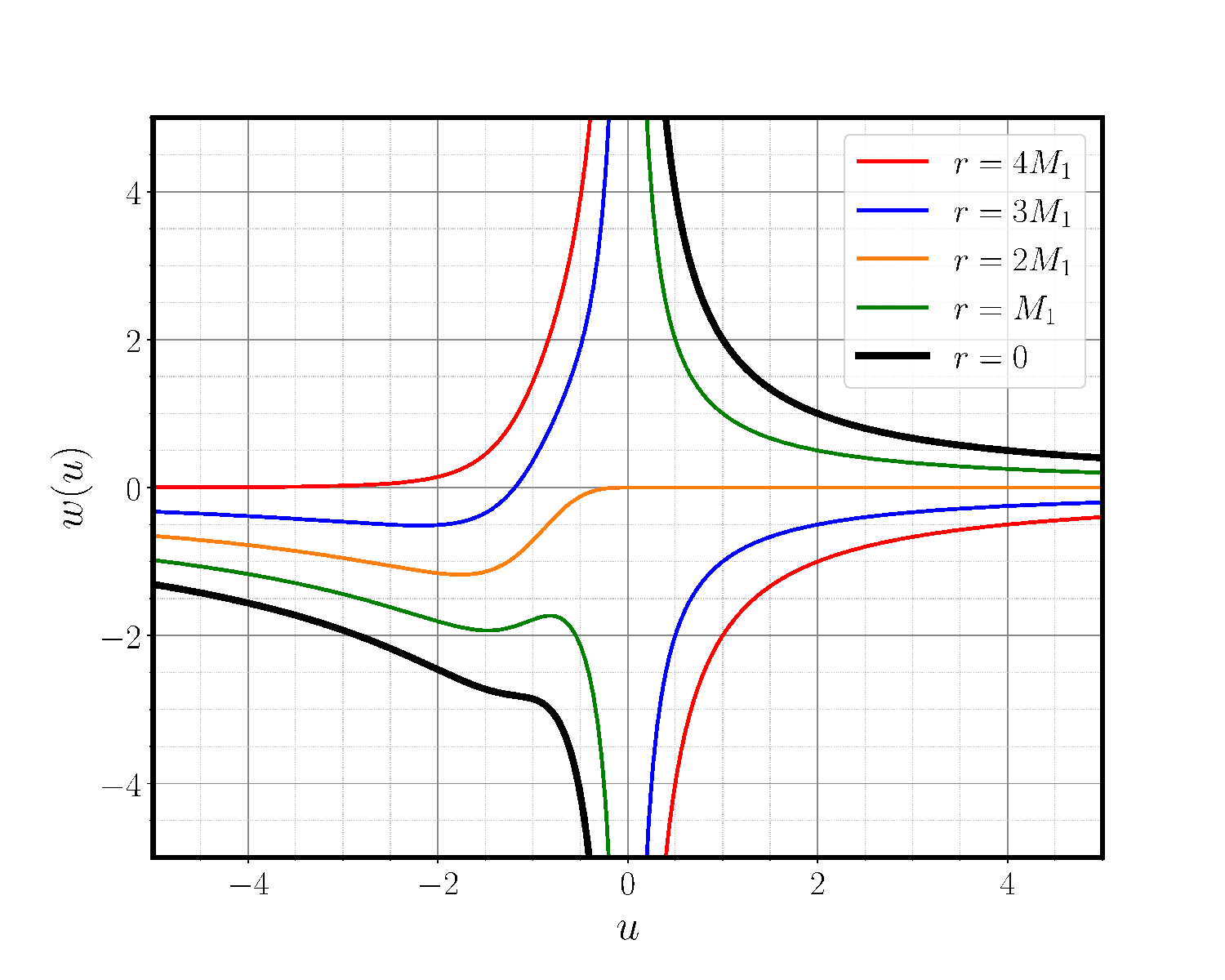}}\quad
  \subcaptionbox{$M_1 = 0.75$}[.32\linewidth][c]{%
    \includegraphics[width=\linewidth]{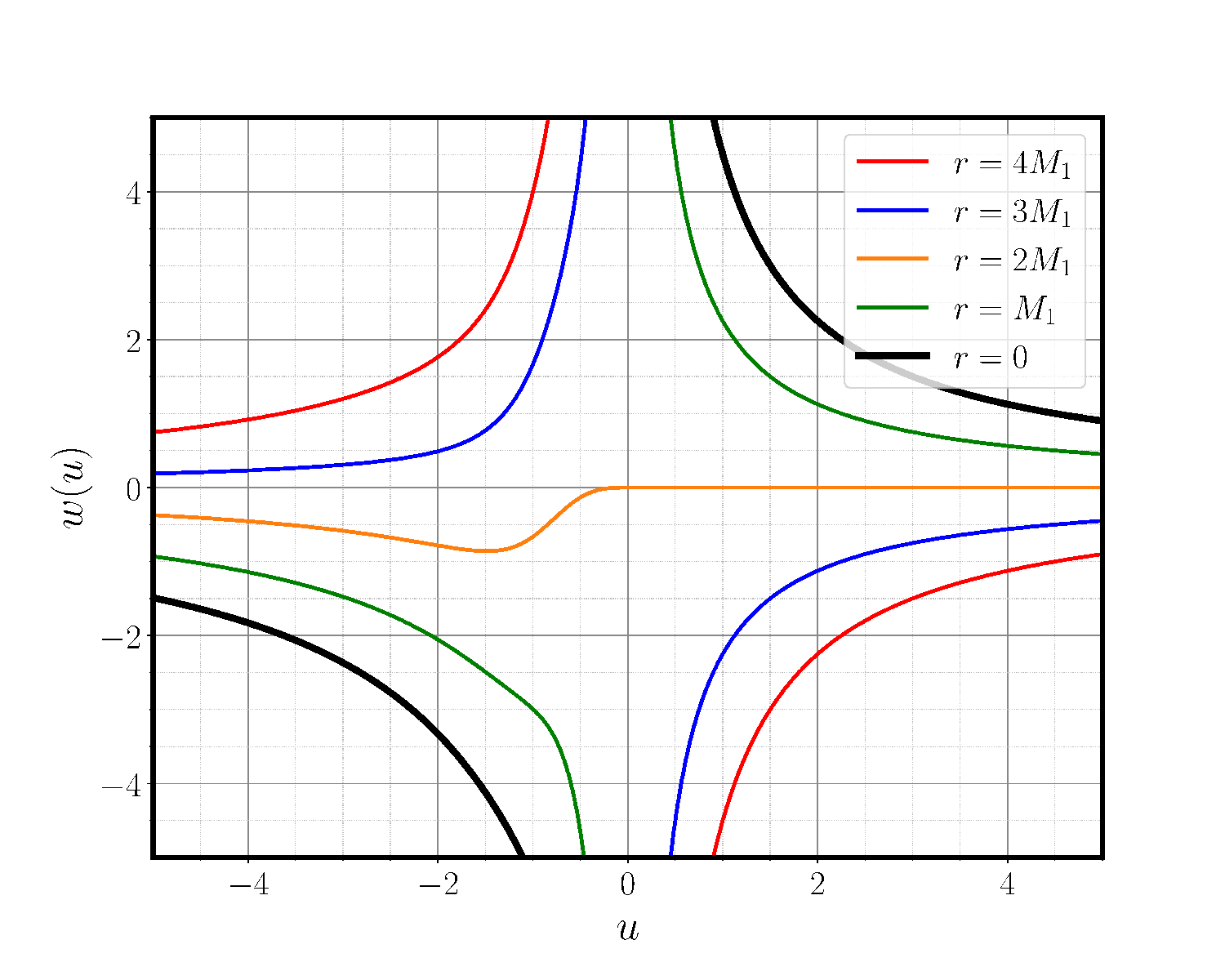}}
  \caption{The top row displays surfaces of constant radius for the choice of $h = +1$, while the bottom row displays the same for $h = -1$. Increasing the value of $M_1$ leads to a less pronounced bulge, while switching the sign of $h$ creates a mirror reflection of the surfaces about the horizontal axis.}
  \label{Ext1_1_ConstR}
\end{figure*}
 It is clear that surfaces computed by the second branch of (\ref{Ext1_1_ConR_eq}) are hyperbolas reminiscent of those in the Schwarzschild vacuum solution (in Kruskal \cite{Kruskal1960} - Szekeres \cite{Szekeres1960} coordinates). However, the situation for the remaining part ($u<0$) is quite different as the relation between the coordinates $u$ and $w$ is not that simple, which results in the appearance of ``bulges" in the surfaces. The bulges are mathematically found by solving the equation, 
 \begin{equation}\label{roots_bulges_ext1}
     0 = \frac{dw}{du} = \frac{-2m^{'}(u)U(u)-\frac{h}{4m(u)}\big(\alpha M_{1}-2m(u)\big)}{U(u)^2}.
 \end{equation}
 Instead of finding the exact numerical values of the roots of (\ref{roots_bulges_ext1}), we graphically, see Fig. \ref{bulges} below, show the approximate locations of the bulges. We note that the surfaces of constant radius $\big(0\leq r(u,w)\leq2M_1\big)$ have two bulges (two roots ), whereas the surfaces $r(u,w)>2M_{1}$ only have one bulge (one root). We also note that with substituting (\ref{Ext1_1_ConR_eq}) in (\ref{roots_bulges_ext1}), we obtain 
 \begin{equation}
     w = -8\frac{m(u)m^{'}(u)}{h},
 \end{equation}
 which gives the $w$-coordinate of the bulge. 
\begin{figure}[htb]
\centering
\includegraphics [width=1\linewidth]{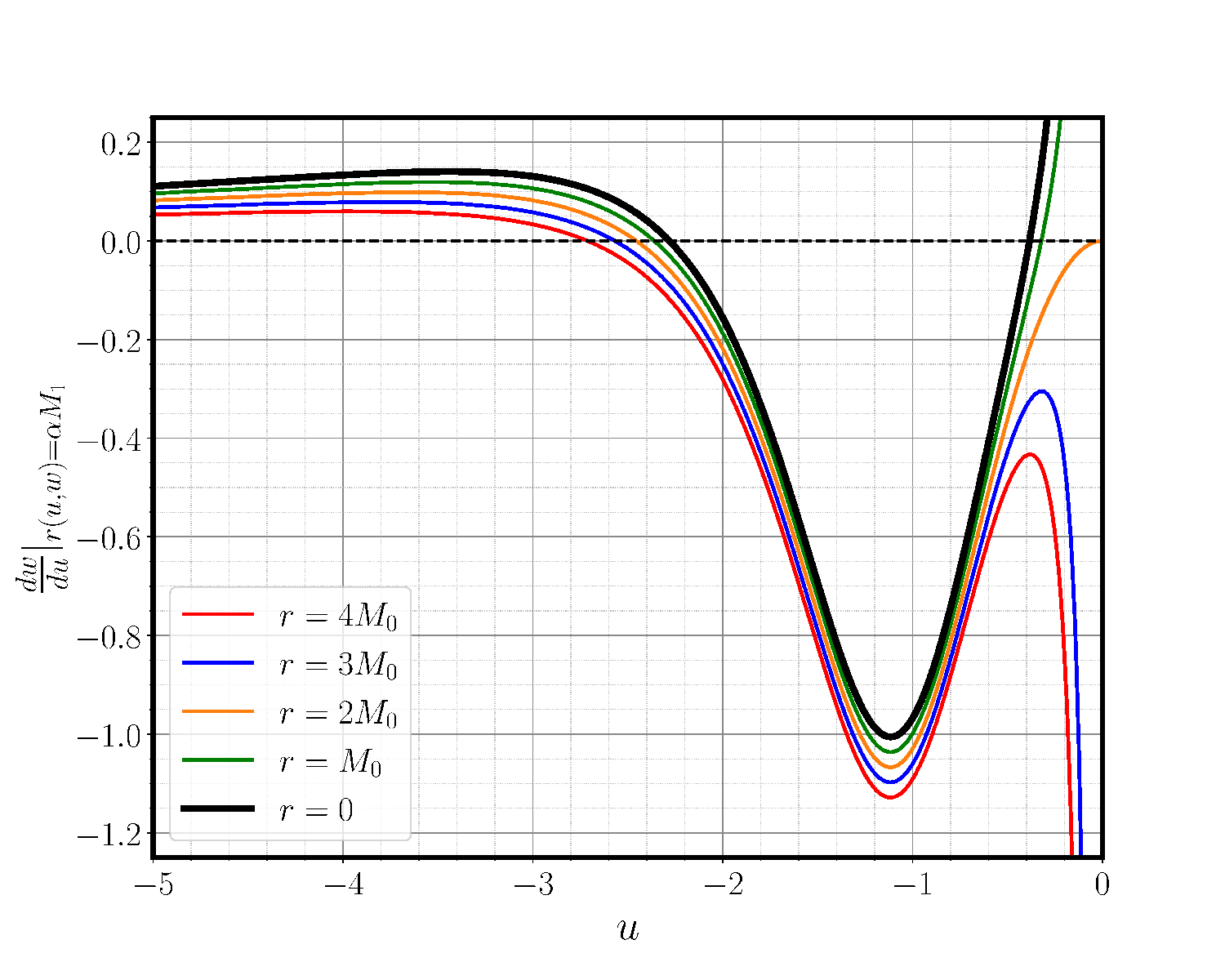}
\caption{The graphical locations of the roots of (\ref{roots_bulges_ext1}) for the choices $h=+1$ and $M_{1} = 0.25$.}
\label{bulges}
\end{figure}
 It is worthy of notice that the surfaces of constant radius tend to eliminate the bulges as $M_1$ increases towards unity (when $M_{1} = 1$ we recover the Schwarzschild vacuum solution). In general, we have demonstrated that, by identifying the sign of the corresponding Lagrangian, the surfaces $r(u,w) = \alpha M_{1}$, for the choice $h = +1(-1)$, are time(space) like in quadrants I and III, and space(time) like in quadrants II and IV. Finally, we want to draw attention to the impact of changing the sign of $h$ on the function $U(u)$ and the surfaces of constant radius. As can be seen from the diagrams (e.g., see Figs. \ref{Ext1_U_func_figs} and \ref{Ext1_1_ConstR}), a change in the sign of $h$ simply produces a mirror reflection of the same graph. Therefore, it can be concluded that the sign of $h$, unlike the EFL coordinates, is equivalent to selecting the time direction and does not introduce any new physics. This will remain true in the following two extensions without any changes to this explanation.
\subsection{Surfaces of Dynamical Radius}
\begin{figure*}[htb]
  \centering
  \subcaptionbox{$M_1 = 0.25$}[.32\linewidth][c]{%
    \includegraphics[width=1\linewidth]{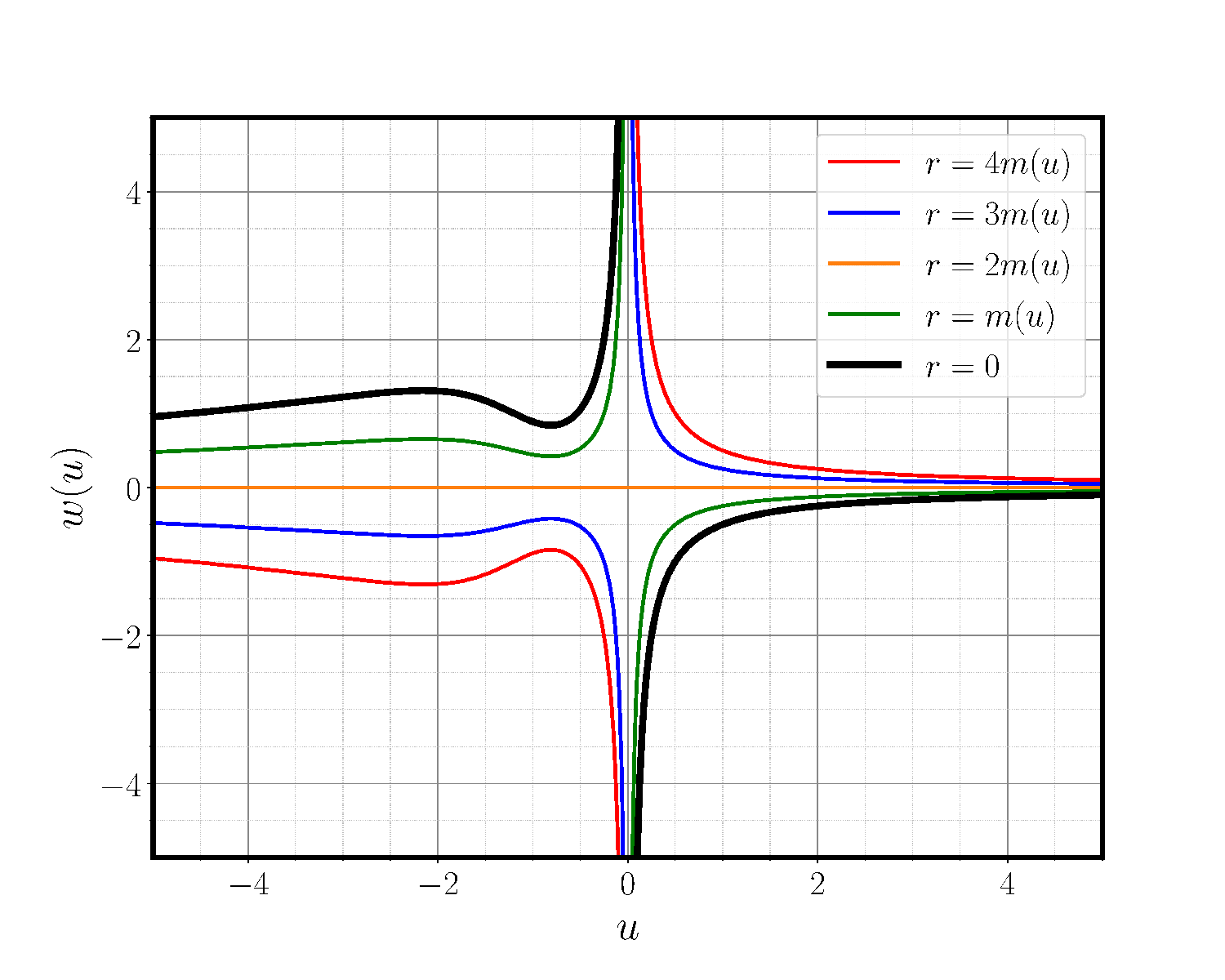}}\quad
  \subcaptionbox{$M_1 = 0.5$}[.32\linewidth][c]{%
    \includegraphics[width=1\linewidth]{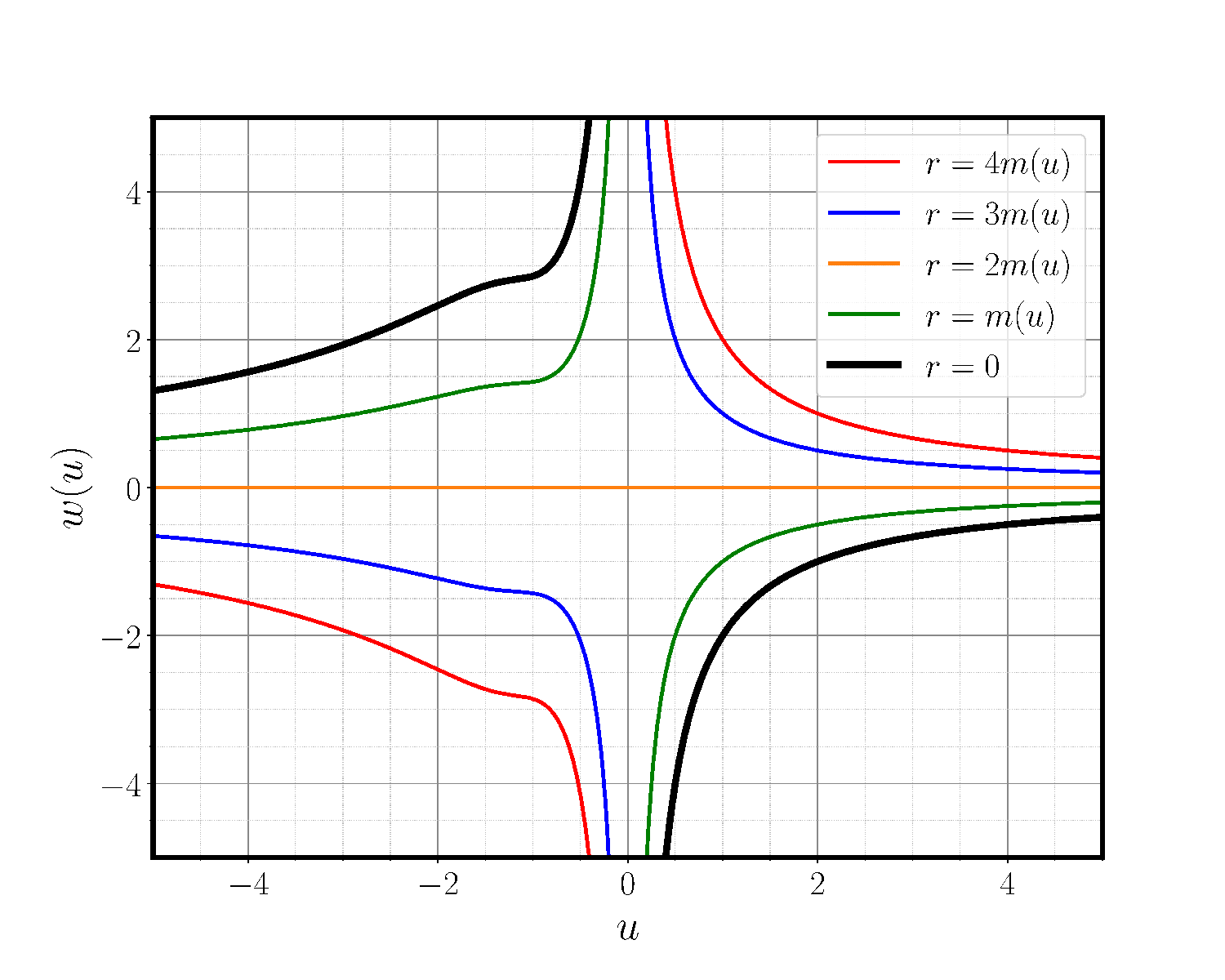}}\quad
  \subcaptionbox{$M_1 = 0.75$}[.32\linewidth][c]{%
    \includegraphics[width=1\linewidth]{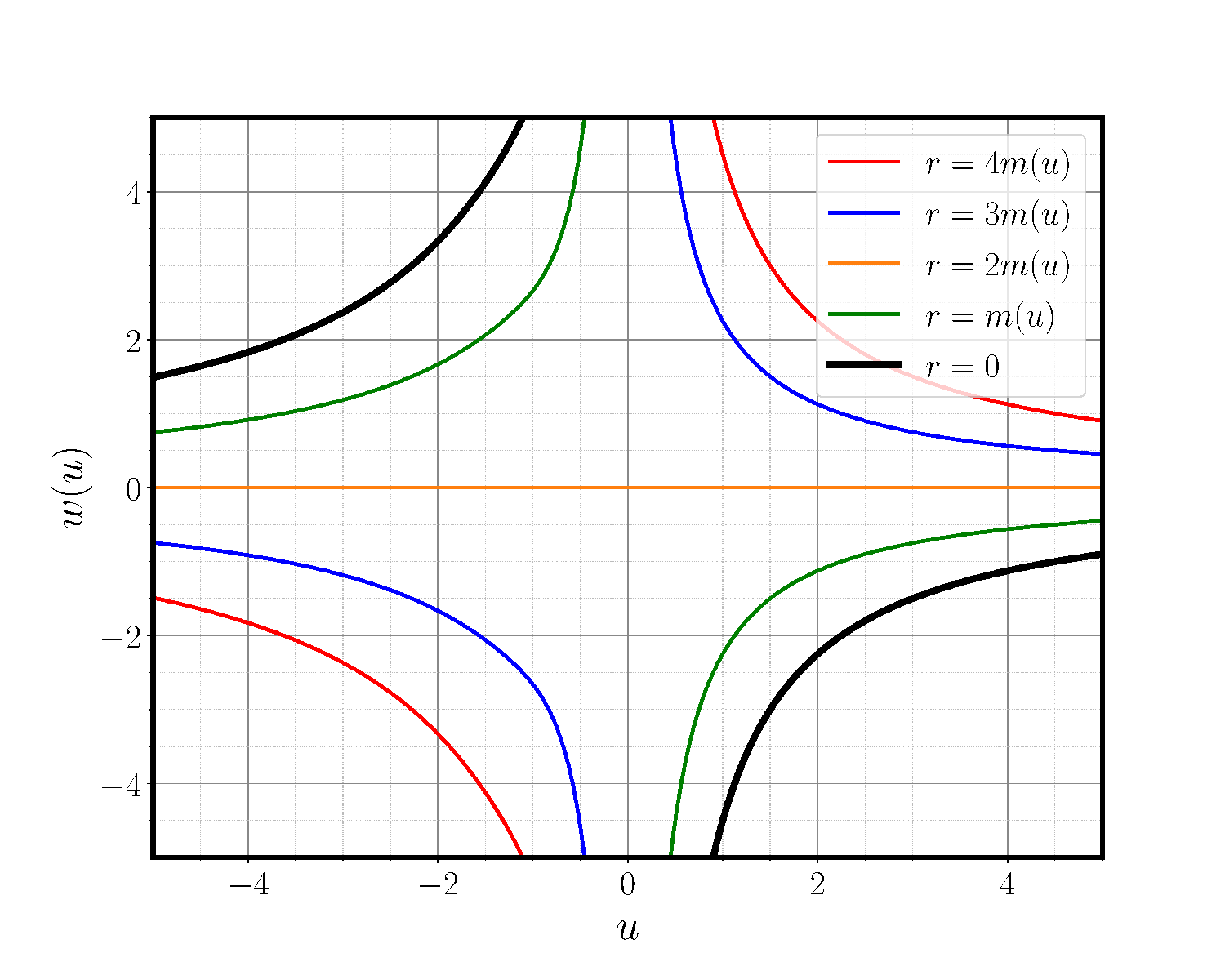}}
  \subcaptionbox{$M_1 = 0.25$}[.32\linewidth][c]{%
    \includegraphics[width=1\linewidth]{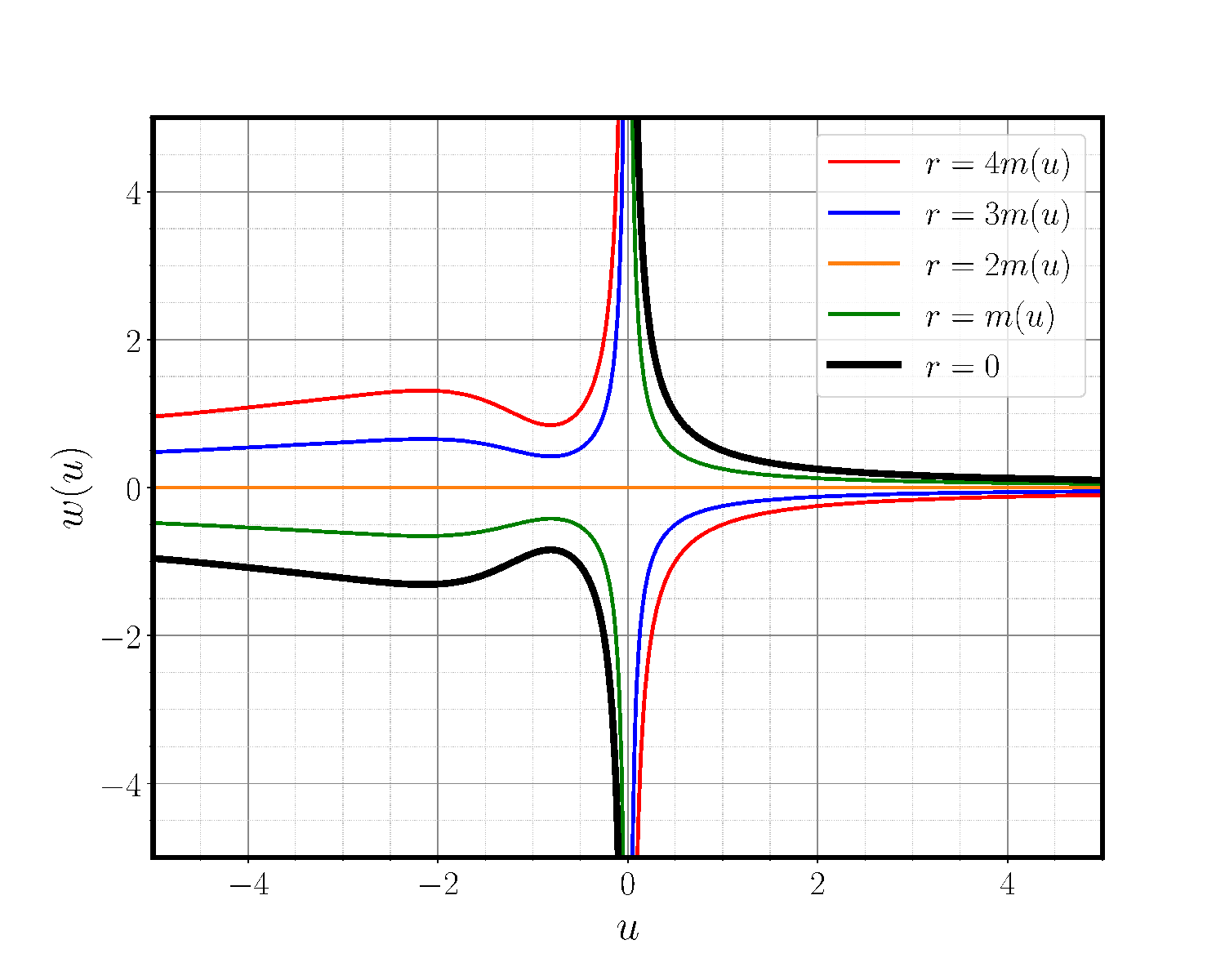}}\quad
  \subcaptionbox{$M_1 = 0.5$}[.32\linewidth][c]{%
    \includegraphics[width=1\linewidth]{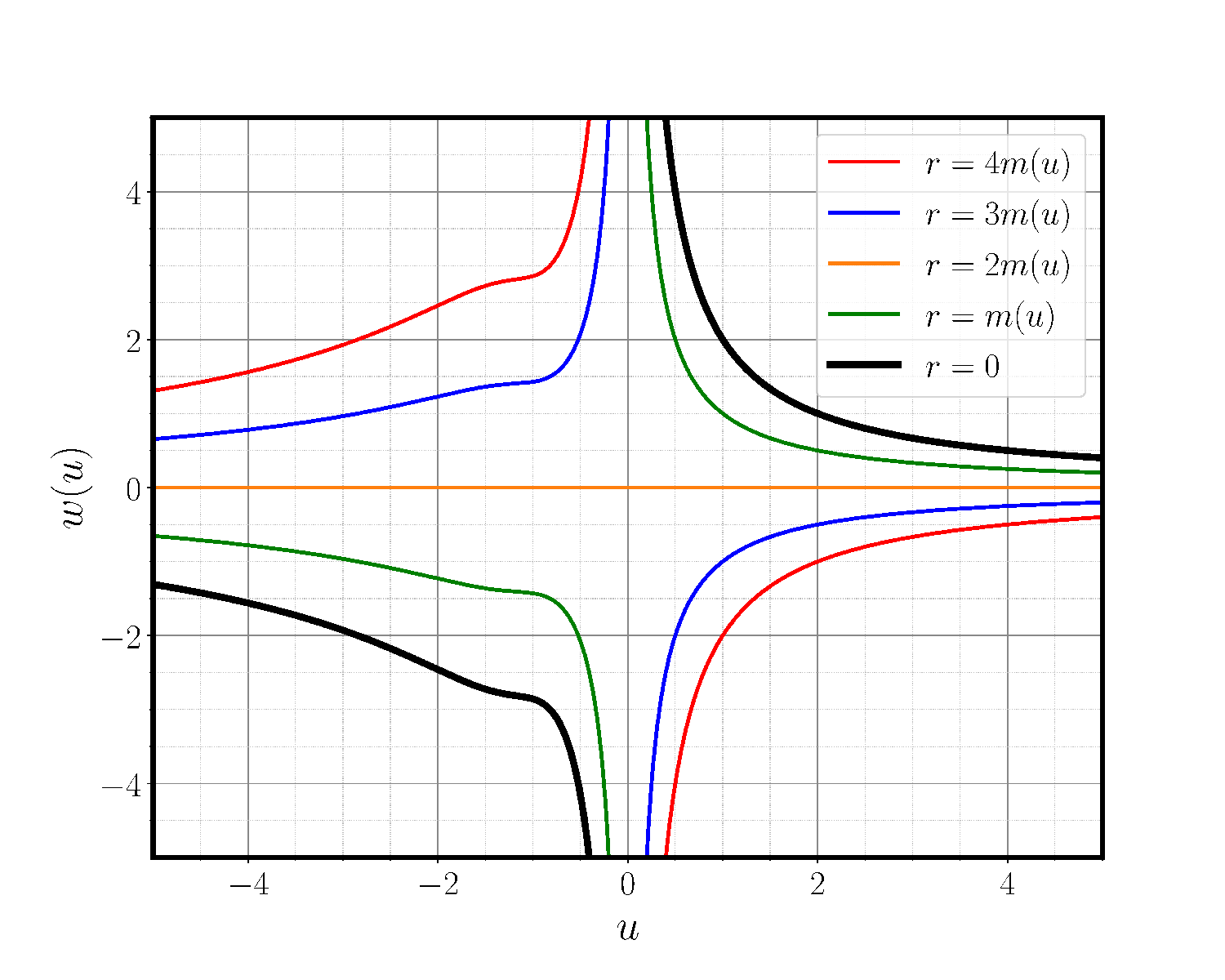}}\quad
  \subcaptionbox{$M_1 = 0.75$}[.32\linewidth][c]{%
    \includegraphics[width=1\linewidth]{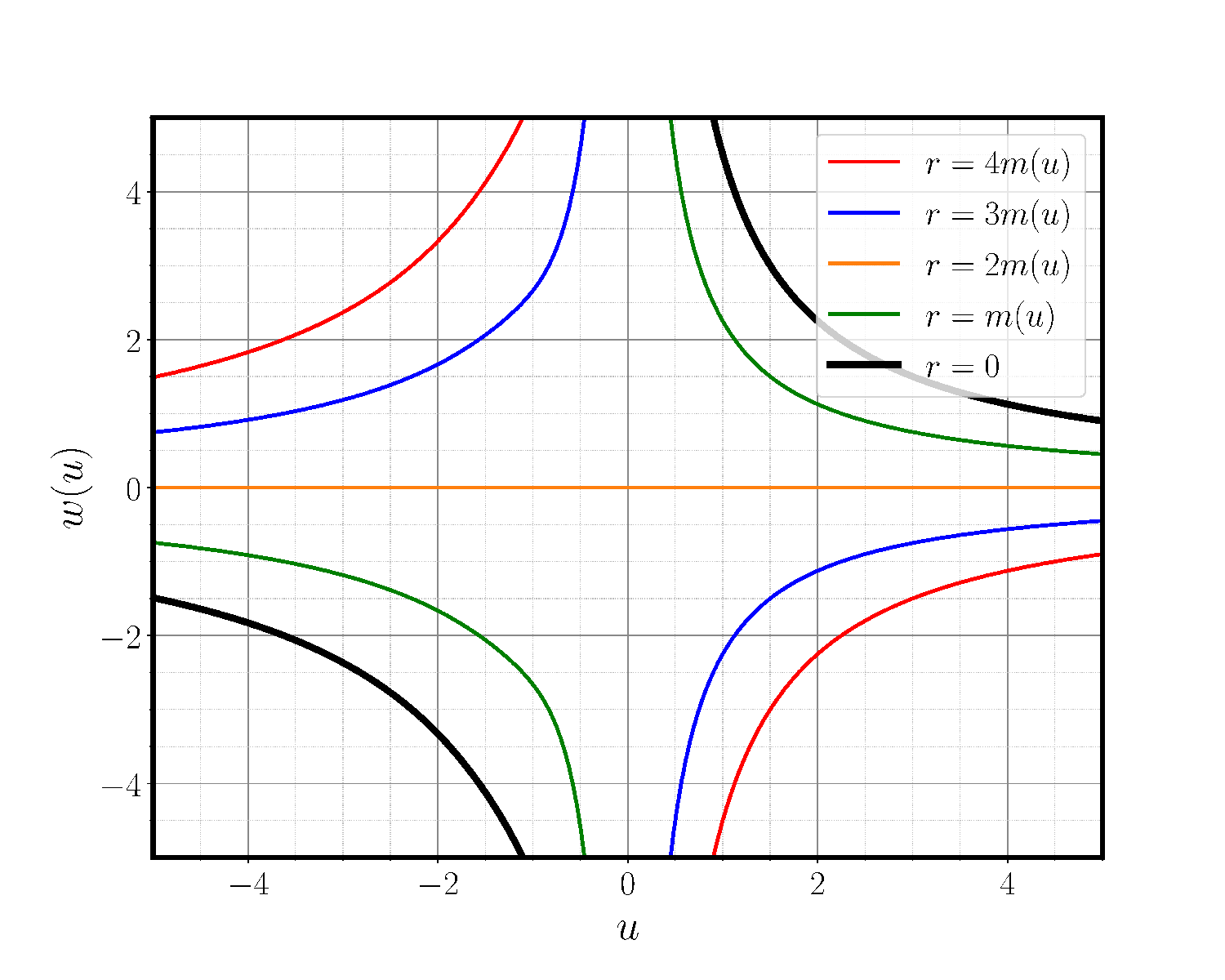}}
  \caption{The top row displays surfaces of dynamical radius for the choice of $h = +1$, while the bottom row displays the same for $h = -1$. Similar to surfaces of constant radius, increasing the value of $M_1$ leads to a less pronounced bulge, while switching the sign of $h$ creates a mirror reflection of the surfaces about the horizontal axis.}
\label{Ext1_1_dynR}
\end{figure*}

 Although the surfaces of constant radius are a great tool for understanding static and stationary spacetimes, they do not appear to have the same significance in dynamical spacetimes. To elucidate, when one looks at the surfaces of constant radius, the apparent horizon in the region $u < 0$ is not represented as a surface of constant radius \cite{paper1}, and the surfaces are not uniformly timelike or spacelike. Thus, we introduce an alternative family of surfaces, which we call surfaces of dynamical radius, $r(u,w) = \alpha m(u)$ with $\alpha \in \mathbb{N}$. This family is represented by 
 \begin{equation}\label{dyn_radius_def}
     w = (\alpha-2) \frac{m(u)}{U(u)},
 \end{equation}
  which is valid for both $u < 0$ and $u>0$ and reduces to the expression for surfaces of constant radius of the Schwarzschild vacuum solution, see (\ref{Ext1_1_ConR_eq}), if the mass function is taken to be constant, $m(u) = M_{1}$. The Lagrangian associated with this family of surfaces is given by 
 \begin{equation}\label{Lag_dyn_radii}
2\mathscr{L}= \left( 2-\alpha+2\alpha^2 \chi(u)\right) \frac{\dot{u}^2}{\alpha U^{2}(u)},
\end{equation}
where $\chi(u)=hm^{'}(u)U(u)>0$. This Lagrangian shows that these surfaces are spacelike when $\alpha \leq 2$, and that they are classified as timelike whenever $\alpha > 2\big( 1+\alpha^2 \chi(u)\big)$. To gain better insight into these surfaces of dynamical radius, graphs are provided in Fig. \ref{Ext1_1_dynR} for three different values of $M_1$ and also considering the impact of changing the sign of $h$. The region $u>0$ is equivalent to surfaces of constant radius, while in the region $u<0$, similar to the surfaces of constant radius, the surfaces come with bulges that attenuate with increasing values of $M_1$ or decreasing values of $u$, as expected with the asymptotically constant mass function. Furthermore, changing the sign of $h$ gives a mirror reflection about the $u$-axis, determining the direction of the propagation of time on the $u-w$ diagram. Studying surfaces of dynamical radius may be more adequate for dynamical black holes, since the apparent horizon is now represented as a surface of dynamical radius in both the regions $u<0$ and $u>0$, unlike the surfaces of constant radius case. Furthermore, there is no discernible alteration in causality as one shifts from quadrant II to quadrant III.
Finally, we note that the location of the bulges may be found by the solution of the equation $\chi(u)=\frac{1}{4}$.
\section{Second Maximal Extension} \label{Sec: Second Extension}
This maximal extension can be envisaged as extending the ingoing Israel metric (it acts as a black hole) to the Schwarzschild metric (it acts as a white hole) in the past. Equivalently, one can think of this extension as Schwarzschild vacuum metric being extended to the ingoing Israel metric. In the following subsection we give an example of a mass function that defines this extension. 
\subsection{The Mass Function and the \texorpdfstring{$U(u)$}{LG} Function}
An example of a mass function $m(u)$ which is designed to fulfill the previously discussed requirements, see Section \ref {mass_requirements}, is given by 
\begin{equation}
  \label{mass_ingoing_full}
   m(u)= \begin{dcases}
          M_{0}, &  u < 0 \\
         \frac{a_{n}u^{2n}+a_{0}}{b_{n}u^{2n}+b_{0}}, & 0\leq u\leq u_{1} \\
         M_{1}, & u\geq u_{1} \\
      \end{dcases}
\end{equation}
where $n\geq 2$, $a_{n}, a_{0}, b_{n},$ and $b_{0}$ are positive real numbers. This mass function is clearly constant for $u<0$ (giving rise to the Schwarzschild vacuum solution with mass $M_0$), increasing in the interval $u\in[0,u_1]$, and then constant again for $u\in[u_1,\infty)$ (yielding another Schwarzschild vacuum solution that is necessarily different from the initial one). However, similarly to what we have done in the mass function characterizing the first maximal extension, we can combine the last two branches of (\ref{mass_ingoing_full}) under the condition that the mass function approaches $M_1$ asymptotically. Thus, the mass function is now written as 
\begin{equation}
  \label{mass_ingoing}
   m(u)= \begin{dcases}
          M_{0}, &  u<0 \\
         \frac{a_{n}u^{2n}+a_{0}}{b_{n}u^{2n}+b_{0}}.  &  u \geq 0 \\
      \end{dcases}
\end{equation}
Since we have demanded that the mass function $m(u)$ asymptotically approaches $M_1$, the following condition must be satisfied:
\begin{equation}
\label{Ext2_param_constr_I}
\lim_{u\rightarrow +\infty} m(u)=\frac{a_{n}}{b_{n}}=M_{1}>M_{0}.
\end{equation}
Another useful condition similar to (\ref{param_constr_II}) is
\begin{equation}
\label{Ext2_param_constr_II}
  \lim_{u\rightarrow 0^{+}}m(u)=\frac{a_0}{b_0} = m(0)=M_{0}.
\end{equation}
Now, we are in a position to express the function $U(u)$ over the entire domain of the coordinate $u$.
   \begin{equation} \label{U_ext2_expression}
     U(u)=\begin{dcases}
      \int_{0}^{u<0} \frac{hdx}{4 M_{0}}, &  u < 0 \\
      \int_{0}^{u>0}\frac{hdx}{4\left(\frac{a_{n}x^{2n}+a_{0}}{b_{n}x^{2n}+b_{0}}\right)}. & u\geq 0 \\
     \end{dcases}
 \end{equation} 
We use the constraints (\ref{Ext2_param_constr_I}) and (\ref{Ext2_param_constr_II}) to introduce a more specific instance of the mass function. For example, we can substitute $n=2$, $a_{2} = a_{0} = b_{2} = M_{0}$, and $b_{0} = 1$ in (\ref{mass_ingoing}), resulting in 
\begin{equation} \label{mass_ingoing_final}
    m(u)= \begin{dcases}
          M_{0}, &  u<0 \\
         \frac{M_{0}(u^{4}+1)}{M_{0}u^{4}+1}. &  u \geq 0 \\
      \end{dcases}
\end{equation}
\begin{figure}[htb]
\centering
  \includegraphics[width=\linewidth]{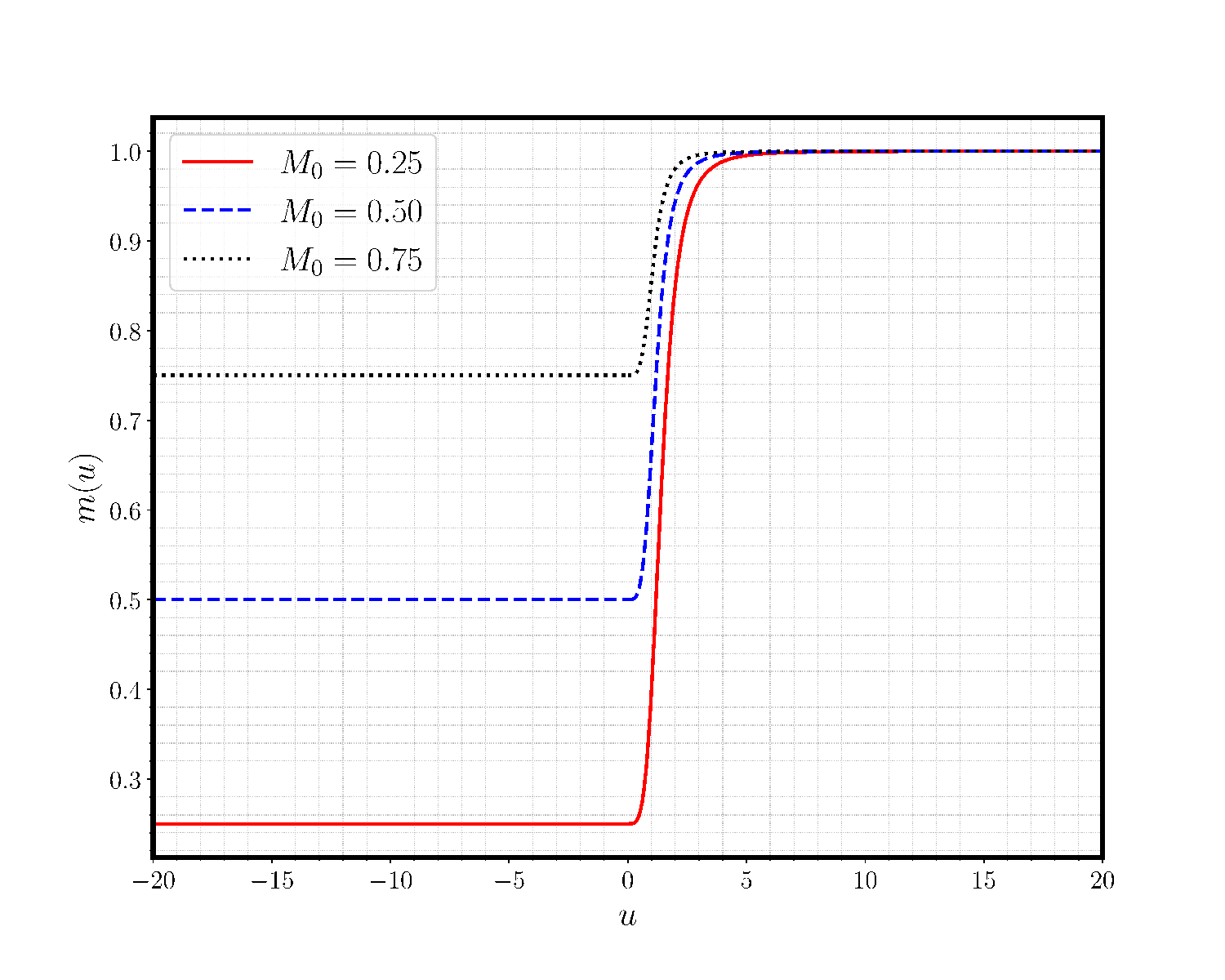}
  \caption{ A plot of the mass function (\ref{mass_ingoing_final}) that gives rise to the second maximal extension to the Vaidya metric. There are three values used for the parameter $M_{0}<M_{1}$, where $M_{1}$ is taken to equal unity.}
  \label{Ext2_m1}
\end{figure}
Accordingly, the expression of the function $U(u)$, with the aid Table. II in \cite{paper1}, is
\begin{equation} \label{Ex2_U1_eq}
      \scriptstyle U(u)=\begin{dcases} 
      \scriptstyle -h \frac{u}{4M_{0}}, &  \scriptstyle u < 0 \\
      \scriptstyle h \left(\frac{u}{4}+\frac{\sqrt{2}\left(M_{0}-1\right)}{16 M_{0}} \left(-\tanh ^{-1} \Gamma-\tan^{-1} \Delta+\tan ^{-1}\Psi\right)\right),  &  \scriptstyle u\geq 0 \\
     \end{dcases}
 \end{equation}
where $h=\pm 1$, $\Gamma = \frac{\sqrt{2} u}{u^2+1}$, $\Delta = \sqrt{2} u+1$, and $\Psi = 1-\sqrt{2} u$.
The behavior of the function $U(u)$ is now distinct depending on the choice of $h$. For $u < 0$, a positive slope, $\frac{-1}{4M_{0}}$, is found when $h = +1$, and this slope becomes negative, $\frac{1}{4M_{0}}$, for $h = -1$. For $u \geq 0$, the function $U(u)$ is greater than 0 when $h = +1$ and less than 0 if $h = -1$. Additionally, there is a similar trend to the first extension, in that the linearity of the function increases with the value of the parameter $M_{0}$ (shown by the dotted black line, in Figs. \ref{2_h_p} and \ref{2_h_n}).
\begin{figure}[htb]
  \begin{subfigure}{0.85\columnwidth}
    \includegraphics[width=\linewidth]{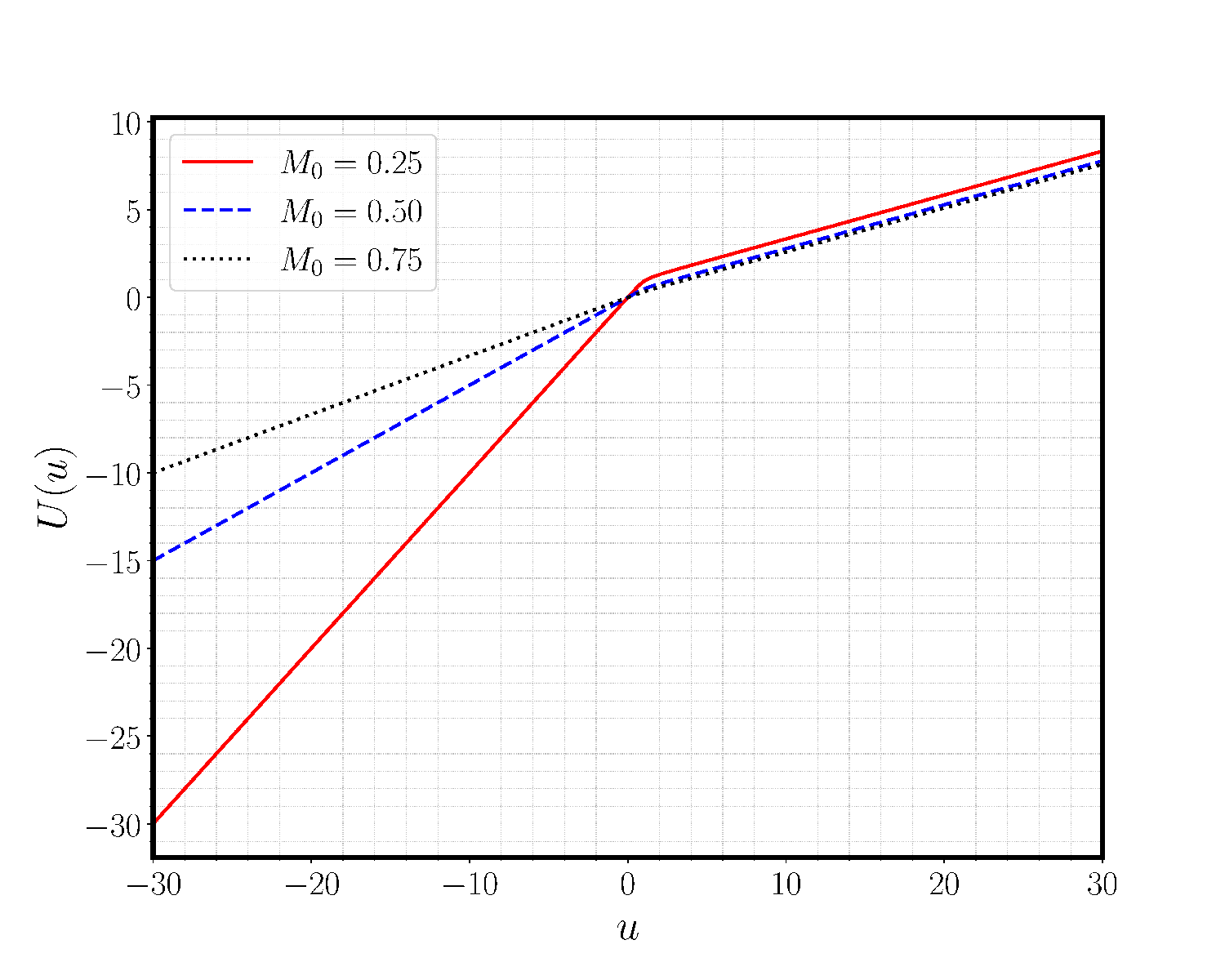}
    \caption{$h = +1$}
    \label{2_h_p}
  \end{subfigure}
  \begin{subfigure}{0.85\columnwidth}
    \includegraphics[width=\linewidth]{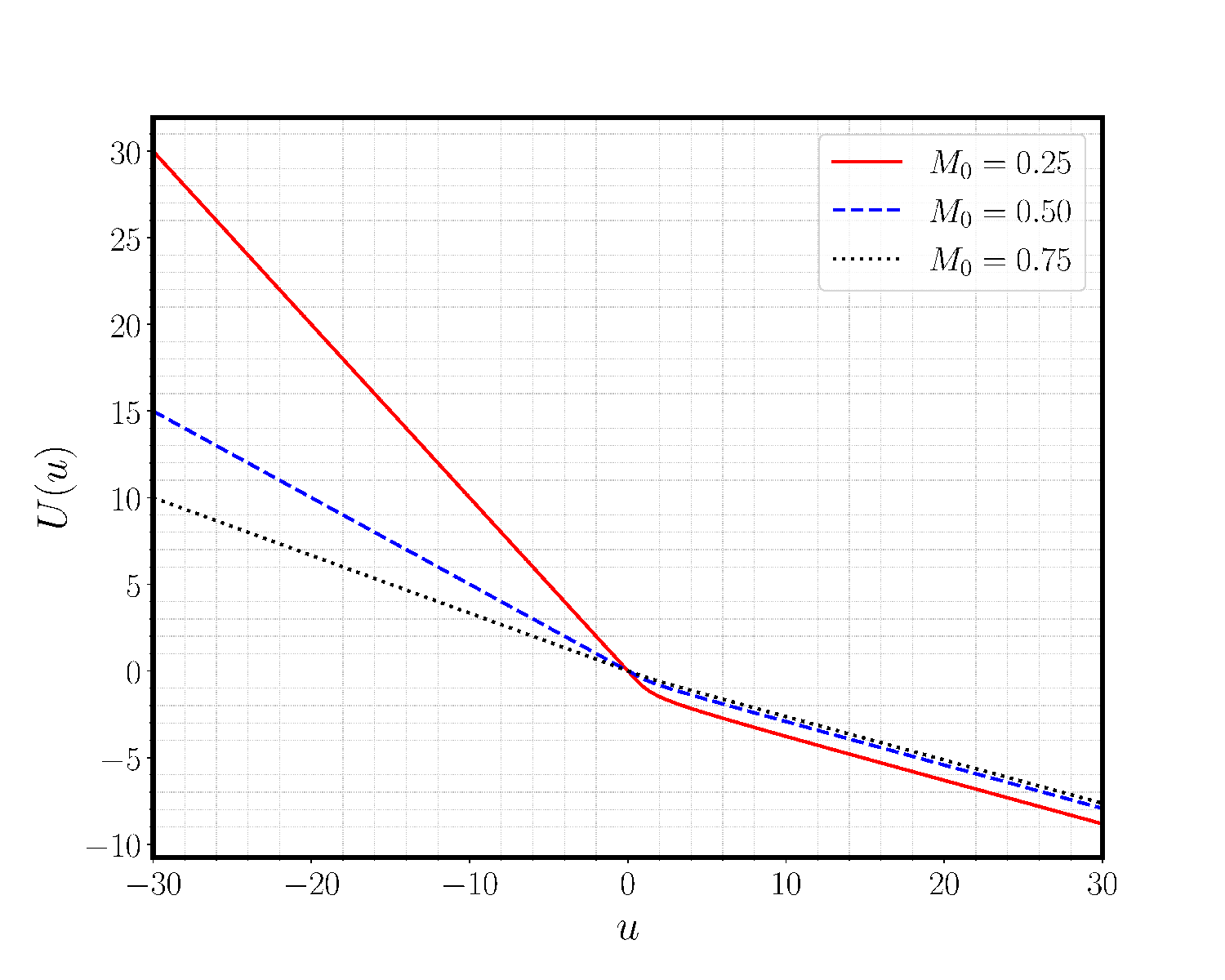}
    \caption{$h = -1$}
    \label{2_h_n}
  \end{subfigure}
  \caption{Two illustrations of the function $U(u)$ (\ref{Ex2_U1_eq}) corresponding to two choices of the function $h$.}
\label{Ext2_U_func_figs}
\end{figure}
\subsection{Surfaces of Constant Radius}
In a similar way to what we did in the first maximal extension, 
by giving explicit expressions for the functions $m(u)$ and $U(u)$ over the entire range of the coordinate $u$, 
we can explore the surfaces of $r = \alpha M_{0}=\text{const}$. Mathematically, these surfaces are described by 
\begin{equation}
\label{Ext2_1_ConR_eq}
    w= \begin{dcases}
         \frac{4(\alpha-2)M_0^2}{hu}, &  u < 0 \\
         \frac{\alpha M_{0}-2m(u)}{U(u)}. & u\geq 0 \\
      \end{dcases}
\end{equation}
In order to examine the causal nature of the surfaces $r = \alpha M_{0}$, we need to determine the sign of the Lagrangian associated with (\ref{Israel_ext_general}), where in this case the mass function is taken to be (\ref{mass_ingoing_final}). Accordingly, The Lagrangian reads
\begin{equation} \label{Ex2_Lagrangian}
     2\mathscr{L}_{r(u,w)=\text{const}}=\begin{dcases}
     \left(-\frac{4w}{\alpha hu}\right)\dot{u}^2 , &  u < 0 \\
     \left(-\frac{w}{\alpha M_{0}U(u)}\right)\dot{u}^2. & u\geq 0 \\
     \end{dcases}
 \end{equation}
Hence, for the choices $h = \pm1$, the surfaces $r = \alpha M_{0}$ are timelike (spacelike) in quadrants I and III, and spacelike (timelike) in quadrants II. We can see that the physical singularity in this extension, the surface $r=0$, is spacelike, and therefore both timelike and null geodesics cannot avoid it. Additionally, for $u\geq 0$ we observe that the apparent horizon (spacelike hypersurface) is not coinciding with the event horizon (null hypersurface). This can be attributed to the fact that the apparent horizon and event horizon are only indistinguishable in static spacetimes, which is not the case for $u\geq0$; this part of the spacetime is considered as dynamical, or simply non-static. Moreover, the surfaces $r = \alpha M_{0}=\text{const}$ are graphically represented by hyperbolas, $u<0$, in the $u-w$ plane due to the simple relation $uw = \text{const}$.
\begin{figure*}[htb]
  \subcaptionbox{$M_1 = 0.25$}[.32\linewidth][c]{%
    \includegraphics[width=1\linewidth]{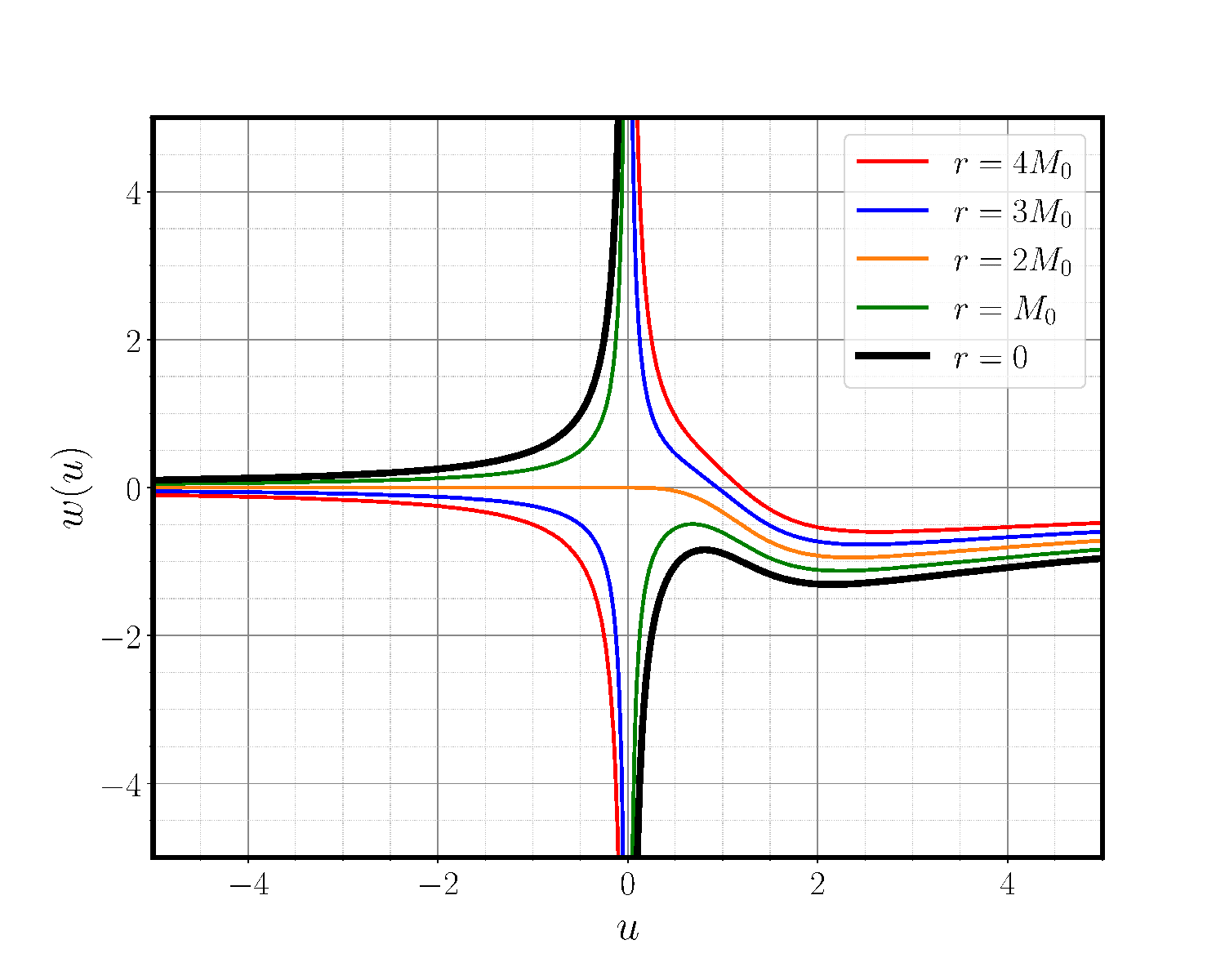}}\quad
  \subcaptionbox{$M_1 = 0.5$}[.32\linewidth][c]{%
    \includegraphics[width=1\linewidth]{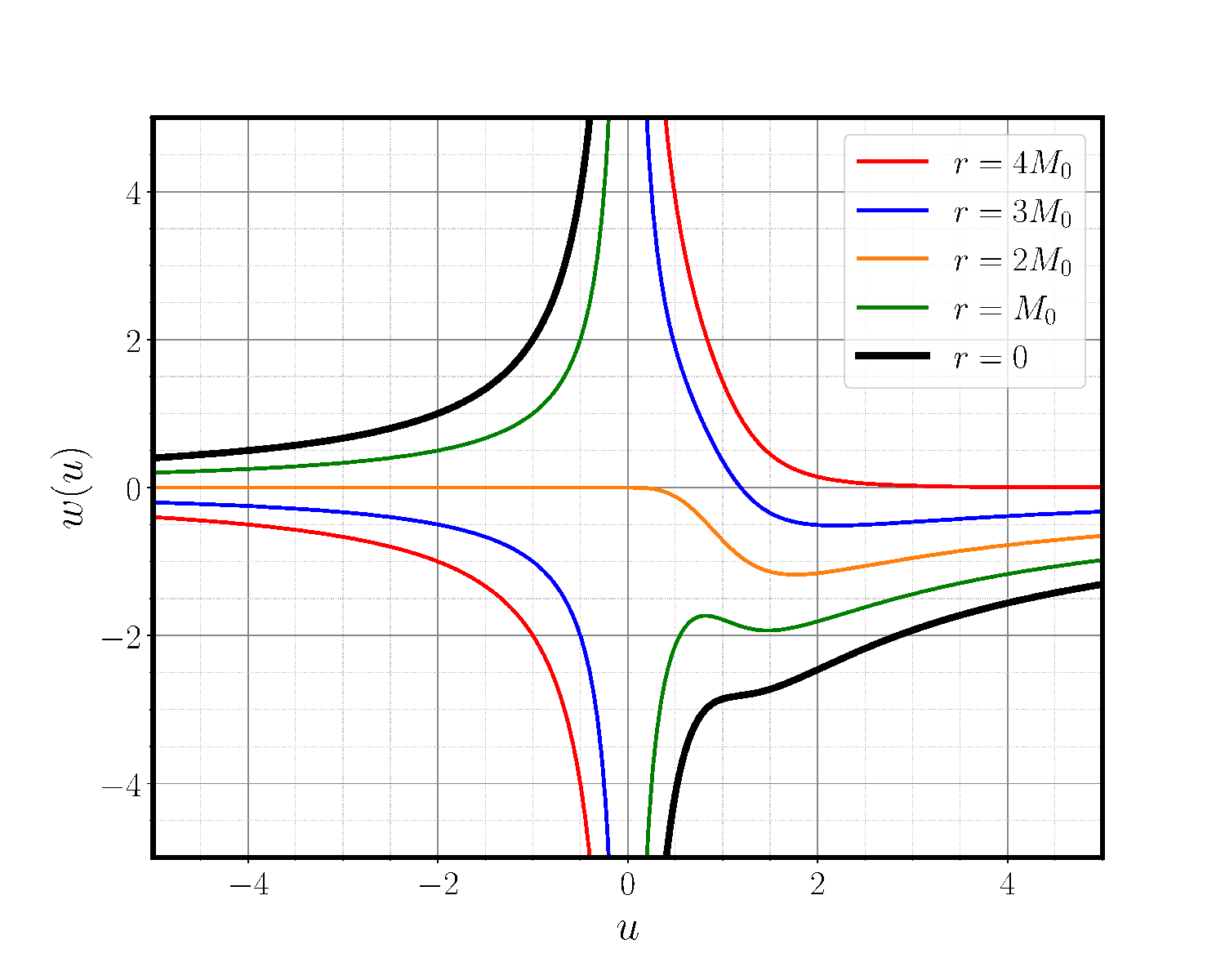}}\quad
  \subcaptionbox{$M_1 = 0.75$}[.32\linewidth][c]{%
    \includegraphics[width=1\linewidth]{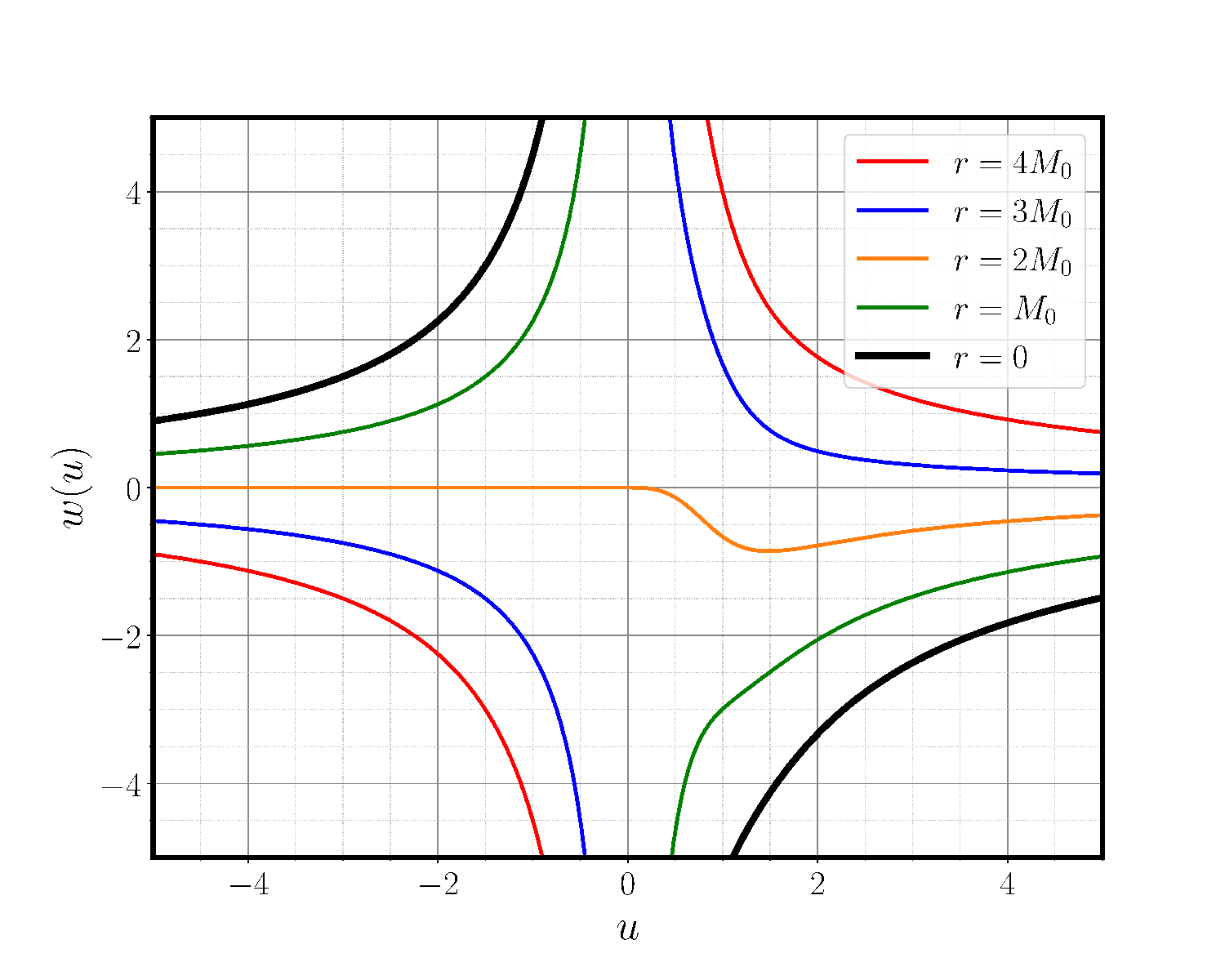}}
  \subcaptionbox{$M_1 = 0.25$}[.32\linewidth][c]{%
    \includegraphics[width=1\linewidth]{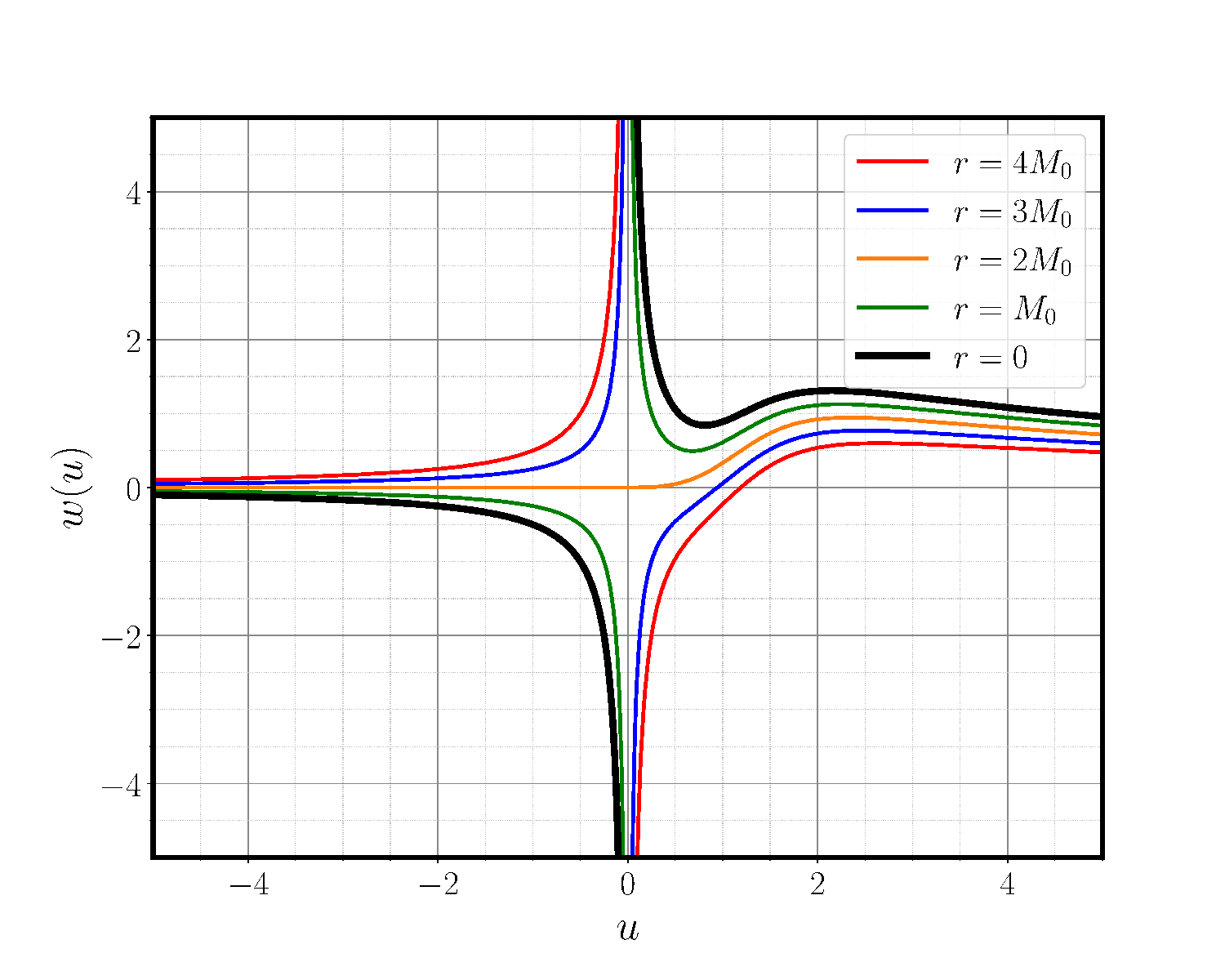}}\quad
  \subcaptionbox{$M_1 = 0.5$}[.32\linewidth][c]{%
    \includegraphics[width=1\linewidth]{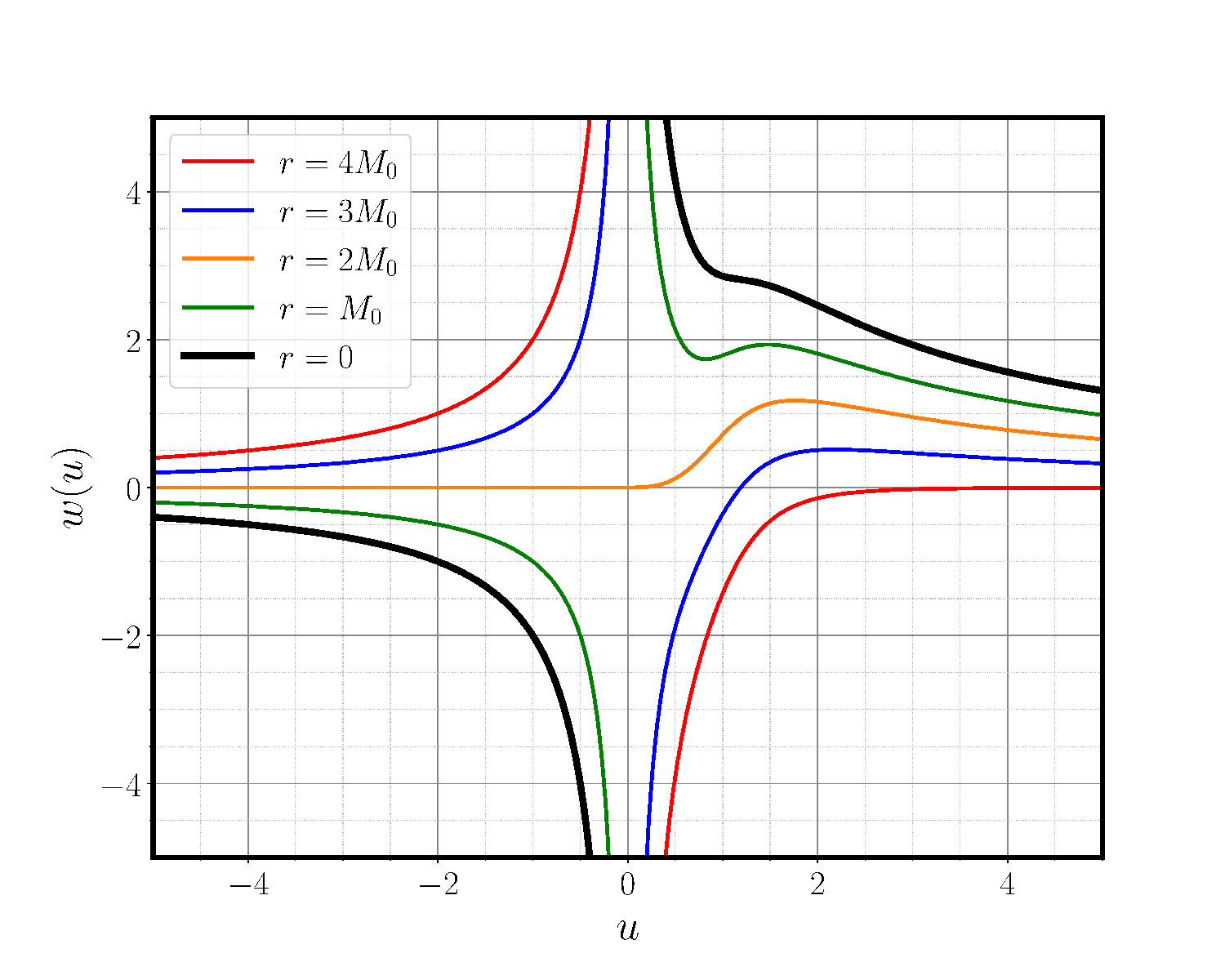}}\quad
  \subcaptionbox{$M_1 = 0.75$}[.32\linewidth][c]{%
    \includegraphics[width=1\linewidth]{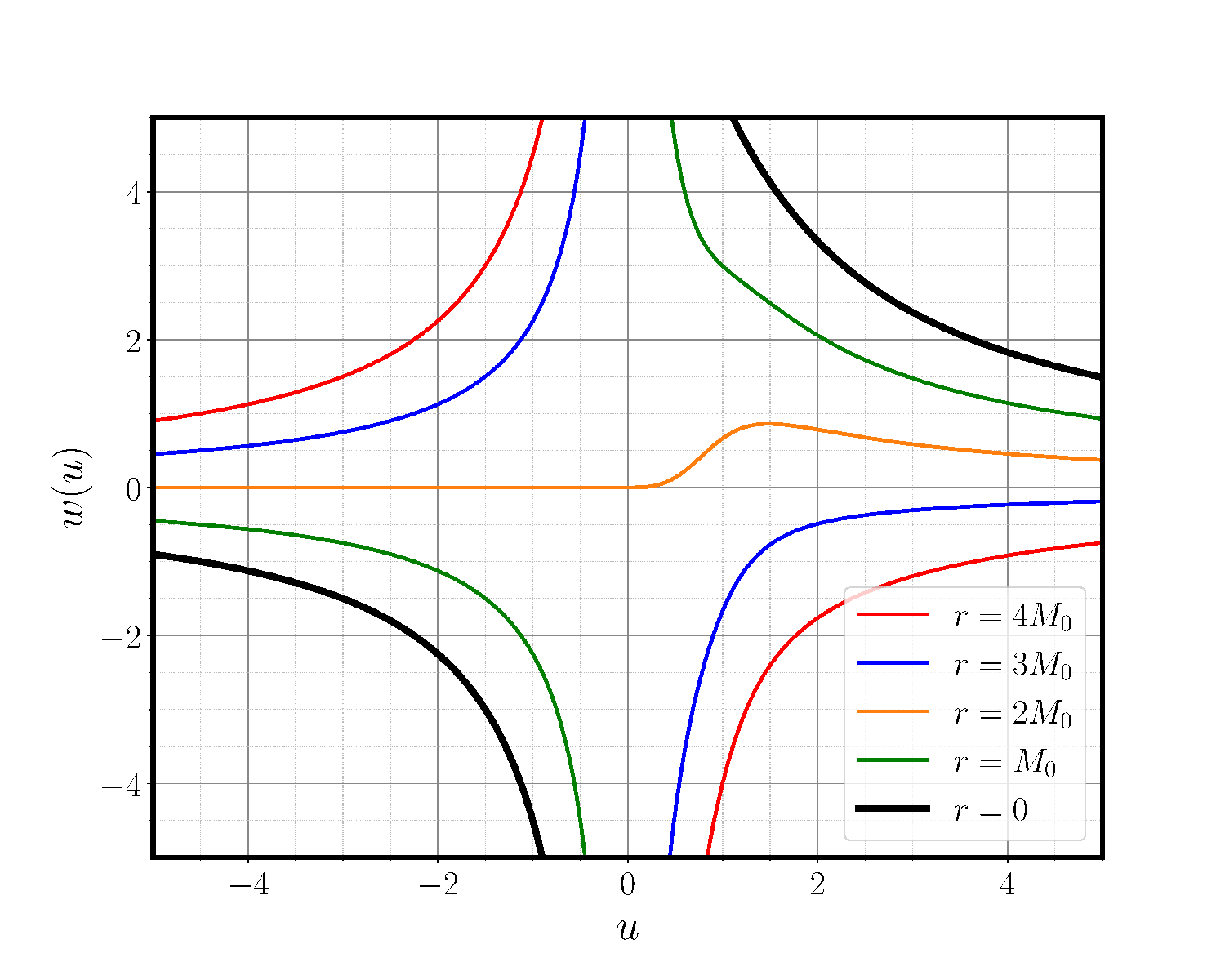}}
  \caption{The top row displays surfaces of constant radius in the second extension for the choice of $h = +1$, while the bottom row displays the same for $h = -1$. Increasing the value of $M_1$ leads to a less pronounced bulge, while switching the sign of $h$ creates a mirror reflection of the surfaces about the horizontal axis.}
  \label{Ext2_1_ConstR}
\end{figure*}
The appearance of bulges in the the curves $r(u,w) = \alpha M_{0}$, see Fig. \ref{Ext2_1_ConstR}, corresponds to the existence of fixed points on the same curves, which mathematically can be characterized by solving 
 \begin{equation}\label{roots_bulges_ext2}
     0 = \frac{dw}{du} = \frac{-2m^{'}(u)U(u)-\frac{h}{4m(u)}\left(\alpha M_{0}-2m(u)\right)}{U(u)^2}, 
 \end{equation}
where the graphical locations of the roots (where the bulges are expected to appear) can be seen from the graph below, Fig. \ref{bulges_ext2}.
\begin{figure}[htb]
\includegraphics [width=1\linewidth]{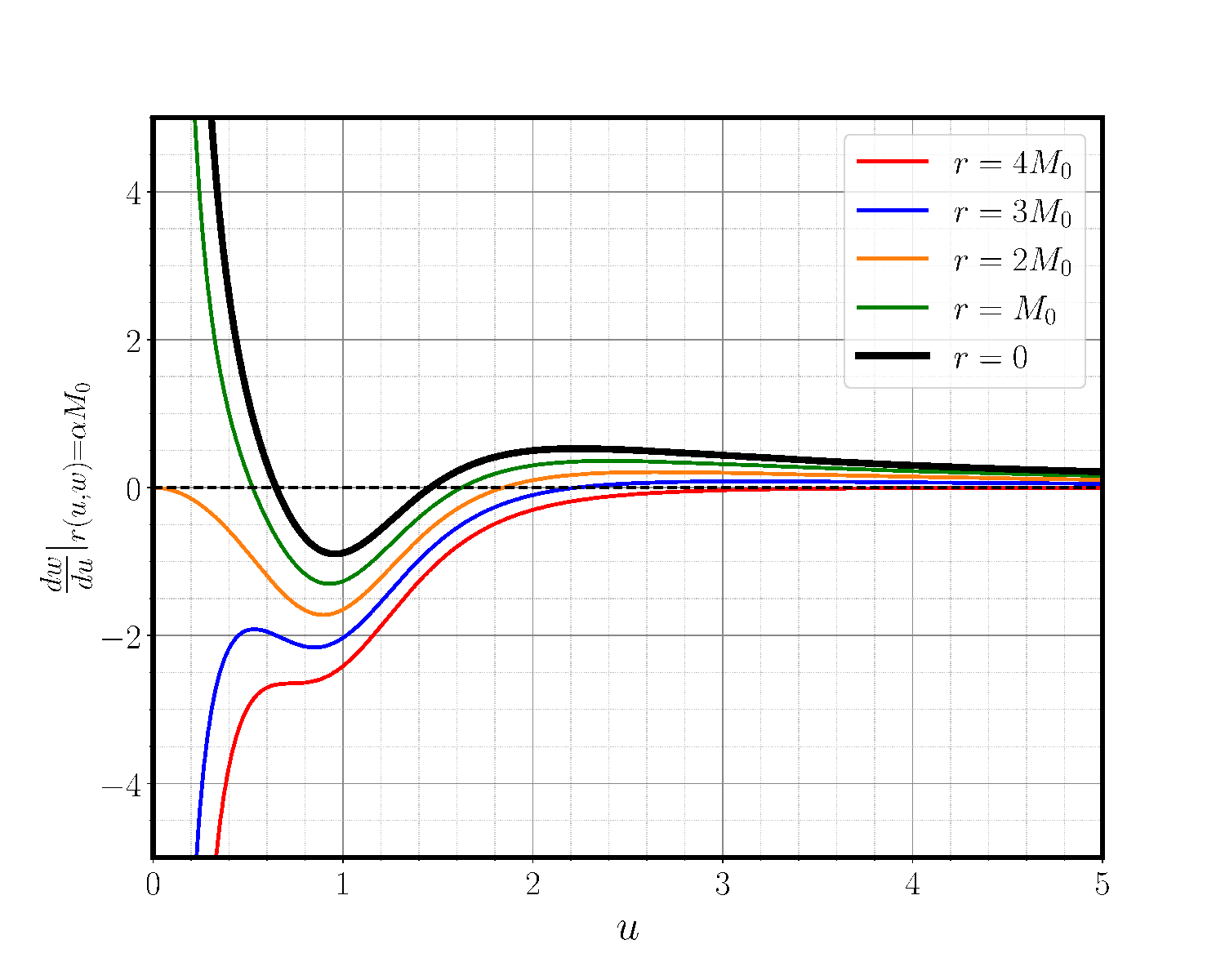}
\caption{The graphical locations of the roots of (\ref{roots_bulges_ext2}) for the choices $h=+1$ and $M_{0} = 0.5$.}
\label{bulges_ext2}
\end{figure}
\subsection{Surfaces of Dynamical Radius}
\begin{figure*}[htb]
  \captionsetup{justification=justified}
  \subcaptionbox{$M_1 = 0.25$}[.32\linewidth][c]{%
    \includegraphics[width=1\linewidth]{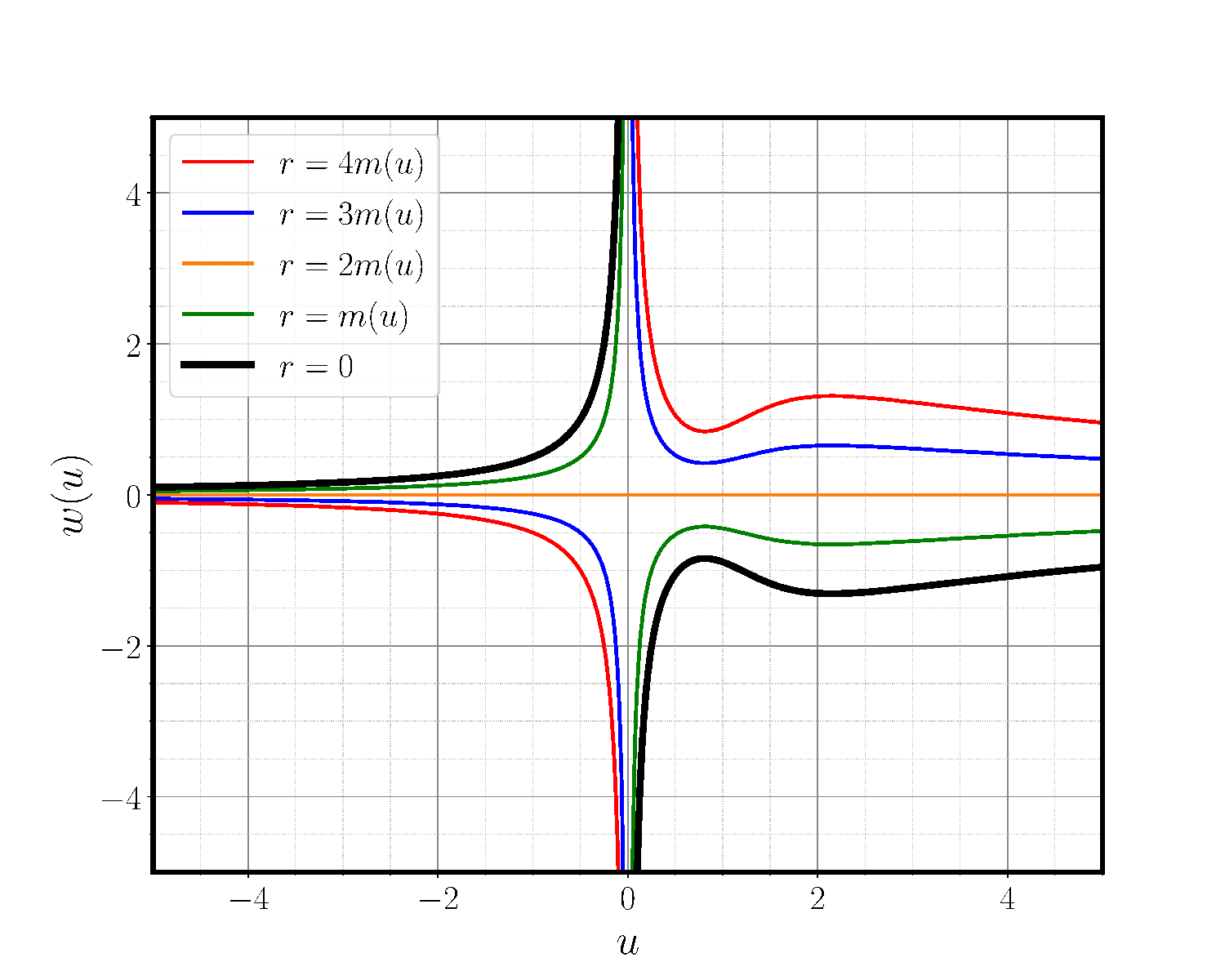}}\quad
  \subcaptionbox{$M_1 = 0.5$}[.32\linewidth][c]{%
    \includegraphics[width=1\linewidth]{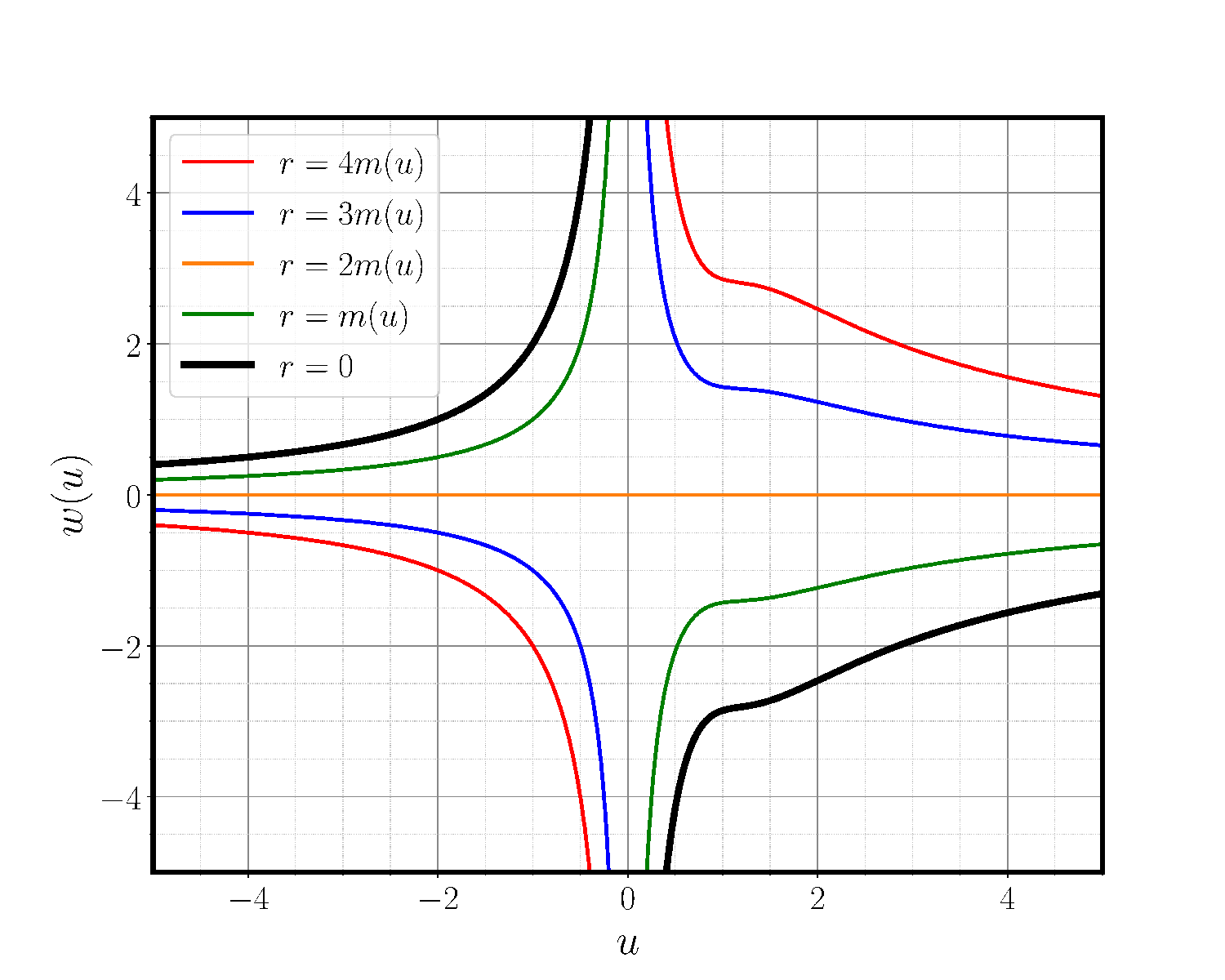}}\quad
  \subcaptionbox{$M_1 = 0.75$}[.32\linewidth][c]{%
    \includegraphics[width=1\linewidth]{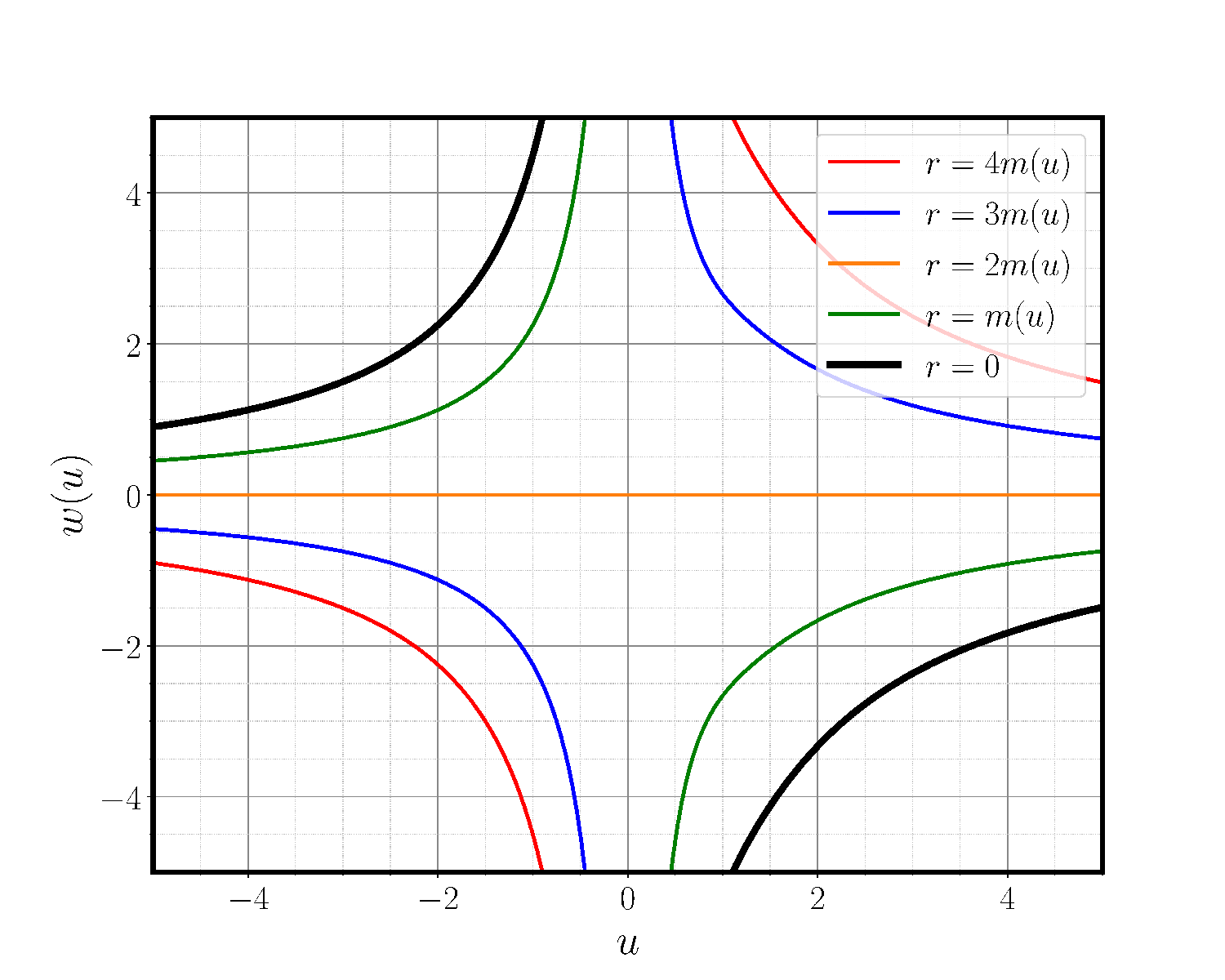}}
  \subcaptionbox{$M_1 = 0.25$}[.32\linewidth][c]{%
    \includegraphics[width=1\linewidth]{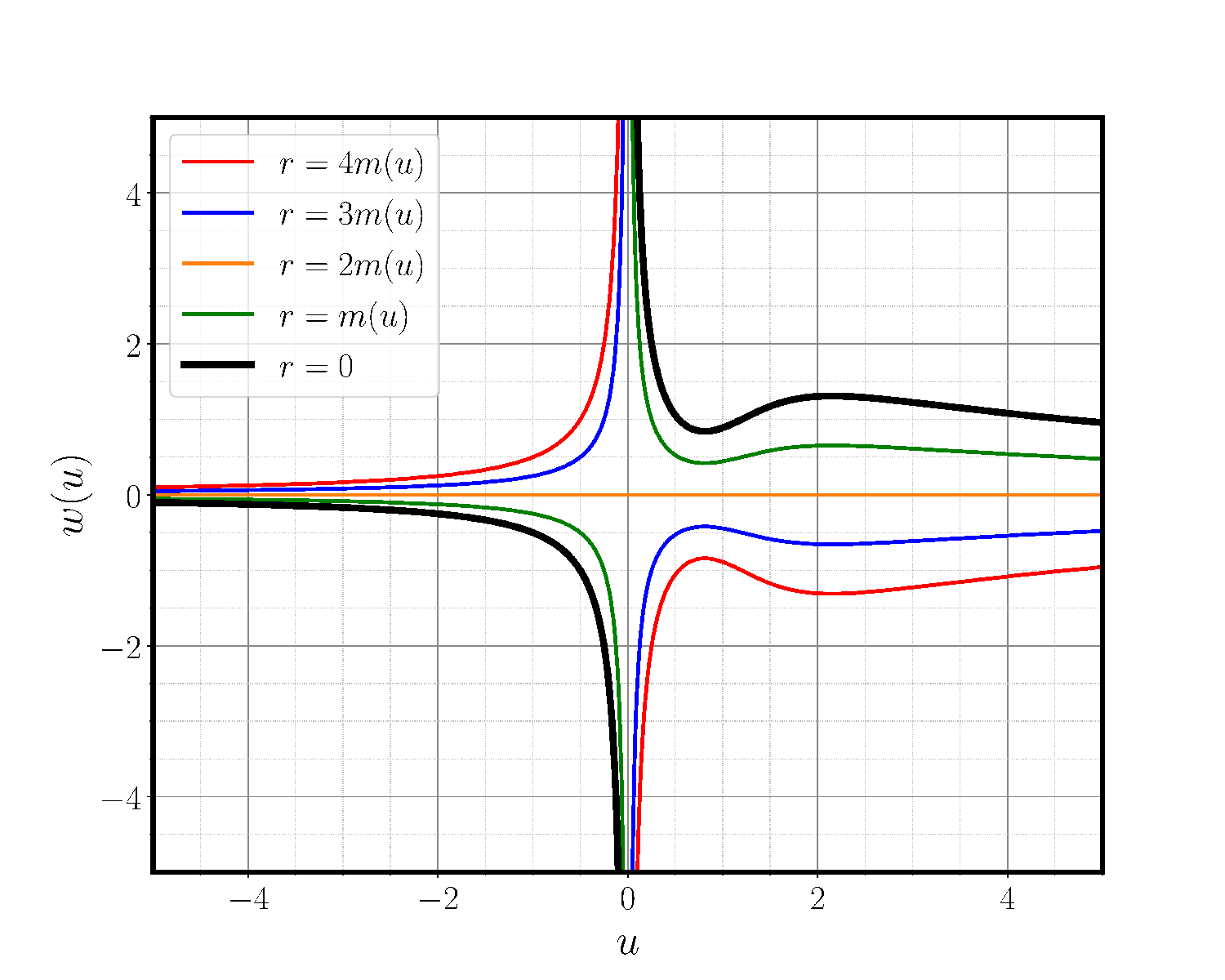}}\quad
  \subcaptionbox{$M_1 = 0.5$}[.32\linewidth][c]{%
    \includegraphics[width=1\linewidth]{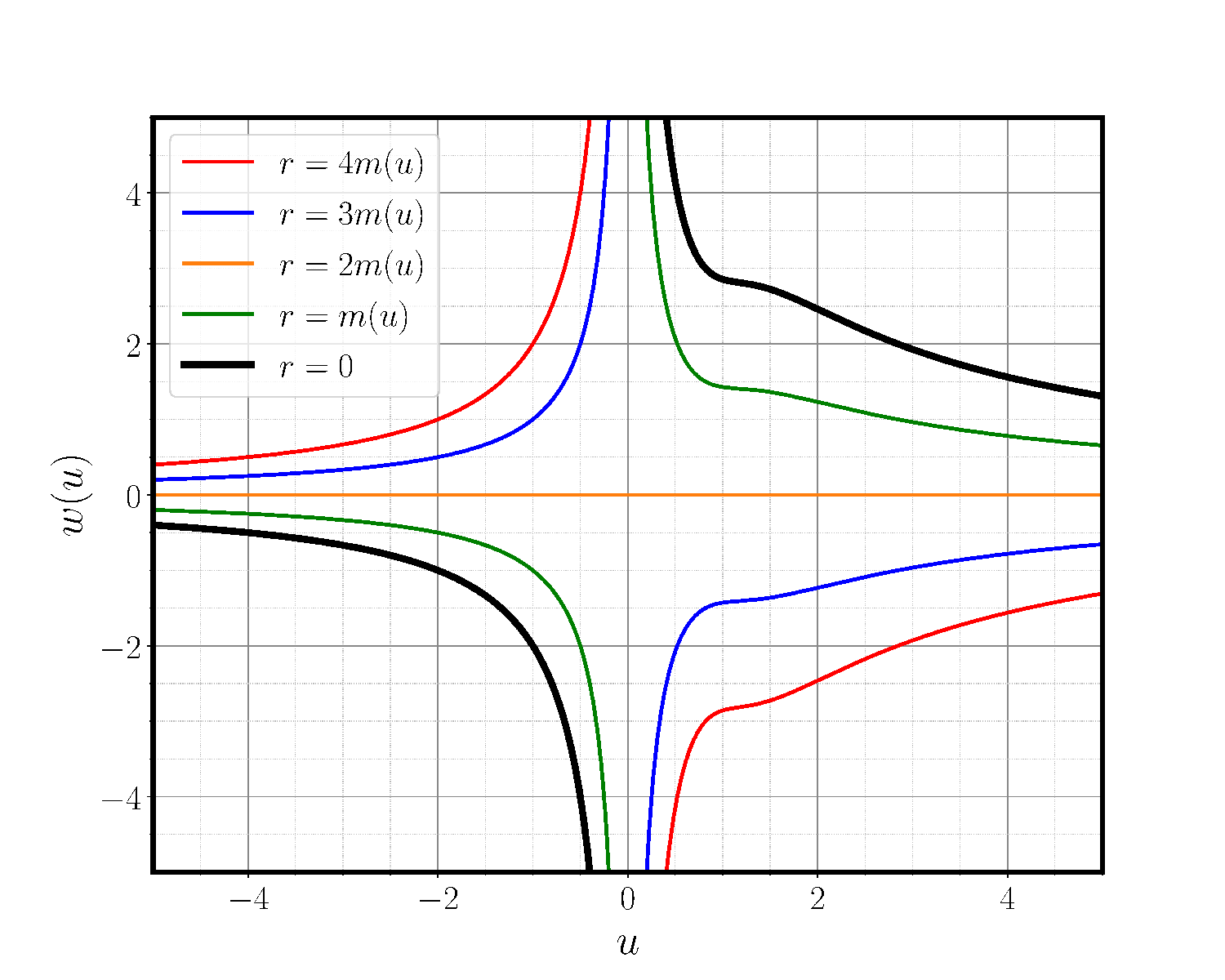}}\quad
  \subcaptionbox{$M_1 = 0.75$}[.32\linewidth][c]{%
    \includegraphics[width=\linewidth]{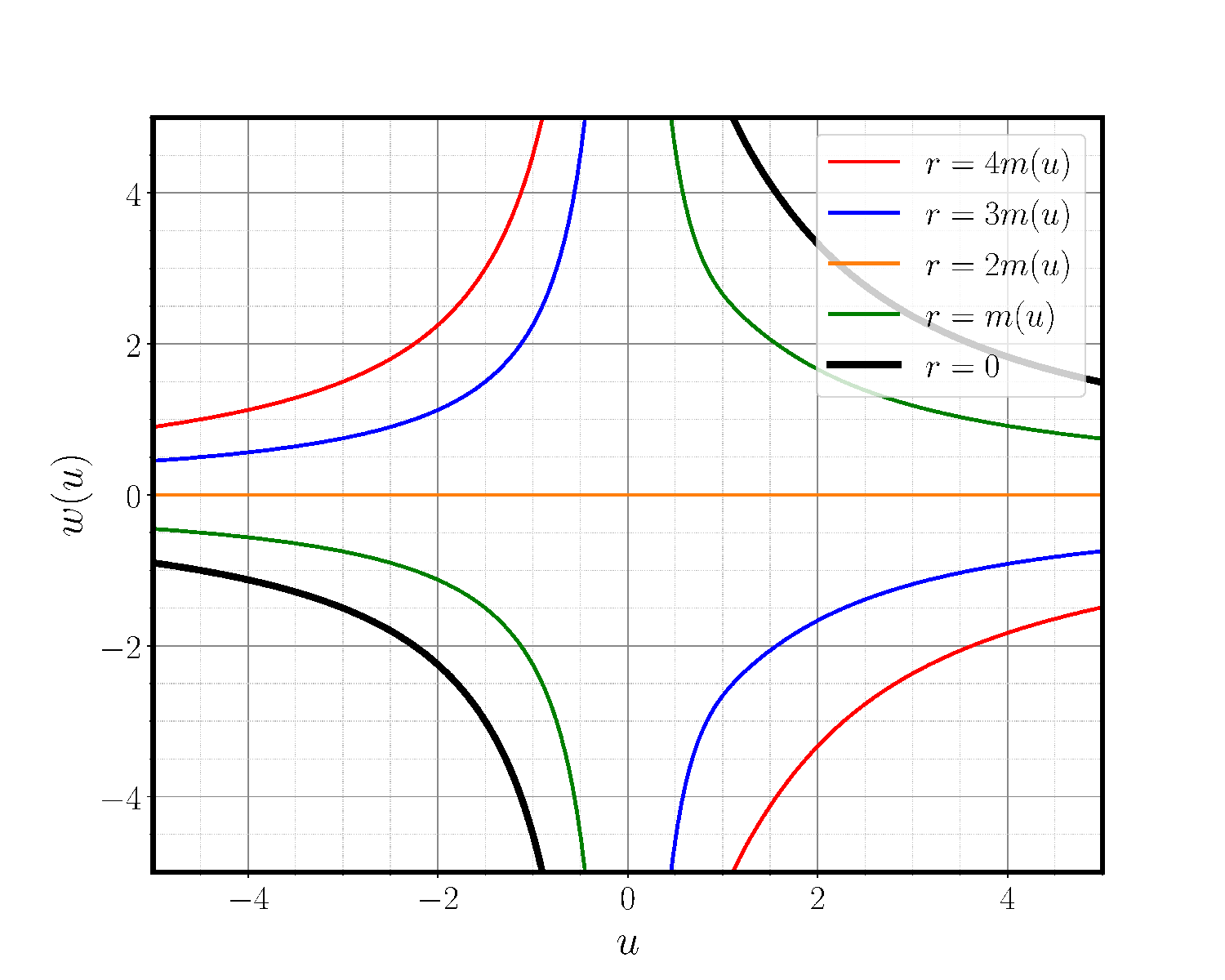}}
\caption{The top row displays surfaces of dynamical radius in the second extension for the choice of $h = +1$, while the bottom row displays the same for $h = -1$. Similar to surfaces of constant radius, increasing the value of $M_1$ leads to a less pronounced bulge, while switching the sign of $h$ creates a reflection of the surfaces.}
\label{Ext2_1_dynR}
\end{figure*}

As has been previously demonstrated in the first extension, the surfaces of constant radius in the second extension are not uniformly timelike or spacelike and the apparent horizon is not represented as a surface of constant radius in the region ($u\geq 0$). Thus, we explore surfaces of dynamical radius to avoid these issues. These surfaces are still specified in (\ref{dyn_radius_def}) and (\ref{Lag_dyn_radii}), and they can be classified as spacelike when $\alpha \leq 2$ or timelike when $\alpha > 2\big( 1+\alpha^2 \chi(u)\big)$. The surfaces are only null when $m^{'}(u)=0$, true in the region $u<0$ and $\alpha=2$ as illustrated in Fig. \ref{Ext2_1_dynR}. We also notice that these surfaces of dynamical radius still form bulges which occur when $\chi(u)=\frac{1}{4}$.
\section{Third Maximal Extension}\label{Sec: Third Extension}
In order to construct this extension, we employ a symmetric mass function around the $u$-axis, thus assuming that the matter content (radiation) is the same when $u<0$ and $u>0$. Nevertheless, the direction of the radiation is dissimilar, being outward-oriented for $u<0$ and inward-directed for $u>0$.
 \subsection{ The Mass Function and the \texorpdfstring{$U(u)$}{LG} Function}
\begin{figure}[htb]
\centering
  \includegraphics[width=1\linewidth]{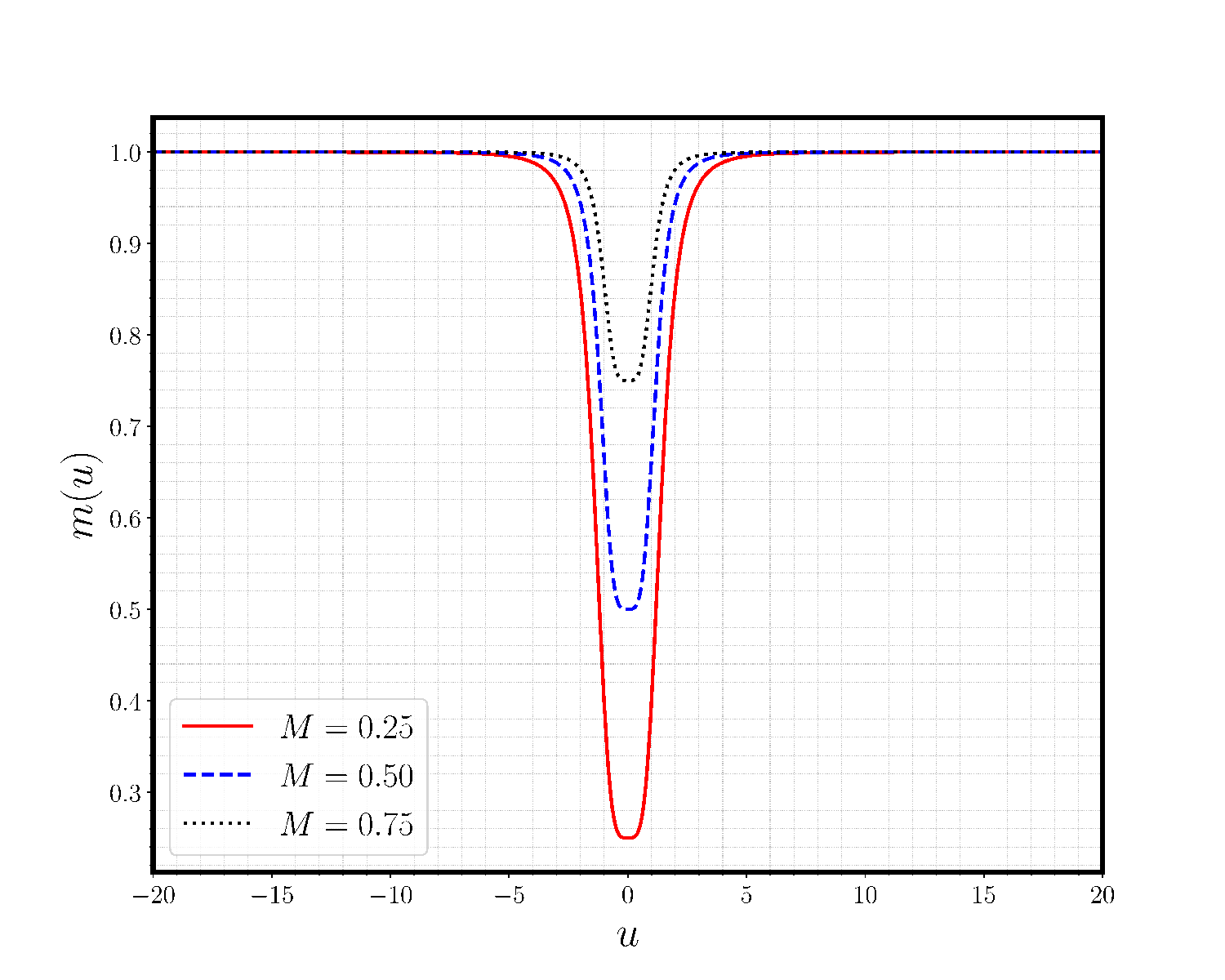}
  \caption{ A plot of the mass function (\ref{mass_out_in_M}) that defines the third maximal extension to the Vaidya metric.}
  \label{Ext3_m1}
\end{figure}

 A particular instance of a mass function that meets all of the criteria outlined in Section \ref {mass_requirements} is
 \begin{equation}
  \label{mass_out_in_full}
   m(u)= \begin{dcases}
          M_{0}, &  u < u_{1} \\
         \frac{a_{n}u^{2n}+a_{0}}{b_{n}u^{2n}+b_{0}}, & u_{1}\leq u < 0 \\
         \frac{c_{n}u^{2n}+c_{0}}{d_{n}u^{2n}+d_{0}}, & 0\leq u < u_{2} \\
         M_{1}, & u\geq u_{2} \\
      \end{dcases}
\end{equation}
 where $n\geq 2$ and $a_{n}, a_{0}, b_{n}, b_{0},c_{n},c_{0},d_{n},$ and $d_{0}$ are positive real numbers. Similarly, we choose to work with the following simpler form of the mass function:
 \begin{equation}
  \label{mass_out_in_1}
   m(u)= \begin{dcases}
   \frac{a_{n}u^{2n}+a_{0}}{b_{n}u^{2n}+b_{0}}, &  u<0 \\
   \frac{c_{n}u^{2n}+c_{0}}{d_{n}u^{2n}+d_{0}}. &  u\geq0 
   \end{dcases}
\end{equation}
It follows that in order for the previous form to be consistent with (\ref{mass_out_in_full}), the following stipulations must be taken into account. First, 
\begin{equation}
     \lim_{u\rightarrow -\infty} m(u) =\frac{a_n}{b_n} =M_{0}.
 \end{equation}
 Second,
 \begin{equation}
     \lim_{u\rightarrow +\infty} m(u) = \frac{c_n}{d_n}=M_{1},
 \end{equation}
 and thus with the aid of these two stipulations, we can choose the mass function to take the form
\begin{equation}
  \label{mass_out_in}
   m(u)= \begin{dcases}
   \frac{M_{0}u^{2n}+a_{0}}{u^{2n}+b_{0}}, &  u<0 \\
   \frac{M_{1}u^{2n}+c_{0}}{u^{2n}+d_{0}}. &  u\geq0 
   \end{dcases}
\end{equation}
Thus, the function $U(u)$ is now defined for the whole range of the coordinate $u$ as
 \begin{equation}
     U(u) = \int_{0}^{u<0}\frac{h dx}{4\left(\frac{M_{0}x^{2n}+a_{0}}{x^{2n}+b_{0}}\right)}+\int_{0}^{u>0}\frac{hdx}{4\left(\frac{M_{1}x^{2n}+c_{0}}{x^{2n}+d_{0}}\right)}.
 \end{equation}
 That said, we now specify a single parameter mass function, see Fig. \ref{Ext3_m1},
 \begin{equation}
  \label{mass_out_in_M}
   m(u)= \frac{M(u^{4}+1)}{Mu^{4}+1},
\end{equation}
 where $n=2$ and $0<M<1$. Moreover, we have demanded that $M_0 = M_1= M$ \footnote{It is to be noted that this demand makes the mass function perfectly symmetric. Alternatively, we could have demanded that $M_0<M_1$ or $M_{0}>M_{1}$.}, $b_4 = d_4 = M$, $a_{0} = c_{0}= M$, and $b_{0} = d_{0} = 1$.
Accordingly, the explicit expression of the function $U(u)$ takes the form
\begin{equation} \label{U_ext3_expression}
   \scriptstyle  U(u) =  \scriptstyle h \left(\frac{u}{4}+\frac{\sqrt{2}\left(M-1\right)}{16 M} \left(-\tanh ^{-1} \Gamma -\tan^{-1}\Delta+\tan ^{-1} \Psi\right)\right),
\end{equation}
where $h=\pm 1$, $\Gamma = \frac{\sqrt{2} u}{u^2+1}$, $\Delta = \sqrt{2} u+1$, and $\Psi = 1-\sqrt{2} u$. It is seen that increasing the value of the parameter $0<M<1$ towards unity, based on the function $U(u)$ in Fig. \ref{3_h_p_n} below, produces a linear behavior similar to the $U(u)$ of the Schwarzschild vacuum solution. When $M=1$, this restores the Schwarzschild vacuum solution in  Israel coordinates.
\begin{figure}[htb]
  \begin{subfigure}{0.85\columnwidth}
    \includegraphics[width=\linewidth]{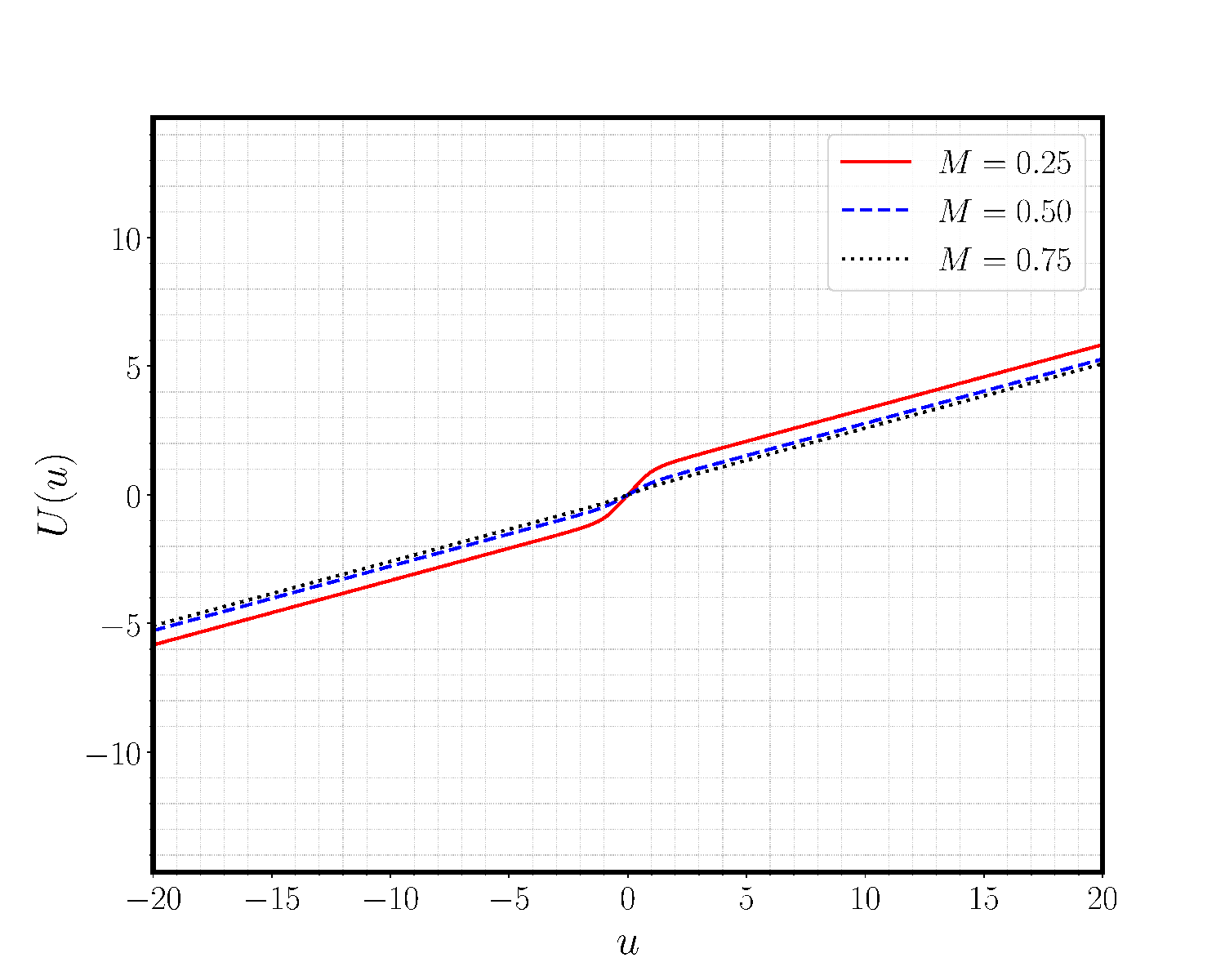}
    \caption{$h = +1$}
    \label{3_h_p}
  \end{subfigure}
  \begin{subfigure}{0.85\columnwidth}
    \includegraphics[width=\linewidth]{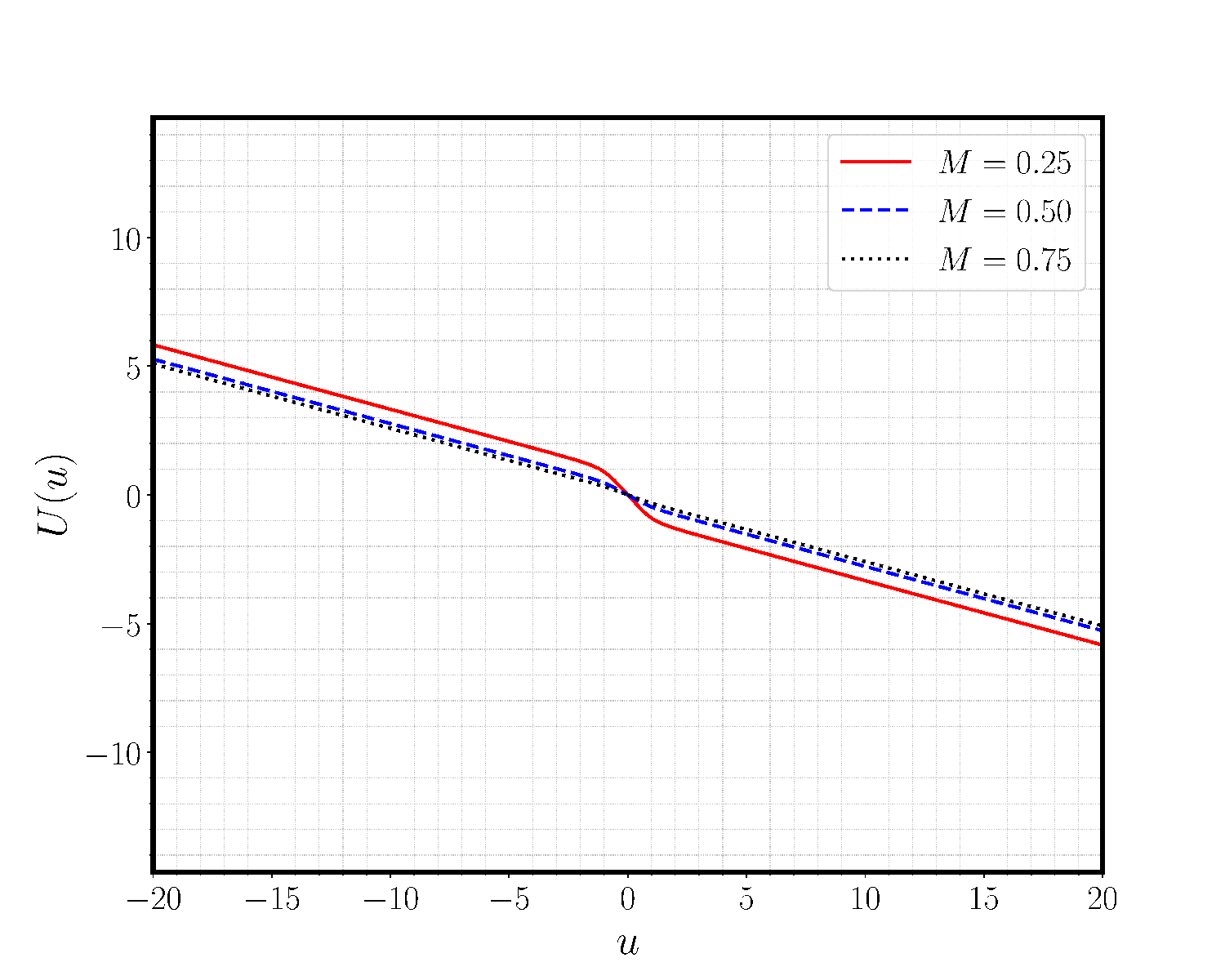}
    \caption{$h = -1$}
    \label{3_h_n}
  \end{subfigure}
  \caption{Two illustrations of the function $U(u)$ (\ref{U_ext3_expression}), that appears in the third extension, corresponding to two choices of $h$.}
\label{3_h_p_n}
\end{figure}
\subsection{Surfaces of Constant Radius}
\begin{figure*}[htb]
  \subcaptionbox{$M = 0.25$}[.32\linewidth][c]{%
    \includegraphics[width=1\linewidth]{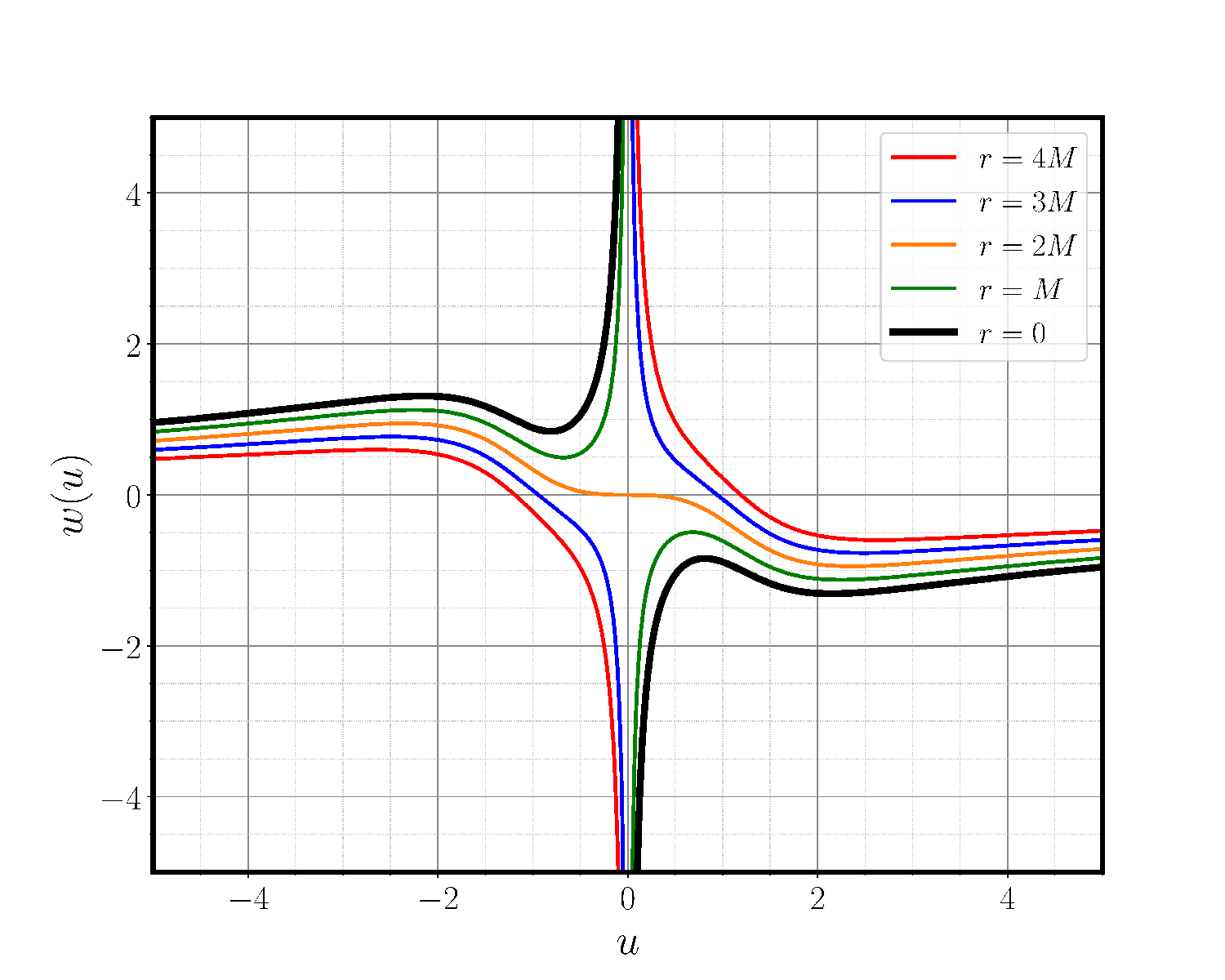}}\quad
  \subcaptionbox{$M = 0.5$}[.32\linewidth][c]{%
    \includegraphics[width=1\linewidth]{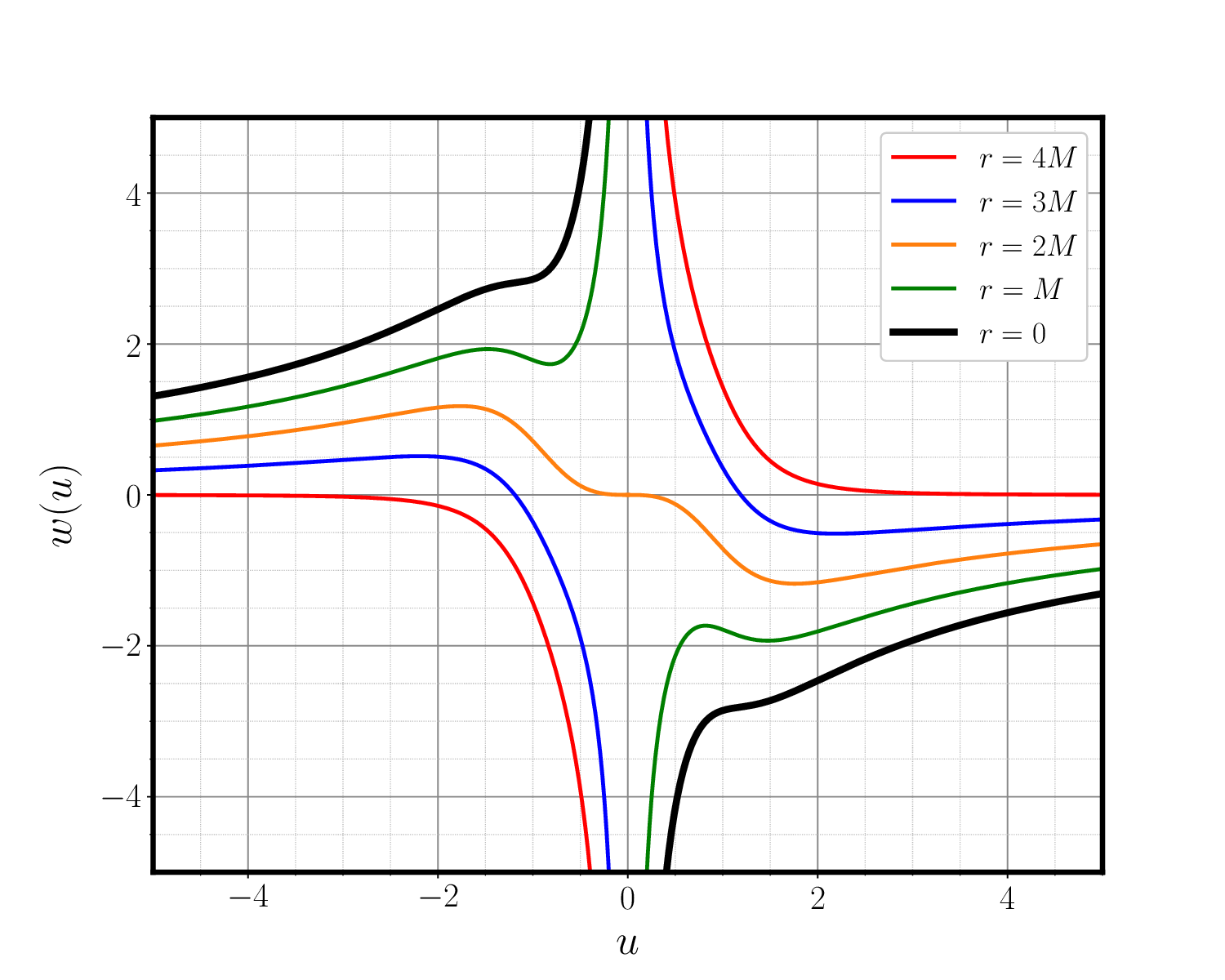}}\quad
  \subcaptionbox{$M = 0.75$}[.32\linewidth][c]{%
    \includegraphics[width=1\linewidth]{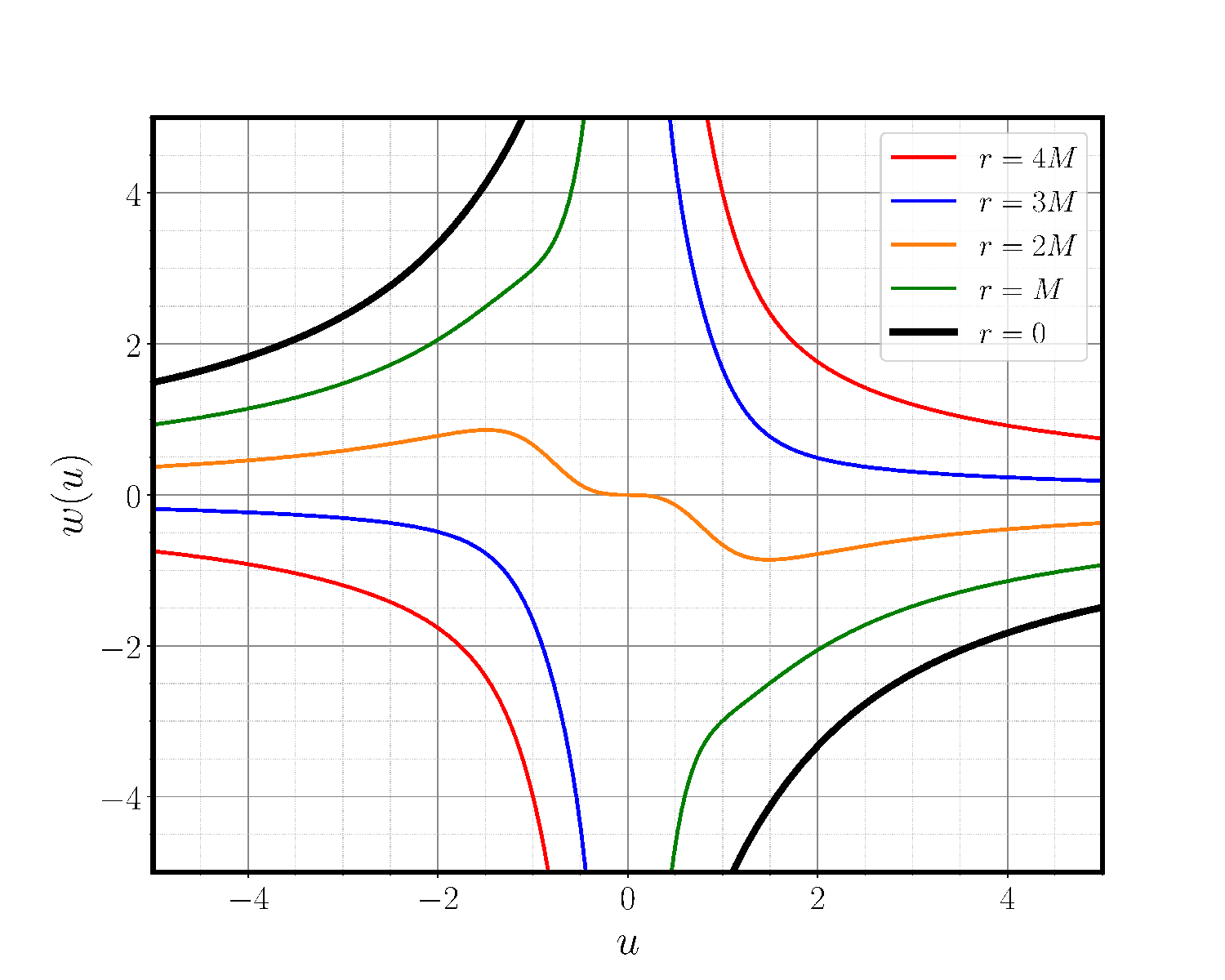}}
  \subcaptionbox{$M = 0.25$}[.32\linewidth][c]{%
    \includegraphics[width=1\linewidth]{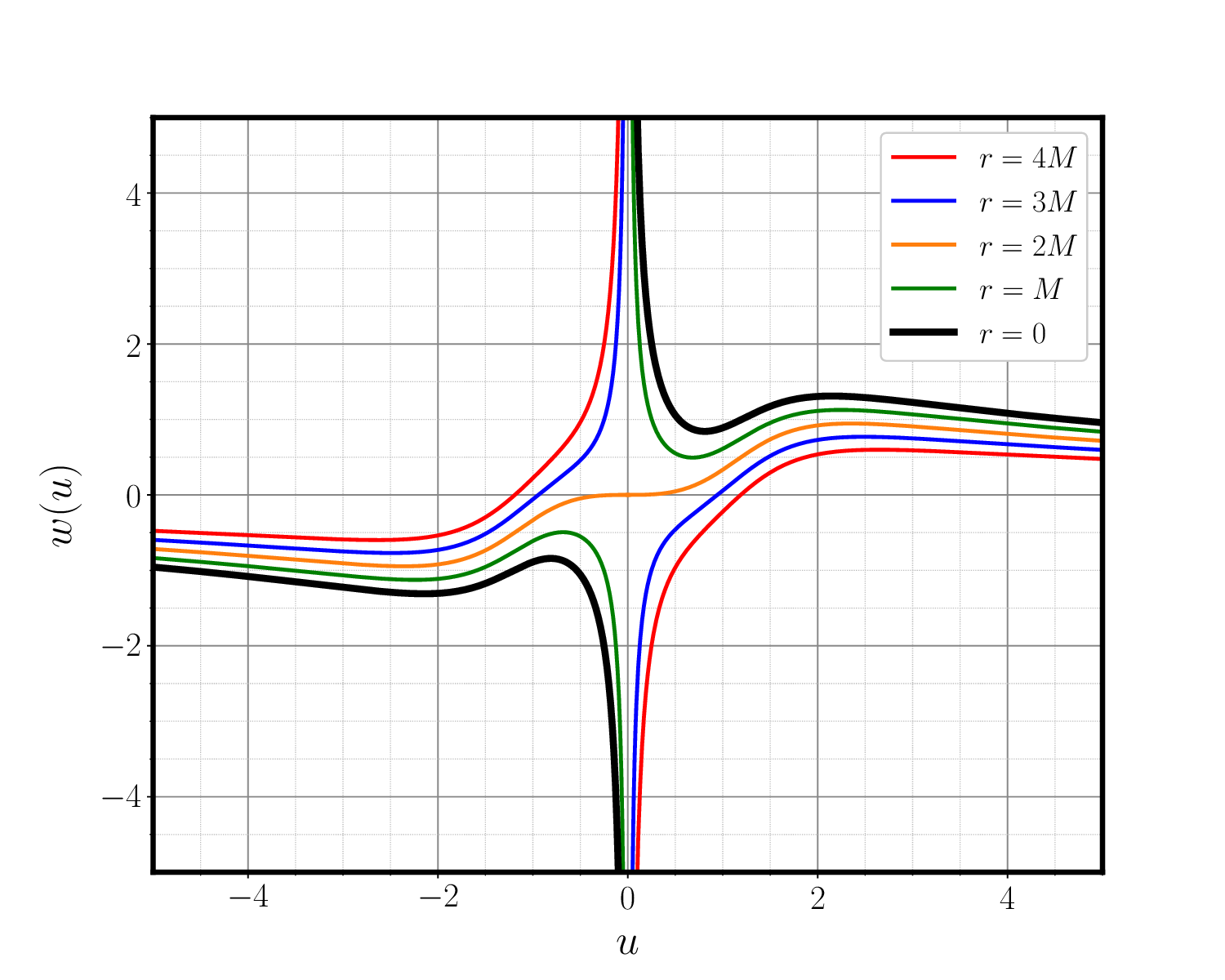}}\quad
  \subcaptionbox{$M = 0.5$}[.32\linewidth][c]{%
    \includegraphics[width=1\linewidth]{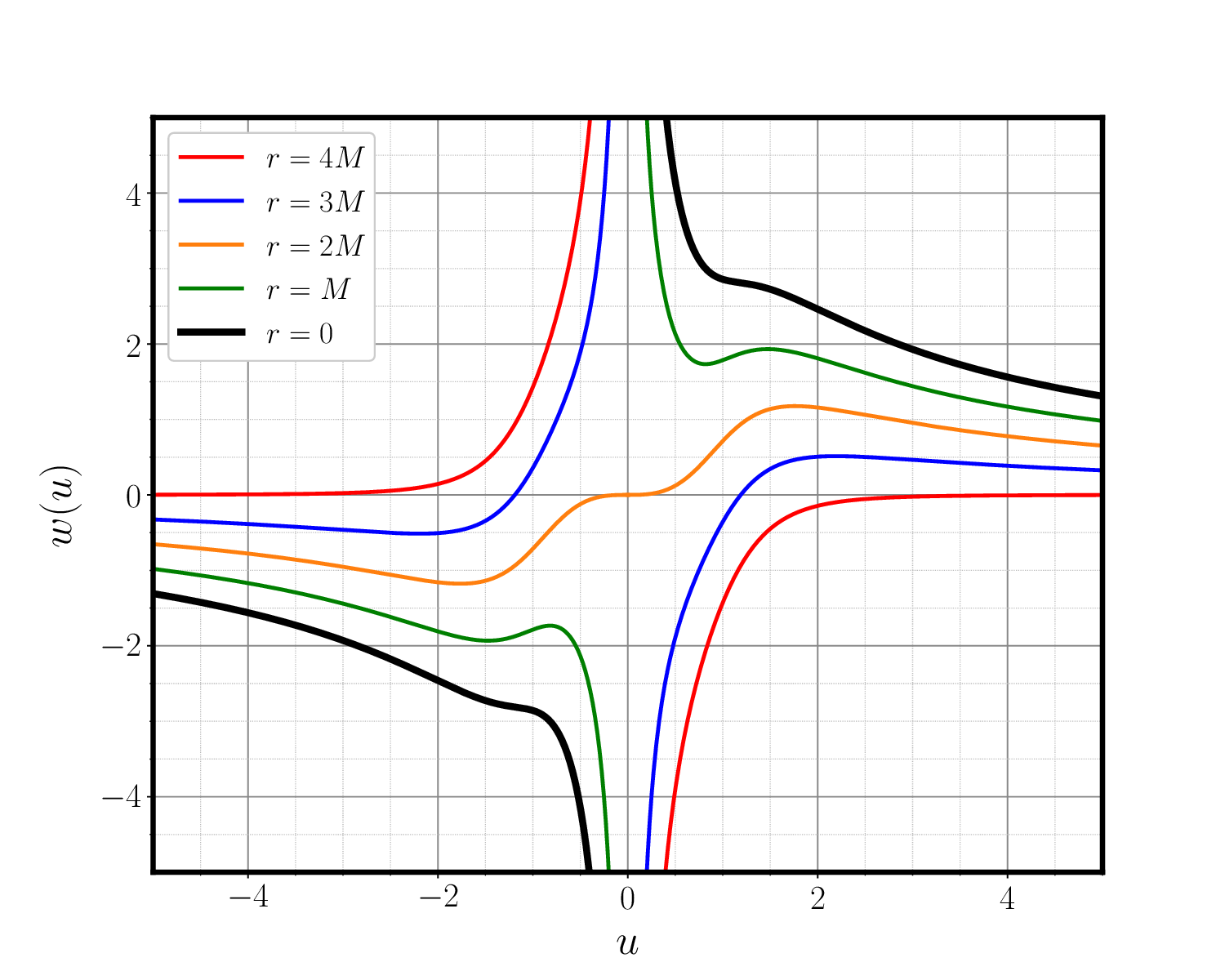}}\quad
  \subcaptionbox{$M = 0.75$}[.32\linewidth][c]{%
    \includegraphics[width=\linewidth]{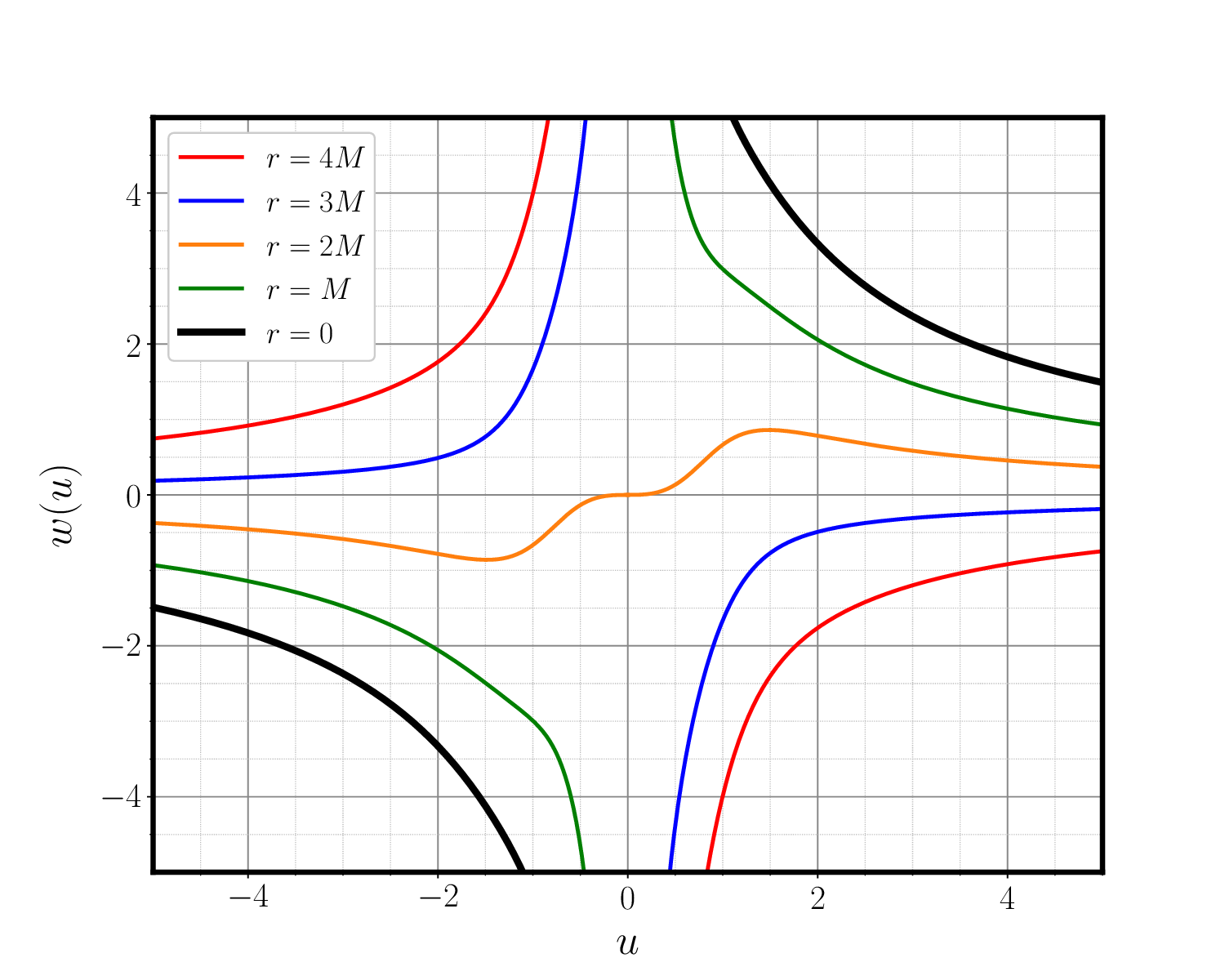}}
\caption{The top row shows surfaces of constant radius for the third extension with $h = +1$, and the bottom row exhibits the same with $h = -1$. As $M$ increases, the bulges become less prominent, whereas reversing the sign of $h$ gives a mirrored version of the surfaces around the horizontal axis.}
\label{Ext3_1_ConstR}
\end{figure*}

The surfaces of constant radius, $r(u,w) = \alpha M$, are explicitly given by 
\begin{equation}
\label{Ext3_1_ConR_eq}
    w= \frac{\alpha M-2m(u)}{U(u)},
\end{equation}
where $\alpha \in \mathbb{N}$. The causality of the surfaces $r(u,w) = \alpha M$ is again determined through the sign of the Lagrangian associated wit this extension
\begin{equation}
    2\mathscr{L}|_{r(u,w) = \alpha M} = \left(-\frac{w}{\alpha M U(u)}\right) \dot{u}^2. 
\end{equation}
We note that, see Fig. \ref{Ext3_1_ConstR}, all the surfaces $r(u,w)=\text{const}$, for the choice $h = \pm 1$, are characterized as: timelike (spacelike) surfaces in quadrants I $(u>0 ~\&~ w>0)$ and III $(u<0 ~\&~ w<0)$, spacelike (timelike) in quadrants II $(u<0 ~\&~ w>0)$ and IV $(u>0 ~\&~ w<0)$. Therefore, The physical singularity, the surface $r=0$, is again a spacelike hypersurface.
Notably, we no longer have the relation $uw=\text{const}$ that yields perfect hyperbolas graphs for the surfaces $r(u,w)=\text{const}$. Therefore, the bulges in the surfaces of constant radius appear to the left and to the right of the $w$-axis. Needless to say that the reason for this state of affairs is that the following equation indeed have roots 
\begin{equation}\label{roots_bulges_ext3}
     0 = \frac{dw}{du} = \frac{-2m^{'}(u)U(u)-\frac{h}{4m(u)}\big(\alpha M-2m(u)\big)}{U(u)^2}. 
 \end{equation}
\begin{figure}[htb]
     \includegraphics[width=1\linewidth]{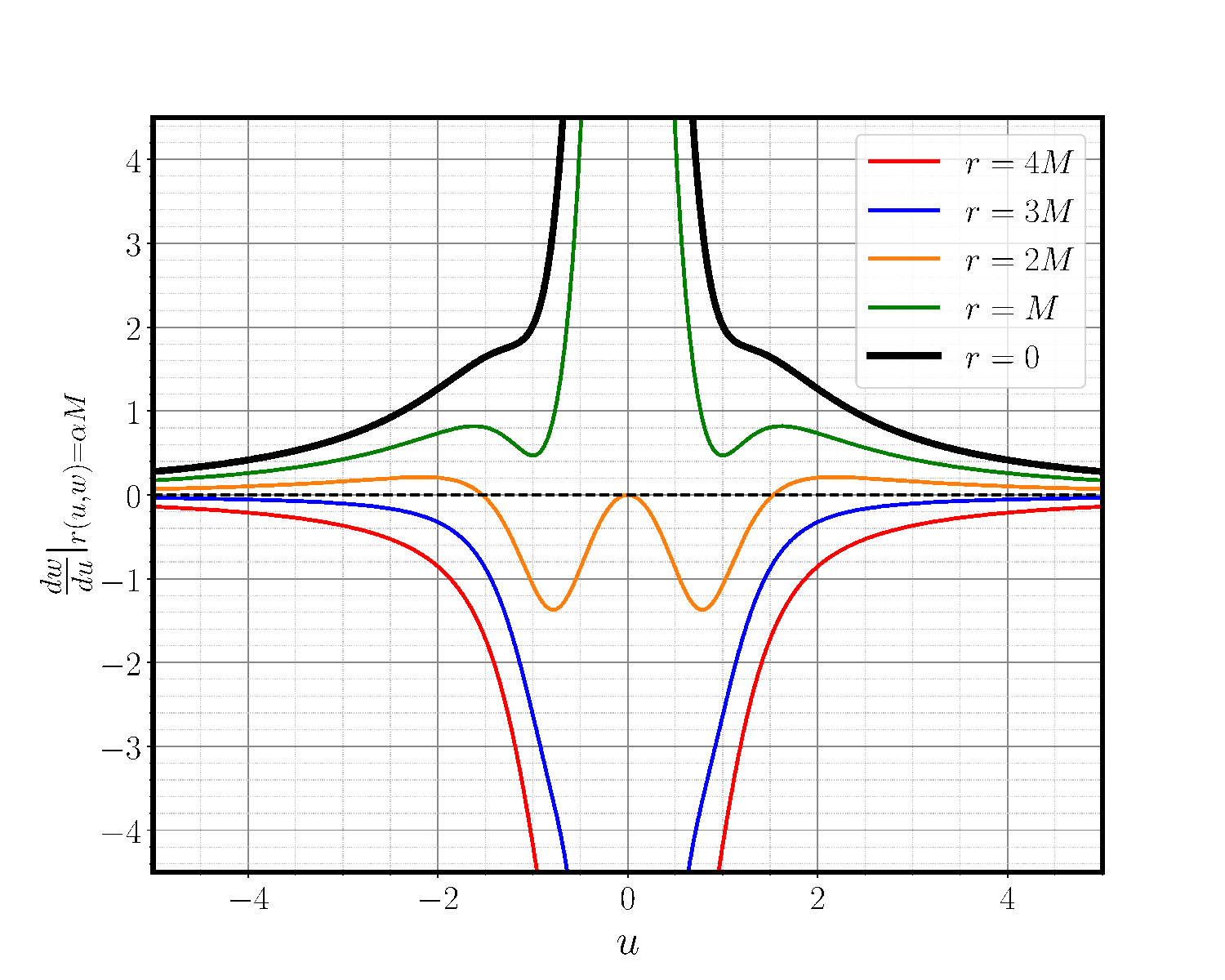}
     \caption{The graphical locations of the roots of (\ref{roots_bulges_ext3}) for the choices $h=+1$ and $M = 0.75$.}
     \label{bulges_ext3}
 \end{figure}
However, we still see the effect of increasing the value of the parameter $M$ on removing most of the bulges that appear with those surfaces. The only bulge, see Figure \ref{bulges_ext3}, that resists the increase in the value of $M$ is the one appearing with the curve $r=2m(u)$. 
\subsection{Surfaces of Dynamical Radius}
\begin{figure*}[htb]
  \subcaptionbox{$M = 0.25$}[.32\linewidth][c]{%
    \includegraphics[width=1\linewidth]{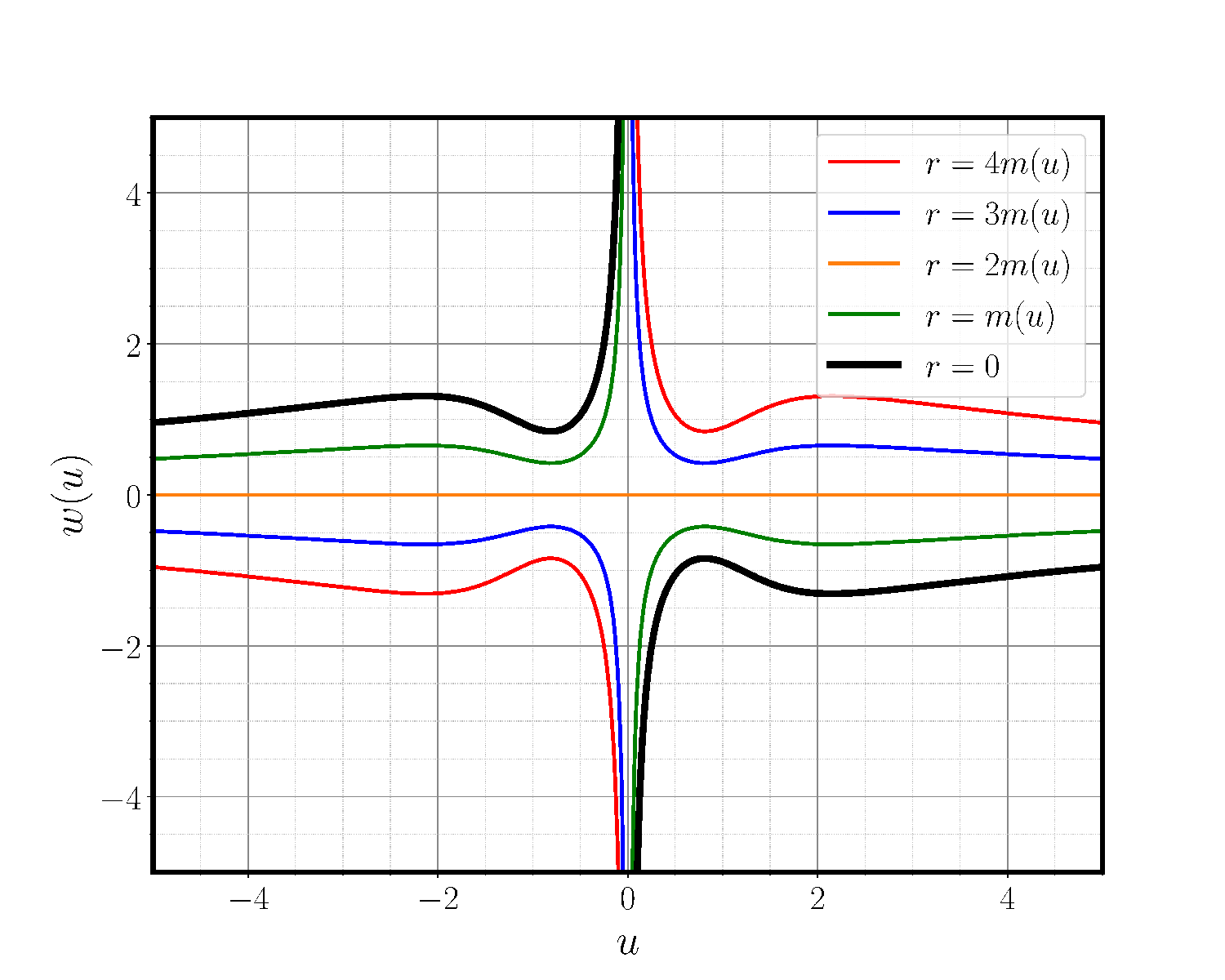}}\quad
  \subcaptionbox{$M = 0.5$}[.32\linewidth][c]{%
    \includegraphics[width=1\linewidth]{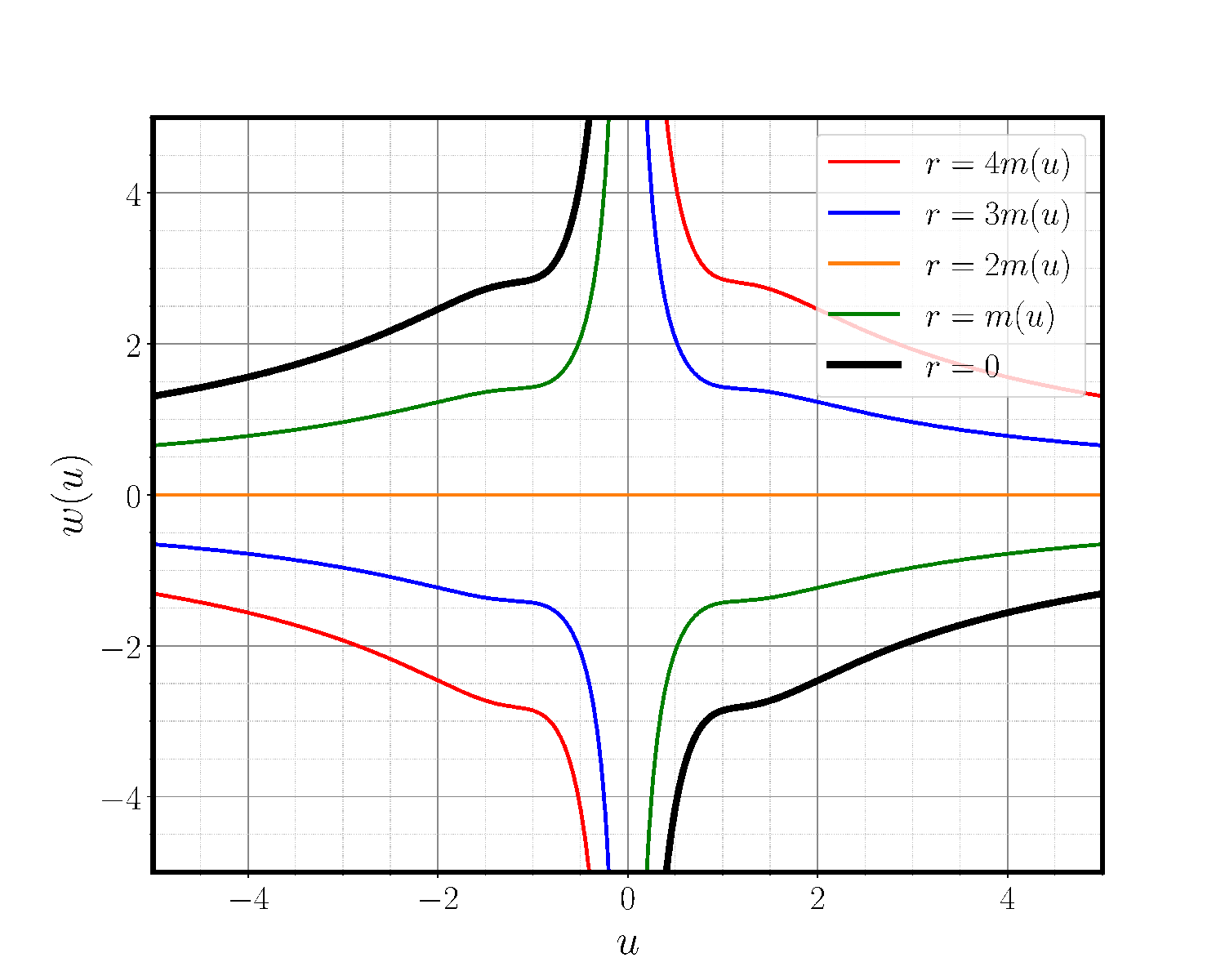}}\quad
  \subcaptionbox{$M = 0.75$}[.32\linewidth][c]{%
    \includegraphics[width=1\linewidth]{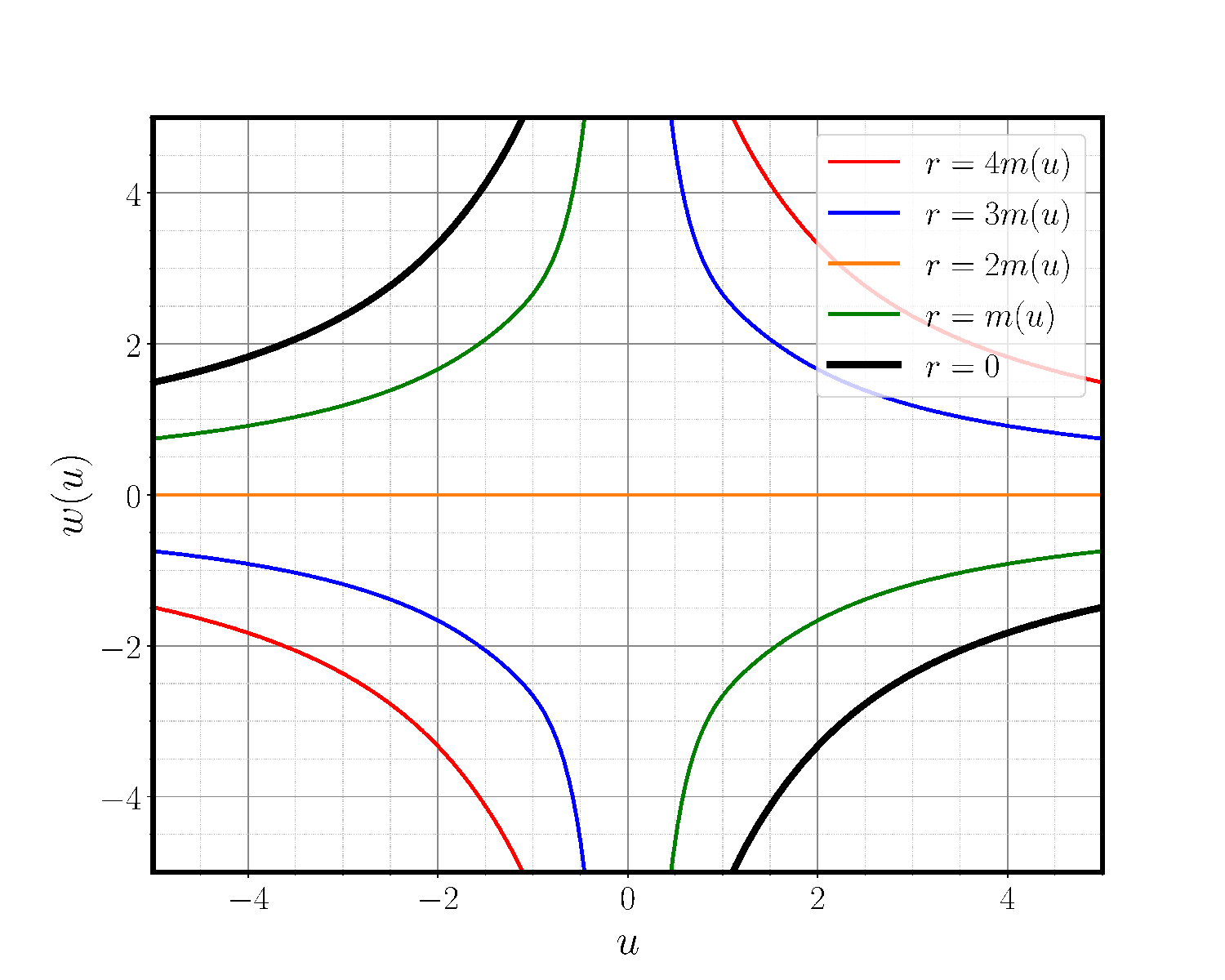}}
  \subcaptionbox{$M = 0.25$}[.32\linewidth][c]{%
    \includegraphics[width=1\linewidth]{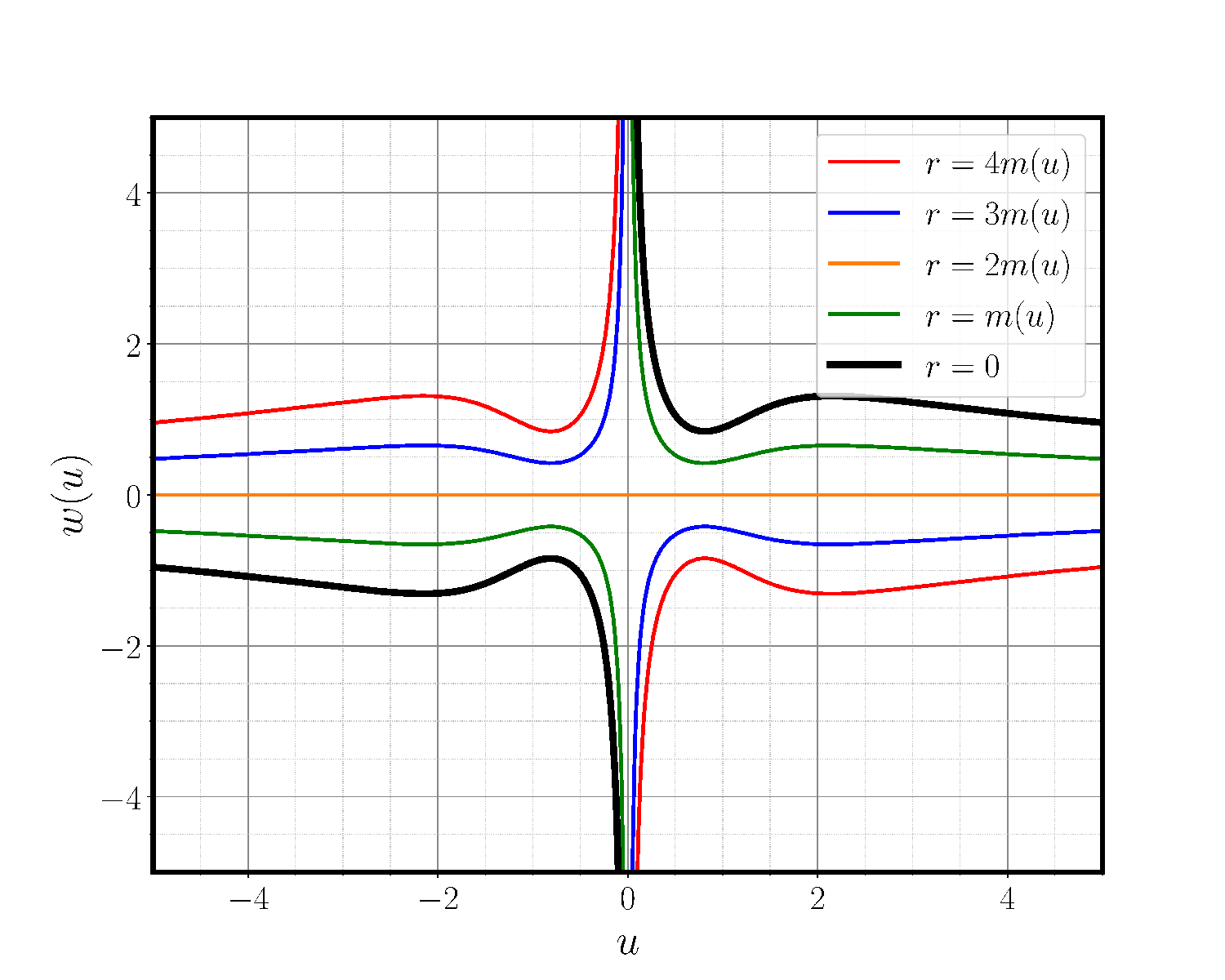}}\quad
  \subcaptionbox{$M = 0.5$}[.32\linewidth][c]{%
    \includegraphics[width=1\linewidth]{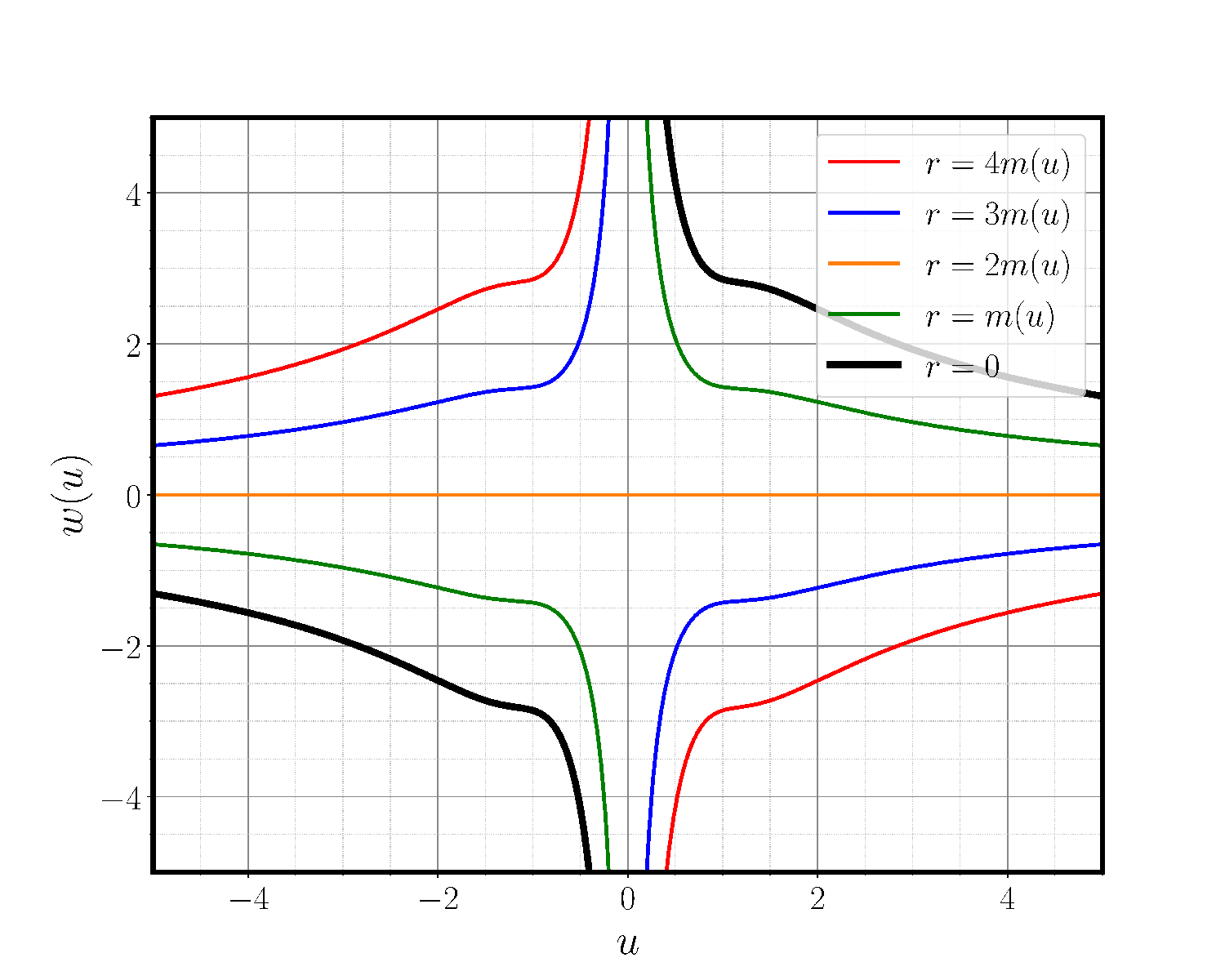}}\quad
  \subcaptionbox{$M = 0.75$}[.32\linewidth][c]{%
    \includegraphics[width=\linewidth]{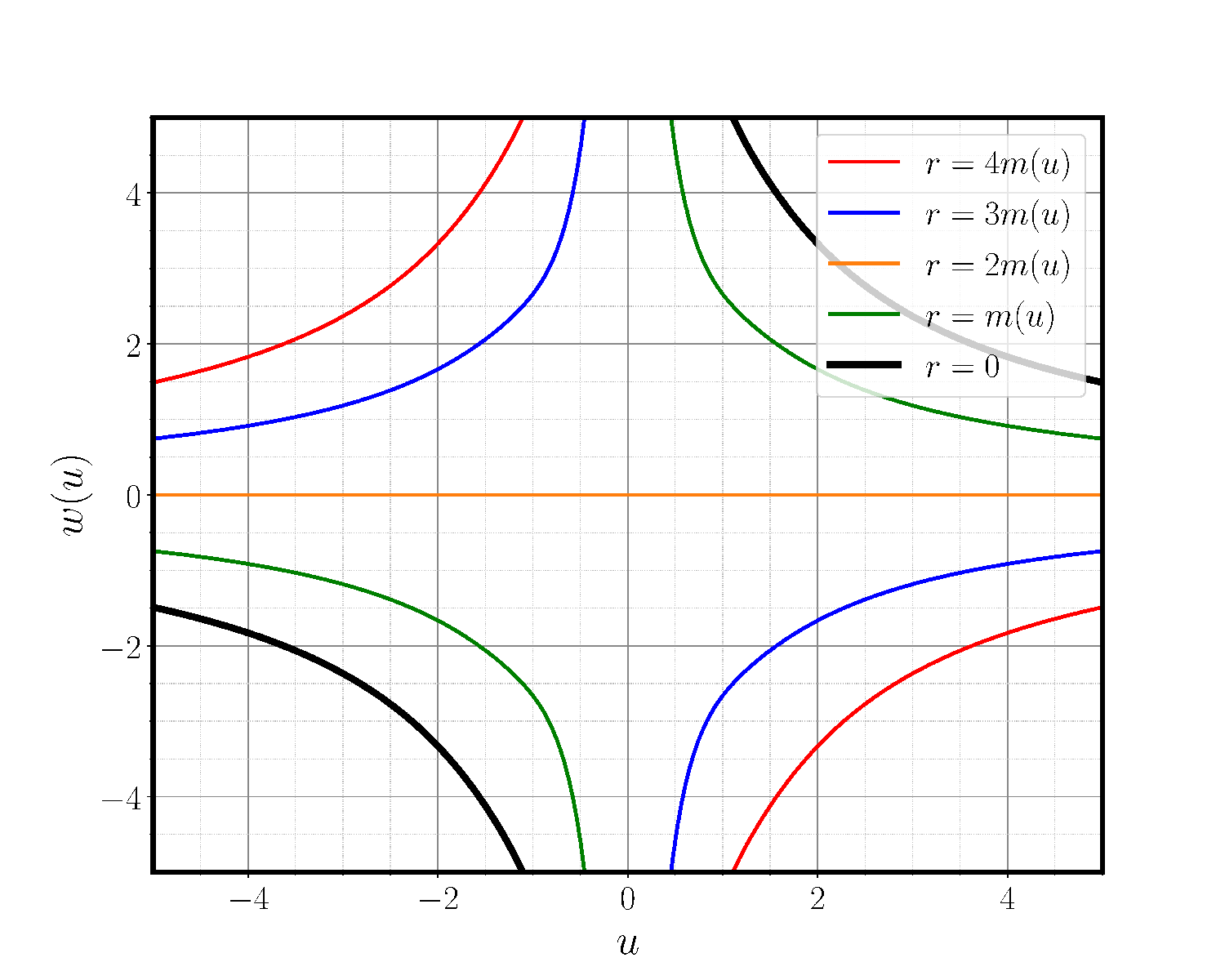}}
\caption{The top row displays surfaces of dynamical radius in the third extension for the choice of $h = +1$, while the bottom row displays the same for $h = -1$. Similar to surfaces of constant radius, increasing the value of $M$ leads to a less pronounced bulge, while switching the sign of $h$ creates a reflection of the surfaces.}
\label{Ext3_1_dynR}
\end{figure*}

\vspace{-2pt}
We once more explore the surfaces of dynamical radius to avoid the causality breaches, i.e., the surfaces are not uniformly timelike or spacelike that are present in the surfaces of constant radius. These surfaces are still specified in (\ref{dyn_radius_def}) and (\ref{Lag_dyn_radii}), and they can be classified as spacelike when $\alpha \leq 2$ or timelike when $\alpha > 2\big( 1+\alpha^2 \chi(u)\big)$. In this extension the surfaces cannot be null since $m^{'}(u)\neq0$, either in the region $u<0$ or in the region $u\geq 0$, see Fig. \ref{Ext3_1_dynR}. Similar to the previous two extension, the surfaces of dynamical radius still form bulges which occur when $\chi(u)=\frac{1}{4}$.
\section{ Completeness of  Israel coordinates } \label{Sec: Completeness}
In this section, we demonstrate the completeness of  Israel coordinates by locally studying the behavior of radial null geodesics in the three extensions and globally by constructing the corresponding Penrose diagrams. The radial $(\theta=\phi=\text{constant})$ null geodesics in the Vaidya metric are solutions to the equation
\begin{equation}\label{Null_eq1}
0 = \left(\frac{w^2}{2m(u)r(u,w)}+\frac{4m^{'}(u)h}{U(u)}\right)du^{2}+2 h du dw.
\end{equation}
It is evident that (\ref{Null_eq1}) admits two solutions. The first solution is the trivial solution $ u = \text{const}$, which represents radial null geodesics affinely parametrized by $w$. The second solution is obtained by integrating
\begin{equation}\label{Null_eq}
\frac{dw}{du} = \frac{-1}{2h} \left(\frac{w^2}{2m(u)r(u,w)}+\frac{4m^{'}(u)h}{U(u)}\right).
\end{equation}
 The previous equation is classified as Abel second type class A for which no known exact solution exists, so solutions must be obtained numerically. Here, we use the package ``\texttt{ODEINT}''\cite {2020SciPy-NMeth} and provide suitable initial conditions, $w(u_0)=w_0$, for each trajectory. 
 \subsection{Radial Null Geodesics in the First Extension}
\begin{figure*}[htb]
  \begin{subfigure}{0.85\columnwidth}
    \includegraphics[width=\linewidth]{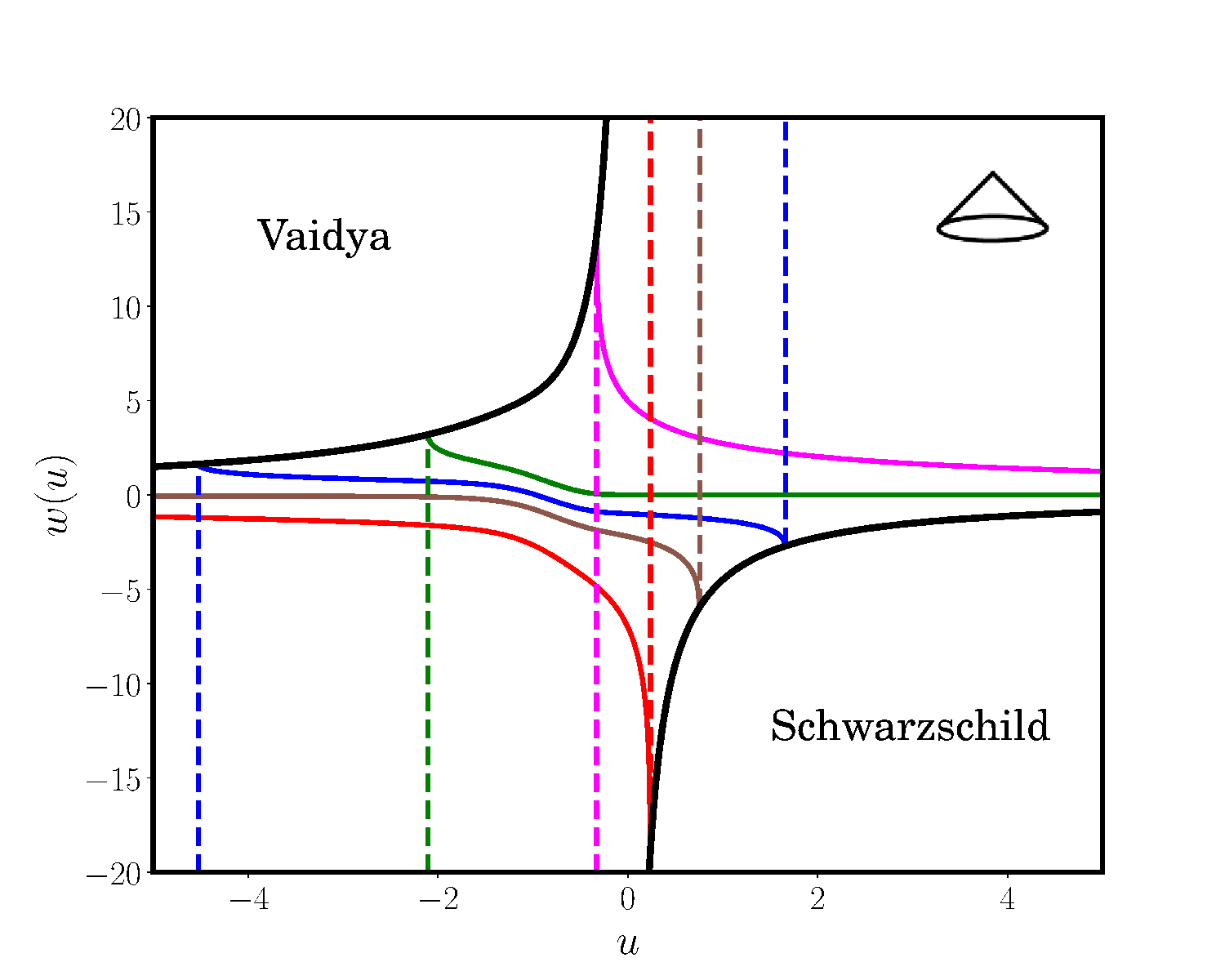}
    \caption{$h = +1$}
    \label{ext1_h_p}
  \end{subfigure}
  \begin{subfigure}{0.85\columnwidth}
    \includegraphics[width=\linewidth]{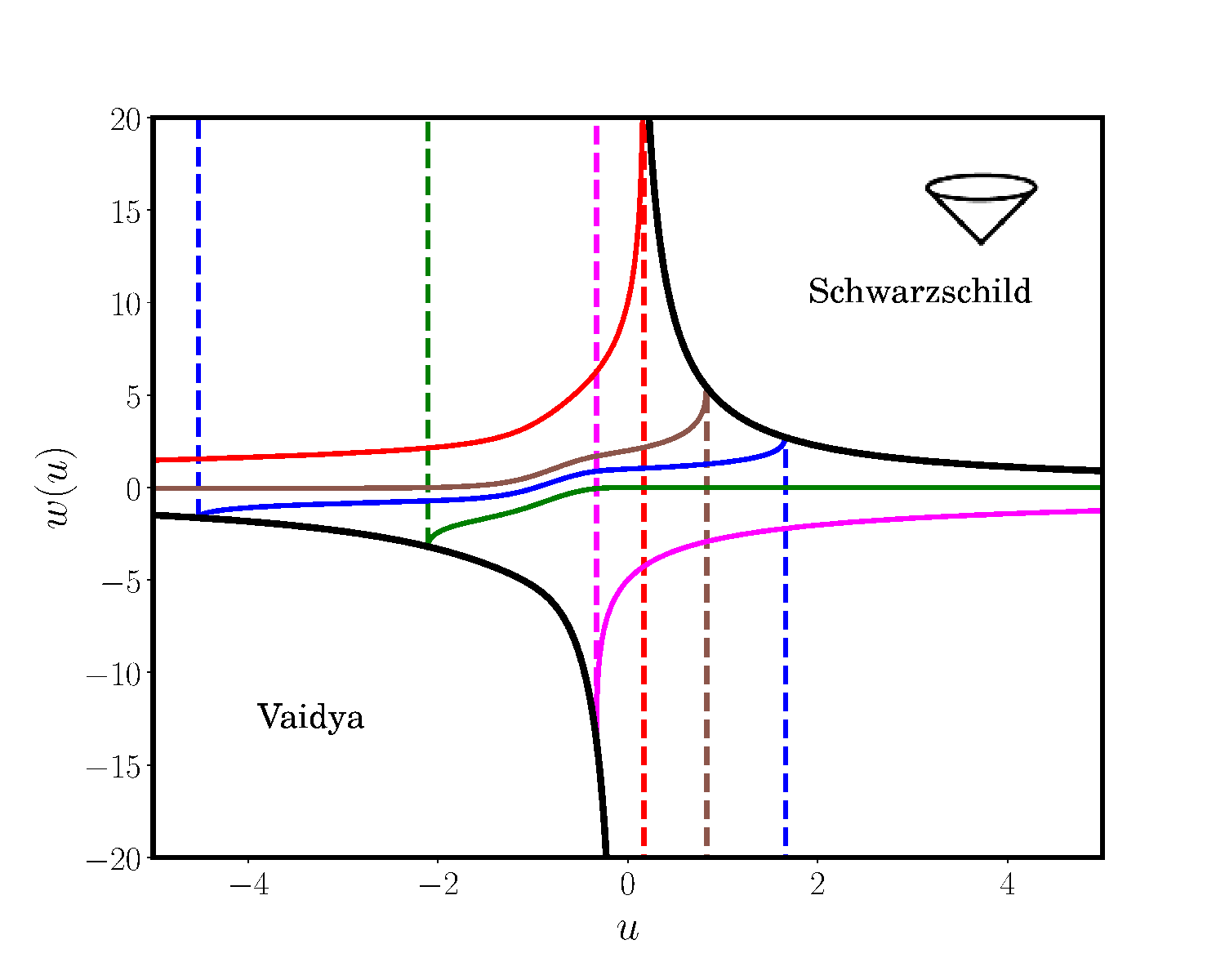}
    \caption{$h = -1$}
    \label{ext1_h_n}
  \end{subfigure}
  \begin{subfigure}{1.2\columnwidth}
\includegraphics[width=\linewidth]{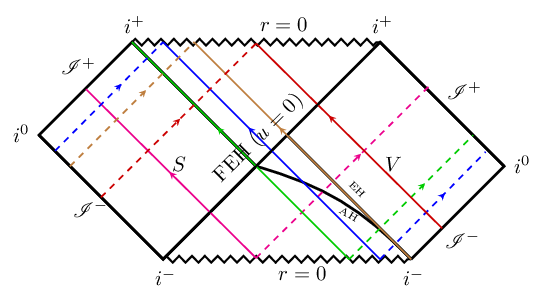}
    \label{ext1_Penrose}
  \end{subfigure}
  \caption{Top: an illustration is presented of the radial null geodesics in the first maximal extension of Vaidya metric for both $h=+1$ (left) and $h=-1$ (right). Solid (non-affinely parametrized) and dashed (affinely parametrized) curves are used to represent the two branches of the radial null geodesics, with black jagged lines indicating $r(u,w)=0$. The future null cone is depicted for orientation. Bottom: the Penrose diagram of the first maximal extension. The future direction on the Penrose diagram is always represented upwards on the page. Additionally, the future event horizon (FEH) of the non-affinely parameterized radial null geodesics is also illustrated on the Penrose diagram.}
  \label{Vaidya_out} 
\end{figure*}

  As the WEC, $8\pi\Phi = \frac{2hm^{'}(u)}{U(u)r(u,w)^2}>0$, indicates, the condition $m'(u)< 0$ implies that when $h=+1$ then $U(u)<0$, whereas for $h=-1$ this results in $U(u) > 0$. To assess both cases, we employ the mass function (\ref{mass_outgoing_F1}) with $M_1 = 0.5$. There is an infinite number of radial null geodesics that can be demonstrated (see Fig. \ref{Vaidya_out}) \footnote{There is a Penrose diagram similar to that in Fig. \ref{Vaidya_out} in \cite{Griffiths:2009dfa}, as depicted in their figure 9.16 (top). However, that Penrose diagram does not take into account studying the validity of  Israel coordinates.}. However, only representatives of all the possible scenarios are considered. The solid magenta curve represents a family of radial null geodesics that originate from Vaidya's white hole singularity and continue to the Schwarzschild part, without crossing the event horizon of the Schwarzschild black hole. In fact, these null geodesics eventually reach $\mathscr{I^{+}}$. The solid green curve is distinct; it comes from Vaidya's white hole and coincides with the Schwarzschild event horizon ($w=0$), before ultimately reaching $i^{+}$ in the future. The solid blue curve is an example of a congruence of null geodesics which originate from Vaidya's singularity, cross Vaidya's apparent horizon ($w=0$), and finally hit Schwarzschild's singularity. The solid brown curve is a unique null geodesic, tracing down the location of Vaidya's event horizon and continuing to hit the spacelike singularity surface $r_\text{Schw} = 0$. Finally, the solid red curve is not unique, but it is the first null geodesic, compared with the aforementioned null geodesics, to emerge from outside Vaidya's event horizon, from $\mathscr{I^{-}}$. This geodesic then continues until it hits the Schwarzschild singularity. The only noticeable difference between the case $h=+1$, see Fig. \ref{ext1_h_p}, and the case $h=-1$, see Fig. \ref{ext1_h_n}, is that the future direction is pointing downward when $h=+1$ and the future direction is pointing upward when $h=-1$. Thus, in this extension, and the other two extensions yet to be discussed, the only change when we switch the sign of $h$ is the direction of the future in the $u-w$ diagrams. We want to emphasize that this was not known until a thorough analysis of all of the constructed extensions was conducted. Now that we have established that there is no difference in the underlying physics between the cases of $h=+1$ and $h=-1$, one can choose either to work with. Moreover, we call the attention to the fact that this case was implicit in \cite{Lake_2006}, but that case was static. 
\subsection{Radial Null Geodesics in the Second Extension}
\begin{figure*}[htb]
  \begin{subfigure}{0.85\columnwidth}
    \includegraphics[width=\linewidth]{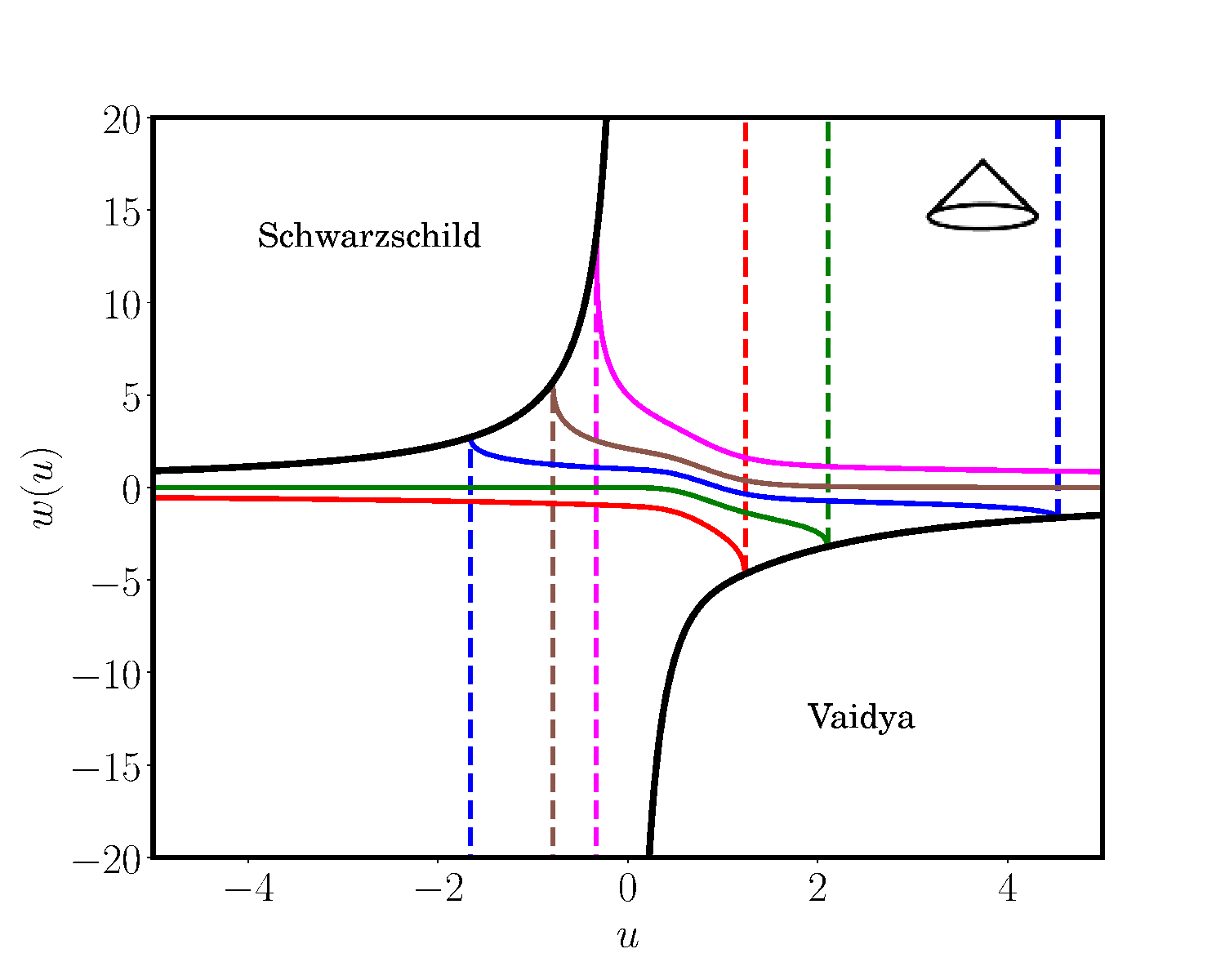}
    \caption{$h = +1$}
    \label{ext2_h_p}
  \end{subfigure}
  \begin{subfigure}{0.85\columnwidth}
    \includegraphics[width=\linewidth]{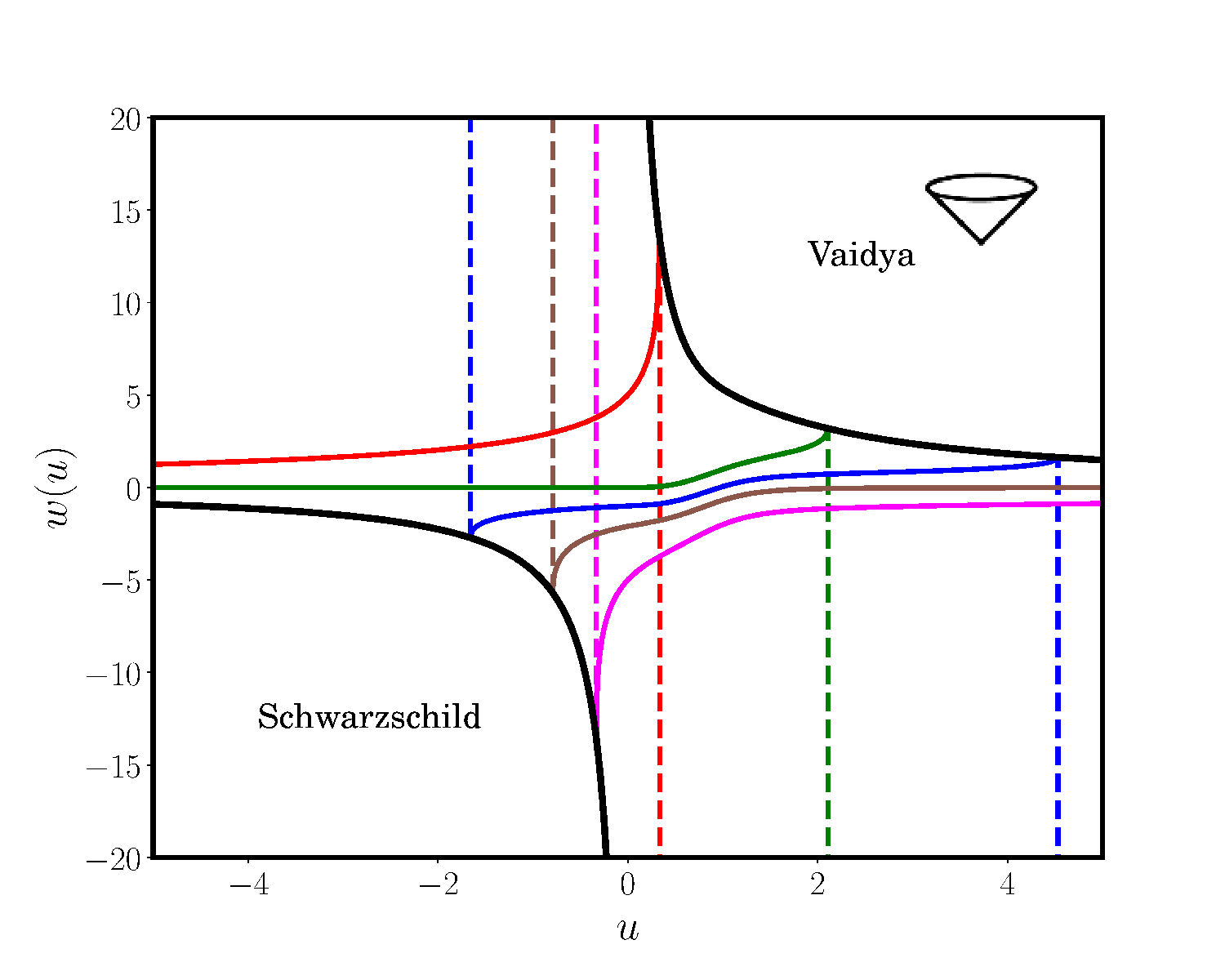}
    \caption{$h = -1$}
    \label{ext2_h_n}
  \end{subfigure}
  \begin{subfigure}{1.2\columnwidth}
\includegraphics[width=\linewidth]{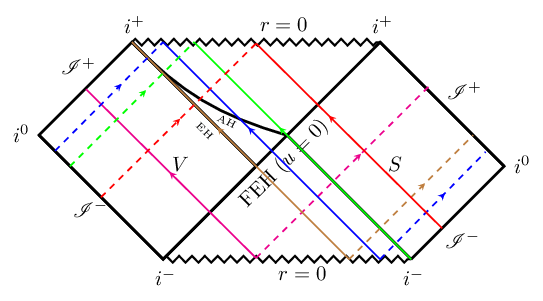}
\label{ext2_Penrose}
  \end{subfigure}
  \caption{Top: an illustration is presented of the radial null geodesics in the second maximal extension of Vaidya metric for both $h=+1$ (left) and $h=-1$ (right). Solid (non-affinely parametrized) and dashed (affinely parametrized) curves are utilized to mark the two branches of the radial null geodesics, with black jagged lines indicating the singularity surface $r(u,w)=0$. The future null cone is depicted for orientation. Bottom: the Penrose diagram of the second maximal extension. The future direction on the Penrose diagram is always represented upwards on the page. Additionally, the future event horizon (FEH) of the non-affinely parameterized radial null geodesics is also illustrated on the Penrose diagram.}
  \label{Vaidya_in} 
\end{figure*}

 In a similar fashion to the outgoing case, two different choices of the functions $h$ and $U(u)$ are considered. Since the ingoing radiation is characterized by $m^{'}(u)>0$, the choice $h = +1$ implies $ U(u) > 0$, and the choice $ h = -1$ results in $U(u) < 0$. Both cases are examined below, using the mass function (\ref{mass_ingoing_final}) with $M_0 = 0.5$. The trajectories of the radial null geodesics that flow from the Schwarzschild white hole and proceed to the ingoing Vaidya section are given in Fig. \ref{Vaidya_in}. Similar to the outgoing case, the solid magenta curve represents a whole class of geodesics which emerges from the Schwarzschild white hole singularity and travels to the ingoing Vaidya section without ever crossing the apparent horizon ($w = 0$). The curve starts from $r_{\text{Schw}} = 0$ and ultimately hits $\mathscr{I^{+}}$ in the Vaidya's section. The solid brown curve is unique in the sense that it is the only null geodesic that evolves from the Schwarzschild's singularity and is asymptotic to the Vaidya dynamic apparent horizon at $i^{+}$. Thus, the previous null geodesic actually never makes it to the surface $r_{\text{Vaidya}} = 0$ in a finite value of its affine parameter. The solid blue curve represents a class of null geodesics that originates at $r_{\text{Schw}} = 0 $ and cannot clearly evade an encounter with $r_{\text{Vaidya}} = 0 $. The solid green curve is another unique geodesic since it is the only one that traces down the location of the event horizon of the Schwarzschild metric, but it separates from the Vaidya's apparent horizon and continues moving to finally hit the singularity in the Vaidya part. Finally, a candidate from the family of the null geodesics which originate at the past null infinity, $\mathscr{I^{-}}$, in the Schwarzschild spacetime is given by the solid red curve. This null geodesic, as shown, cannot elude an encounter with the ingoing Vaidya's singularity surface, $r(u,w) = 0$.
\subsection{Radial Null Geodesics in the Third Extension}
\begin{figure*}[htb]
  \begin{subfigure}{0.85\columnwidth}
    \includegraphics[width=\linewidth]{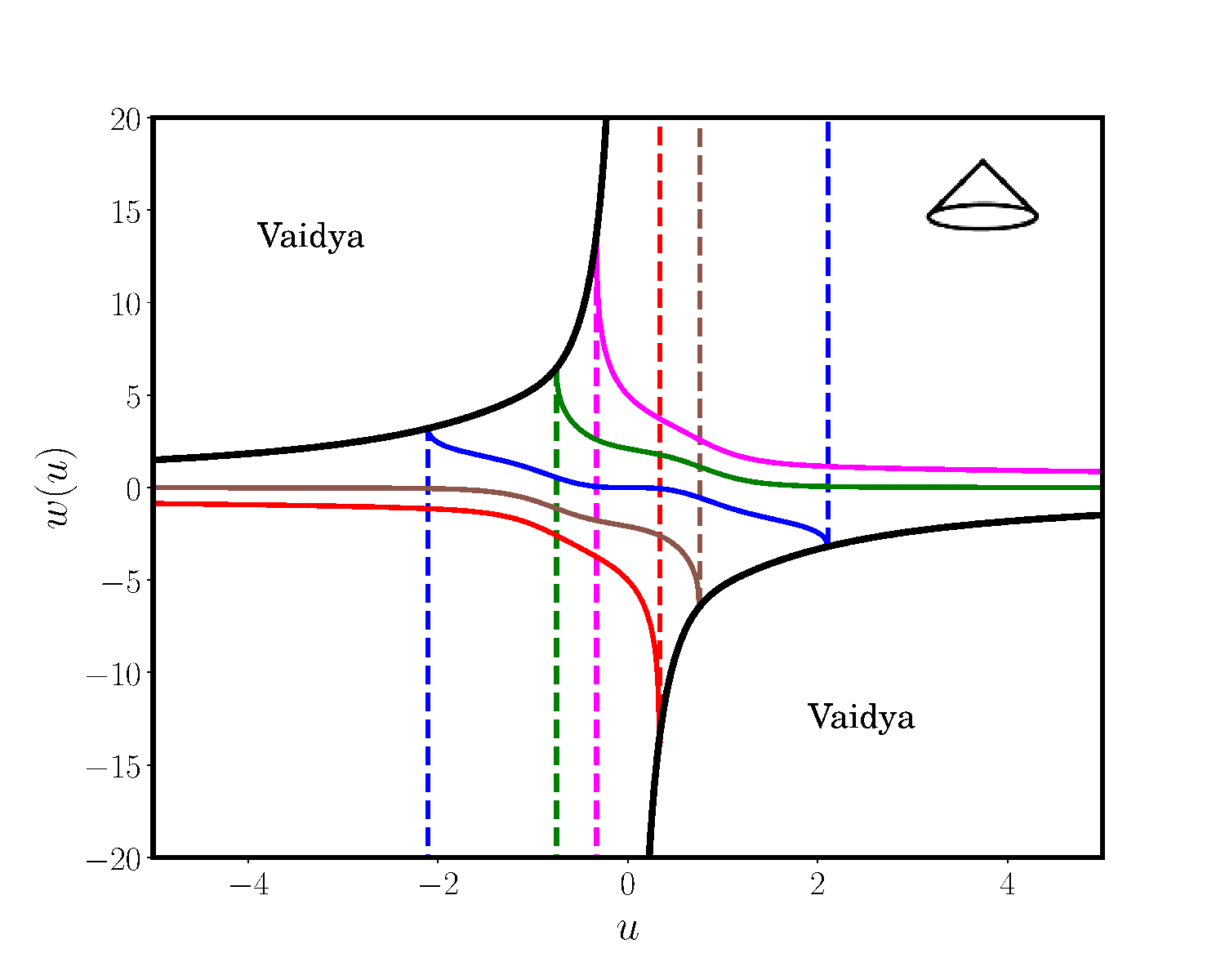}
    \caption{$h = +1$}
    \label{ext3_h_p}
  \end{subfigure}
  \begin{subfigure}{0.85\columnwidth}
    \includegraphics[width=\linewidth]{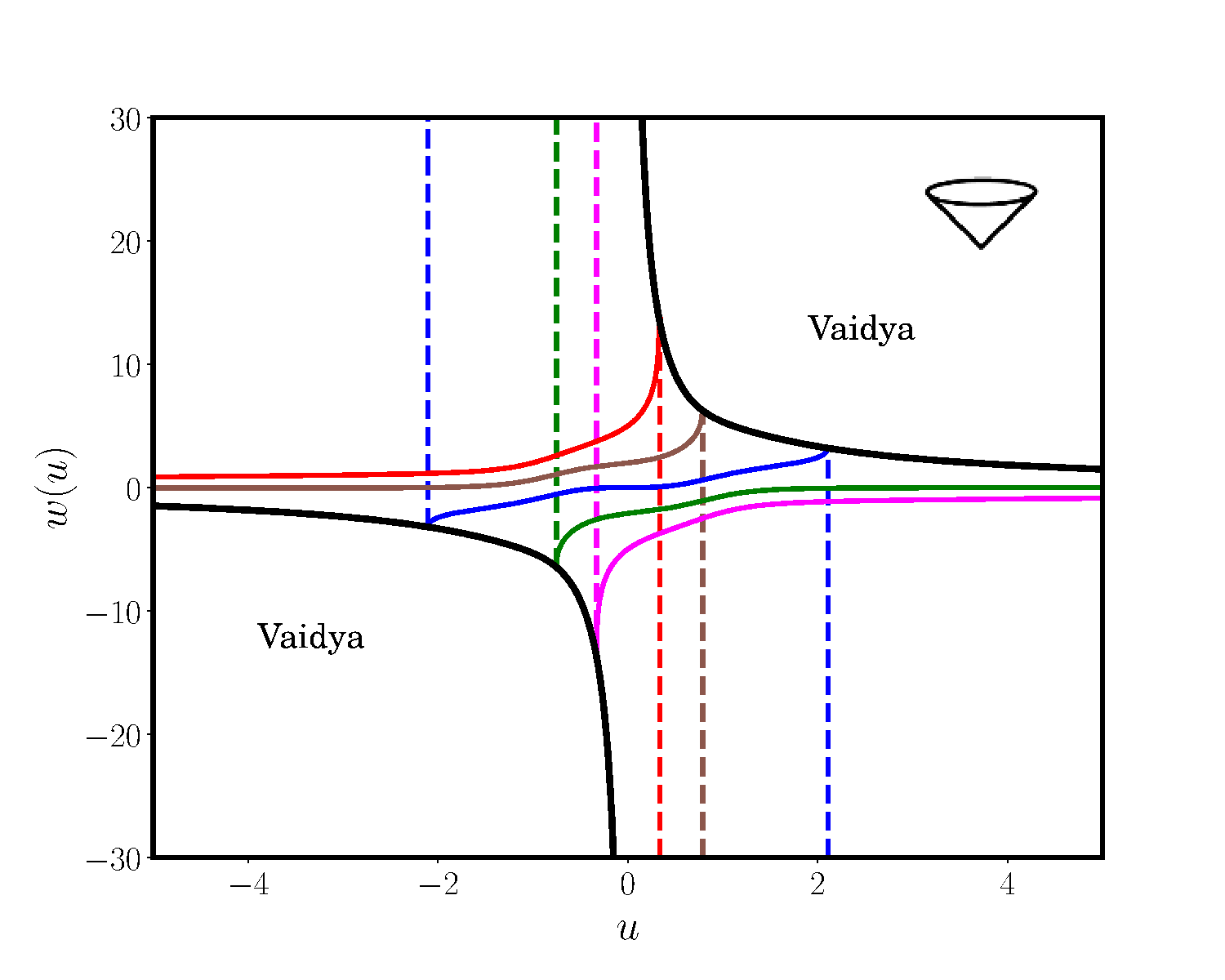}
    \caption{$h = -1$}
    \label{ext3_h_n}
  \end{subfigure}
  \begin{subfigure}{1.2\columnwidth}
    \includegraphics[width=\linewidth]{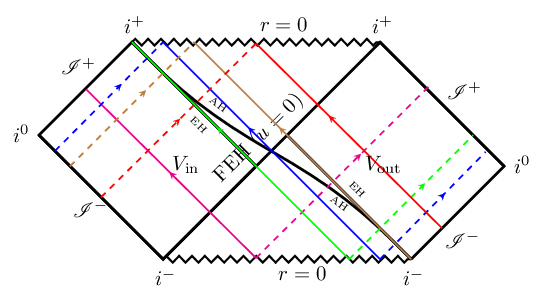}
    \label{ext3_Penrose}
  \end{subfigure}
  \caption{Top: an illustration is presented of the radial null geodesics in the third maximal extension of Vaidya metric for both $h=+1$ (left) and $h=-1$ (right). Solid (non-affinely parametrized) and dashed (affinely parametrized) curves are utilized to mark the two branches of the radial null geodesics, with black jagged lines indicating the singularity surface $r(u,w)=0$. The future null cone is depicted for orientation. Bottom: the Penrose diagram of the third maximal extension. The future direction on the Penrose diagram is always represented upwards on the page. Additionally, the futures event horizon (FEH) of the non-affinely parameterized radial null geodesics is also illustrated on the Penrose diagram.}
  \label{Vaidya_out_in} 
\end{figure*}

Having exhausted the information from the previous two extensions, we now turn our attention to exploring the behavior of the radial null geodesics existing in the outgoing Vaidya metric extended to the ingoing version. Similar to the previous two extensions, we choose address both the cases $h=+1$ and the case $h=-1$, where the mass function that gives rise to this extension is chosen to be (\ref{mass_out_in_M}), with $M=0.75$.  The trajectories of the radial null geodesics that flow from the outgoing Vaidya metric and proceed to the ingoing Vaidya metric are given in Fig. \ref{Vaidya_out_in} \footnote{There is a Penrose diagram similar to that in Fig. \ref{Vaidya_out_in} in \cite{Griffiths:2009dfa}, as shown in their figure 9.16 (bottom).}. The solid magenta curve represents a family of radial null geodesics which originate from the outgoing Vaidya singularity and extend to the ingoing Vaidya section without crossing the surface $w=0$, and instead continuing to $\mathscr{I^+}$. The solid green curve is unique as it is the only curve that originates from the outgoing Vaidya singularity and traces the location of the event horizon of the ingoing Vaidya section. This geodesic is also asymptotic to the ingoing Vaidya apparent horizon at $i^{+}$. Many radial null geodesics move from the surface $r_{\text{Vaidya, out}} = 0$ to the surface $r_{\text{Vaidya, in}} =0$, yet only the solid blue curve is considered. The solid brown curve is the only one which starts from $i^{-}$, combines with the outgoing Vaidya apparent horizon, and then reaches the singularity surface in the ingoing Vaidya part. The solid red curve represents the first class of radial null geodesics to come from outside the causal boundary in the outgoing Vaidya part. This class of geodesics is unable to avoid encountering the ingoing Vaidya singularity surface, $r_{\text{Vaidya, in}}=0$. At the end, we need emphasize the fact that all the maximal extensions that have been discussed in this paper were assumed to be asymptotically flat, which explains the existence of the $\mathscr{I}^+$ and $\mathscr{I}^-$ in the out(in) going Vaidya regions in the Penrose diagrams (see Figs. \ref{Vaidya_out}, \ref{Vaidya_in}, and \ref{Vaidya_out_in}). In fact, the existence of asymptotically flat regions in the Vaidya metric is not always guaranteed as the metric is more likely to be matched to other solutions that represents other regions in the spacetime. However, for the purpose of studying the various categories of radial null geodesics, we assume that the Vaidya metric becomes asymptotically flat. 
\section{summary and discussion} \label{Sec: Summary}
We have studied the issue of finding maximally extended Vaidya manifolds, leveraging the fact that it reduces to choosing a mass function that is defined along the entire range of coordinates. Consequently, we have derived a set of criteria that the mass function must satisfy in order to be a valid extension of the manifold. Utilizing these requirements, we have constructed three mass functions that give different interpretations of a maximally extended Vaidya manifold. Moreover, we have examined the qualitative features of the mass function, the $U(u)$ function, and the causality of the surfaces of constant (dynamical) radius, which is vital for gaining an understanding of the nature of crucial parts of the manifold such as the apparent horizon and the singularity hypersurface.
Since we have claimed that the Vaidya manifold in Israel coordinates is null geodesically complete, we have shown this to be true via studying the behavior of the radial null geodesics in the three constructed maximal extensions. The equation of the radial null geodesics is an Abel second type class A, and as we have already stated before that this equation does not have a general solution. We have thus developed a numerical scheme to solve this equation and obtain the radial null geodesics. The diagrams of these geodesics demonstrate that the manifold is geodesically complete as the radial, null geodesics terminate at a true singularity or continue their motion indefinitely. To further bolster the notion that Israel coordinates are globally valid, the Penrose diagrams for the three extensions were provided, with the trajectories of radial null geodesics plotted on them. Furthermore, we prove that the null junction conditions (see appendix \ref{Junction_Null}) are satisfied, and that there is no jump or discontinuity present on the joining surface $u=0$ between the outgoing Israel metric and the Schwarzschild metric when $h=+1$.
\newpage
\section{Acknowledgement}
This work was supported (in part) by a grant from the Natural Sciences and Engineering Research Council of Canada (to KL). 
\bibliography{paperII}
\appendix
\section{Null Junction Conditions}\label{Junction_Null}
 In general, if two regions $\mathcal{V}^{-}$, endowed with the metric $g^{-}_{\alpha\beta}(x^{\mu}_{-})$, and $\mathcal{V}^{+}$, endowed with the metric $g^{+}_{\alpha\beta}(x^{\mu}_{+})$, of a spacetime are joined on a null hypersurface $\Sigma$, there must be some stipulations to ascertain that the resulting metric $g^{-}_{\alpha\beta} \cup g^{+}_{\alpha\beta}$ is a valid solution to the Einstein field equations.
\subsection{The Null Junction Conditions of the First Maximal Extension} 
 In this appendix, the null junction conditions shall be developed \footnote{Here, we follow the same presentation given in \cite{Poisson2004}, which is in turn an adaptation of the Barrabes-Israel formulation \cite{PhysRevD.43.1129}.} for the case $h=+1$ and $U(u)<0$. In the course of our development of these conditions the metric in $\mathcal{V}^{-}$ is taken to be the outgoing Vaidya metric, while the metric outside $\mathcal{V}^{+}$ is given by the Schwarzschild metric. As seen either from $\mathcal{V}^{-}$ or from $\mathcal{V}^{+}$, the null hypersurface $\Sigma$ is given by the equation $u=0$. Thus, the intrinsic metric to $\Sigma$ is given by $ds^2_{\Sigma} =r(u,w)^2d\Omega^{2}_{2} $. Given that $\theta^{A} = (\theta,\phi)$ and that the null generators of $\Sigma$ are affinely parametrized by $w$, $w = \lambda$, the intrinsic metric, on both sides, now assumes the form
\begin{equation}
\sigma_{AB} d\theta^{A}d\theta^{B} = 4M_{1}^{2}(d\theta^{2}+\sin^{2}\theta d\phi^{2}).
\end{equation}
As seen from $\mathcal{V}^{-}$, the null hypersurface $\Sigma$ is  explicitly given by $u = 0$, $w=\lambda$, $\theta = \theta$, and $\phi = \phi$. This gives rise to the tangent vectors $k^{\alpha}\partial_{\alpha} = \partial_{w}$, $e^{\alpha}_{\theta} \partial_{\alpha}= \partial_{\theta}$, and $e^{\alpha}_{\phi} \partial_{\alpha}= \partial_{\phi}$, and the basis is completed by $N_{\alpha}dx^{\alpha} = -\frac{g_{uu,Vaidya}}{2} du-dw$. Based on this setup, the vanishing component of the transverse curvature tensor \cite{Poisson2004} is 
\begin{equation}
C^{-}_{\lambda \lambda} = 0,
\end{equation}
in accordance with the fact that $\lambda = w$ is an affine parameter on the $\mathcal{V}^{-}$ side of $\Sigma$. The non-vanishing components are given as
\begin{equation}
C^{-}_{AB} = \frac{-w}{8 M_{1}^{2}}\sigma_{AB}.
\end{equation}
As seen from $\mathcal{V}^{+}$, the parametric equations are  $u = 0$, $w=\lambda$, $\theta = \theta$, and $\phi = \phi$. The basis vectors are $k^{\alpha}\partial_{\alpha} = \partial_{w}$, $e^{\alpha}_{\theta} \partial_{\alpha}= \partial_{\theta}$, and $e^{\alpha}_{\phi} \partial_{\alpha}= \partial_{\phi}$, and $N_{\alpha}dx^{\alpha} = -\frac{w^2}{uw+8M_1^2} du-dw$. The components of $C^{+}_{AB}$ are found to be identical to those of $C^{-}_{AB}$. The fact that $C^{+}_{\lambda\lambda} =0$ completes the argument that $\lambda$ is an affine parameter on both sides of $\Sigma$. The angular components of the transverse curvature are found to be continuous across $\Sigma$. Thus, we conclude that there is no jump across the surface $u=0$, and the resulting metric from joining the outgoing Vaidya metric to the Schwarzschild metric is a true solution to the field equations. Finally, we would like to emphasize that the null junction conditions for the other two extensions can be derived in a similar manner to the one presented here.  
\end{document}